\documentclass[11pt,a4paper,twoside, openright]{book} 

\usepackage{natbib}
\usepackage[swedish,english]{babel} 
\usepackage[applemac]{inputenc} 
\usepackage[T1]{fontenc}
\usepackage{mathptmx} 
\usepackage{helvet}

\usepackage{ThesisSU} 
\ifpdf
   \usepackage[pdftex]{graphicx}
  \else
    \usepackage[dvips]{graphicx}
\fi 

\bibpunct{(}{)}{,}{a}{}{;}
\usepackage{amsmath, amssymb}
\usepackage{fancyheadings}                      

\newcommand{\wll}{$\lambda \lambda$}
\newcommand{\wl}{$\lambda$}

\def\kms{\hbox{km\,s$^{-1}$}}
\def\msun{\hbox{M$_\odot$}}

\DeclareMathAlphabet{\mathsc}{OT1}{cmr}{m}{sc}
\def\testbx{bx}%
\DeclareRobustCommand{\ion}[2]{%
\relax\ifmmode
\ifx\testbx\f@series
{\mathbf{#1\,\mathsc{#2}}}\else
{\mathrm{#1\,\mathsc{#2}}}\fi
\else\textup{#1\,{\mdseries\textsc{#2}}}%
\fi}

\newcommand{\topTextDoc}{Doctoral Dissertation 2011}
\newcommand{\myInst}{Department of Astronomy}
\newcommand{\myInstB}{Stockholm University}
\newcommand{\instAdress}{SE-106 91 Stockholm}

\newcommand{\authorSurname}{Jerkstrand} 
\newcommand{\authorFirstName}{Anders} 
\newcommand{\dissertationTitle}{Spectral modeling of nebular-phase supernovae} 
\newcommand{\dissertationSubtitle}{}
\newcommand{\yearOfPublication}{2011} 
\newcommand{\placeOfPublication}{Stockholm} 
\newcommand{\ISBN}{978-91-7447-407-7} 
\newcommand{\distributor}{Department of Astronomy, Stockholm University} 

\newcommand{\printer}{US-AB}  
\newcommand{\printingdate}{Stockholm 2011} 
\newcommand{\iso}[1]{${}^{#1}$}
\newcommand{\e}[1]{\times 10^{#1}}
\newcommand\abstracttext{%
\noindent
\section*{Abstract}

Massive stars live fast and die young. They shine furiously for a few million years, during which time they synthesize most of the heavy elements in the universe in their cores. They end by blowing themselves up in a powerful explosion known as a supernova (SN). During this process, the core collapses to a neutron star or a black hole, while the outer layers are expelled with velocities of thousands of kilometers per second. The resulting fireworks often outshine the entire host galaxy for many weeks.

The explosion energy is eventually radiated away, but powering of the newborn nebula continues by radioactive isotopes synthesized in the explosion. The ejecta are now quite transparent, and we can see the material produced in the deep interiors of the star. To interpret the observations, detailed spectral modeling is needed. This thesis aims to develop and apply state-of-the-art computational tools for interpreting and modeling SN observations in the nebular phase. This requires calculation of the physical conditions throughout the nebula, including non-thermal processes from the radioactivity, thermal and statistical equilibrium, as well as radiative transport. The inclusion of multi-line radiative transfer, which we compute with a Monte Carlo technique, represents one of the major advancements presented in this thesis.

On February 23 1987, the first SN observable by the naked eye since 1604 exploded, SN 1987A. Its proximity has allowed unprecedented observations, which in turn have lead to significant advancements in our understanding of SN explosions. 
As a first application of our model, we analyze the \iso{44}Ti-powered phase  (t $\gtrsim 5$ years) of SN 1987A.
We find that a magnetic field is present in the nebula, trapping the positrons that provide the energy input, and resulting in strong iron lines in the spectrum. We determine the \iso{44}Ti mass to $1.5_{-0.5}^{+0.5}\e{-4}$ M$_\odot$. From the near-infrared spectrum at an age of 19 years, we identify strong emission lines from explosively synthesized metals such as silicon, calcium, and iron. We use integral-field spectroscopy to construct three-dimensional maps of the ejecta, showing a morphology suggesting an asymmetric explosion.  

The model is then applied to the close-by and well-observed Type IIP SN 2004et, analyzing its ultraviolet to mid-infrared evolution. Based on its Mg I] 4571 \AA, Na I 5890, 5896 \AA, [O I] 6300, 6364 \AA, and [Ne II] 12.81 $\mu$m nebular emission lines, we determine its progenitor mass to be around $15$ M$_\odot$. We confirm that silicate dust, SiO, and CO have formed in the ejecta. 

Finally, the major optical emission lines in a sample of Type IIP SNe are analyzed. We find that most spectral regions in Type IIP SNe are dominated by emission from the massive hydrogen envelope, which explains the relatively small variation seen in the sample. We also show that the similar line profiles seen from all elements suggest extensive mixing occurring in most hydrogen-rich SNe. 

} 

\newcommand\abstracttextSpikblad{%
\noindent
\textbf{Abstract:}
} 

\renewcommand{\listofpapers}
{\chapter*{List of Papers}
\noindent{\timesTen This thesis is based on the following publications:

} \vspace{13pt}
    \begin{romanlist}
\item {\bf The 3-D structure of SN 1987A's inner ejecta} \\
	Kjaer~K., Leibundgut~B., Fransson~C., Jerkstrand~A., Spyromilio, J., 2010, {\em A\&A}, 517, 51 \\
\vspace{3mm}
\item {\bf The \iso{44}Ti-powered spectrum of SN 1987A} \\
	Jerkstrand~A., Fransson~C., Kozma~C., 2011, {\em A\&A}, 530, 45 \\
\vspace{3mm}
\item {\bf The progenitor mass of the Type IIP supernova 2004et from late-time spectral modeling} \\
        Jerkstrand~A., Fransson~C., Maguire~K., Smartt~S., Ergon~M., Spyromilio~J., 2011, To be submitted to {\em A\&A}. \\
\vspace{3mm}
\item {\bf Constraining the properties of Type IIP supernovae using nebular-phase spectra} \\
	Maguire~K., Jerkstrand~A., Smartt~S., Fransson~C., Pastorello~A., Benetti~S., Valenti~S., Bufano~F., Leloudas~G., 2011, Accepted for publication in {\em MNRAS}.\\
  \end{romanlist}
\vspace{13pt}
\noindent {\timesTen The articles are referred to in the text by their Roman numerals.}}

\newcommand{\dedication}%
{\newpage
\thispagestyle{empty}
\vspace*{\stretch{3}}
\begin{flushright}		
		{\fontfamily{pzc}\Large\selectfont
        \emph{Till farfar\\
        }}

\end{flushright}
\vspace*{\stretch{1}}} 

\begin{document}

\thispagestyle{empty}
\frontmatterSU
\thispagestyle{empty}
\listofpapers
\thispagestyle{empty}

\tableofcontents

\mainmatter
\chapter{Introduction}

\begin{figure}[htb]
\centering
\includegraphics[width=0.8\linewidth]{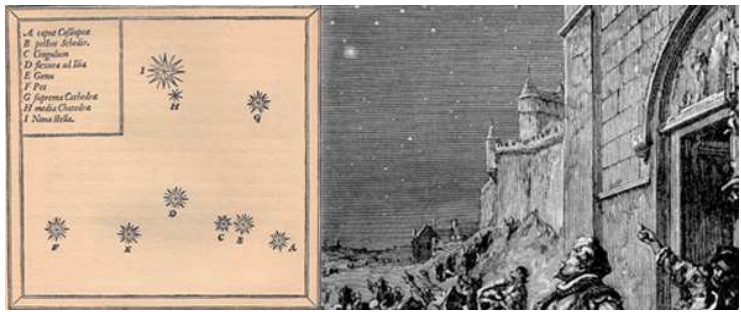}
\caption{The great Tycho supernova of 1572. Drawing from 'Astronomie Populaire', by C. Flammarion.}
\label{fig:tycho}
\end{figure}

Supernovae (SNe) are explosions of stars that have ceased their fusion processes. They represent astrophysical laboratories where a large and diverse set of physical processes take place. By studying them, we can learn about the formation of neutron stars and black holes, test theories of stellar evolution and nucleosynthesis, and understand the chemical enrichment histories of galaxies. We can determine their contribution to the dust and cosmic rays in the universe, and refine their use as standard candles for cosmological distance measurements.

SNe emit radiation at all wavelengths, from radio to gamma-rays. The diverse set of observations is matched by an equally diverse set of models and theories needed for their interpretation. Initially, the expanding fireball is hot and opaque, and models for this phase use radiation hydrodynamics to simulate the coupled evolution of radiation and matter, providing information on the explosion energy, the ejecta mass, and the radius of the progenitor. Scattering models may be applied to the outer atmospheric layers to determine their density profiles and composition. Later on, as the SN expands and cools, its inner parts become visible, glowing from input by radioactive elements produced in the explosion. The matter is now further from thermodynamic equilibrium, and models have to consider a large set of physical processes to compute the state of the gas. In this phase, we can determine the morphology of the SN ejecta, the amount of elements produced by the nucleosynthesis, the amount of mixing and fragmentation that has occured in the explosion, and possibly observe signatures of the compact object formed at the center.

This thesis deals with modeling of such late phases. The research field is still in its infancy, as SNe are dim and hard to observe at late times. For many years, the early phases of the nearby SN 1987A have been the only observations studied in depth. But due to recent advancements in observational surveys, the number of SNe with spectral data in the later phases is increasing. Combined with developments in atomic data calculations and easy-access parallel computing, the field is moving rapidly forward. We are now coming to a point where accurate spectra from realistic explosion models can be computed and compared with an increasing number of well-observed events, allowing constraints to be placed on which type of stars explode as which type of SNe. In the coming years, the development of several robotic wide-field surveys will further increase the inflow of data, and theorists will have their hands full to analyze them.

The specific aims of this thesis have been to develop state-of-the-art computational models for nebular-phase SNe, and apply these models to derive properties of hydrogen-rich core-collapse SNe. We initially investigate the very late phases of the famous SN 1987A, which has mainly been analysed in its early phases before. We then investigate a recently obtained sample of other Type IIP spectra, with special focus on the interesting object SN 2004et. 

To elucidate the nature of SNe, we have to understand several subjects in physics and astronomy. Chapter 2 deals with the evolution of the massive stars that are their progenitors, providing the underpinning for what SN ejecta actually contain. While the study of SN may eventually come to \emph{revise} stellar evolution theory, these are the current ideas and the perceived uncertainties in them. Chapter 3 provides a broad overview of SNe, including a historical perspective and observational aspects such as classification. Chapters 4 and 5 contain a rather detailed description of the fundamental theory incorporated into the computational model developed. In Chapter 4, I describe how the physical conditions in the SN are determined by a balance between radioactive, thermal, and radiative processes. Chapter 5 describes the treatment of the radiation field, which represents a major advancement in the modeling compared to earlier models developed in Stockholm and elsewhere. Finally, Chapter 6 provides a summary of the results that we have obtained by application of the model.


Most things in the universe evolve \emph{slowly}. SNe are different, providing entertainment on a time-scale more
suitable for the human mind. They evolve into something new in about the same time it takes you to write a paper about what they were before. I hope to have captured some of their fascination in this thesis, and that you will enjoy reading it.\\
\\
A.R.J.\\
Stockholm, November 1, 2011

\chapter{Evolution of massive stars}
\label{SE}
\begin{centering}
\small{
\emph{'Astronomy? Impossible to understand and madness to investigate!'}}\\
\end{centering}
\begin{flushright}
\small{\emph{Sophocles, 420 B.C.}}\\
\end{flushright}
Stars are gravitationally confined fusion reactors. Inside them, light elements are converted into heavier ones, releasing energy that provides pressure support against the gravity and makes the star shine. Low-mass stars like the sun take a few billion years to deplete their supply of hydrogen, after which gravitational contraction occurs until helium fusion begins. During this burning process, helium is converted to oxygen and carbon. In the next contraction stage, the densities become so high that the star stabilizes due to electron degeneracy pressure, and a \emph{white dwarf} is formed.

\begin{figure}[htb]
\centering
\includegraphics[width=0.5\linewidth]{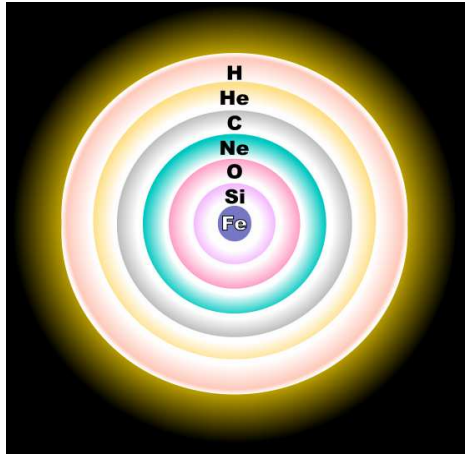}
\caption{Schematic final structure of a massive star that has synthesized elements up to iron (Fe).}
\label{fig:onion}
\end{figure}

Massive stars ($M\gtrsim 8~$\msun), on the other hand, never reach that stabilizing point. They ignite also their central supplies of carbon and oxygen, and after that also heavier elements until the core has been transformed to iron. The final structure of the star is onion-like, with layers made up by the ashes of the various burning stages (Fig. \ref{fig:onion}). In this chapter, I review the major aspects of the lives of massive stars up until the formation of the iron core. Specific stellar evolution models (\citet[][H89]{Hashimoto1989}, \citet[][WW95]{WW95}, \citet[][T96]{Thielemann1996}, \citet[][H04]{Hirschi2004}, \citet[][N97]{Nomoto1997}, \citet[][WHW02]{Woosley2002}, \citet[][WH07]{Woosley2007}) are occasionally referenced.

\section{Fusion}
\label{sec:fusion}
Fusion reactions are mediated by the nuclear force, which has an effective range of $R\sim 10^{-13}$ cm. Nuclei must therefore come within this distance to each other for reactions to occur. Such close encounters are opposed by the repulsive electric force between the positively charged nuclei. We may attempt to estimate the critical temperature required for fusion to occur by equating the kinetic energy to the electrostatic potential energy, and solving for $T$:
\begin{equation}
E_{kin} \sim E_{pot} \rightarrow k_BT \sim \frac{Z^2q^2}{R}~,
\end{equation}
where $Z$ is the proton number, $q$ is the charge unit, and $k_B$ is Boltzmann's constant. 
For hydrogen ($Z=1$), we obtain $T\sim 10^{10}$ K. At the beginning of the 20th century, calculations showed that stars were not this hot inside (the sun is only $\sim 10^7$  K at its center), and fusion was therefore rejected as their power source. However, two effects make fusion effective even at $\sim 10^3$  times lower temperature; some nuclei have much higher kinetic energies than $kT$, and secondly quantum mechanical tunneling allows reactions to occur also for sub-energetic particles. It was therefore not until the 1930s, when quantum mechanics was thoroughly understood, that fusion was established as the power source of stars.

\subsection{Hydrogen burning}
\label{sec:Hburn}
The net effect of hydrogen fusion, starting at $T\sim 10^7$ K, is
\begin{equation}
4~\mbox{\iso{1}H} \rightarrow \mbox{\iso{4}He}~.
\end{equation}
In addition, two positrons and two neutrinos are created. The path taken is not four protons simultaneously colliding into one helium nucleus, but rather a sequence of successive buildups. In massive stars, the dominant path is the \emph{CNO cycle}, in which primordial abundances of carbon, nitrogen, and oxygen are used as catalysts. The main branch is
\begin{equation}
\mbox{\iso{12}C(p,$\gamma$)\iso{13}N(e$^+$$\nu_e$)\iso{13}C(p,$\gamma$)\iso{14}N(p,$\gamma$)\iso{15}O(e$^+$$\nu_e$)\iso{15}N(p,\iso{4}He)\iso{12}C}~.
\end{equation}
For a complete cycle, \iso{12}C is recycled in the last step. However, while the cycle is running, the abundances of the involved nuclei are altered. The reason is that the various reactions occur at different rates, with
the $^{14}\mbox{N}(\mbox{p},\gamma)^{15}\mbox{O}$ reaction being significantly slower than the other ones. In equilibrium, the number of $^{14}\mbox{N}$ nuclei is therefore enhanced and the number of $^{12}\mbox{C}$ nuclei is suppressed. Another reaction occurring is $^{16}\mbox{O}(p,\gamma)^{17}\mbox{F}(e^+\nu_e)^{17}\mbox{O}(p,^4 \mbox{He})^{14}\mbox{N}$, which converts $^{16}\mbox{O}$ to $^{14}\mbox{N}$. The net result is that the carbon and oxygen abundances are suppressed during CNO burning, whereas the nitrogen abundance is enhanced. These shifts in abundances are retained when fusion reactions eventually cease.

Table \ref{table:Hburn} shows the composition of the H-burning ashes from a few different models. We can see the alteration of the original C, N, and O abundances. This alteration explains why oxygen can be a coolant in the outer (unburnt) hydrogen envelopes of SNe, but not in the He zones (\textbf{paper IV}).

\begin{table}[htb]
\centering
\begin{tabular}{ c|c|c|c}
\hline
\emph{El.} & \emph{WH07-15} & \emph{WHW02-15} & \emph{H04-20} \\
\hline
He & 0.98  & 0.98 & 0.98 \\
C &  $1.9\e{-4}$ (0.08) & n/a & $2.5\e{-4}$ (0.10) \\
N &  $9.0\e{-3}$ (11) &  0.011 (14) & n/a \\
O & $1.8\e{-4}$ (0.03) & n/a & $3.0\e{-4}$ (0.05)\\
H & $\sim 10^{-8}$ & n/a & n/a\\
\hline
\end{tabular}
\caption{Composition (mass fractions) of the hydrogen-burning ashes from a few stellar evolution models. The mass of the star is given after the reference. The change relative to solar abundances for the CNO elements are given in parenthesis. All other elements have their natal mass fractions.}
\label{table:Hburn}
\end{table}

During the hydrogen-burning phase, the star is said to be on the \emph{main sequence}. Massive stars will belong to classes O or B, with surface temperatures of $20,000-40,000$ K, and luminosities of $10^3-10^5$ L$_\odot$. The time-scale for this phase is about 
\begin{equation}
t_{\rm MS} \sim 40 \left(\frac{M}{10~\msun}\right)^{-2.5} \mbox{million~years}~.
\end{equation}

\subsection{Helium burning}
\label{sec:heburn}
As hydrogen is eventually exhausted in the core, the star contracts and heats up.
At $T\sim 10^8$ K, helium fusion begins, occurring by the triple-$\alpha$ reaction,
\begin{equation}
3~^4\mbox{He} \rightarrow \mbox{\iso{12}C}~,
\end{equation}
and also by
\begin{equation}
\mbox{\iso{12}C} + \mbox{\iso{4}He} \rightarrow \mbox{\iso{16}O}~,
\end{equation}
which explains why \iso{12}C and \iso{16}O are the major carbon and oxygen isotopes on earth and in the universe. 
The rate of the last reaction is still only known within a factor two or so, and is one of the major uncertainties in stellar evolution models, both for the carbon/oxygen yields, and for the yields of heavier nuclei produced from them in later burning stages (WHW02, WH07). Table \ref{table:oc} shows the composition of the \emph{C/O zone} formed by helium burning from a few different models. It is from this zone we can see emission from molecular CO in SNe (\textbf{paper III}).

\begin{table}[htb]
\centering
\begin{tabular}{c|c|c|c}
\hline
\emph{El.} & \emph{H89-20} & \emph{WH07-12} & \emph{H04-20}\\
\hline
O & 0.76 & 0.79 & 0.65 \\
C & 0.21 & 0.18 & 0.30\\
Ne & 0.015 & 0.011 & $5.0\e{-3}$\\
\hline
\end{tabular}
\caption{Composition (mass fractions) of the helium-burning ashes, taken from a few stellar evolution models (only the three most common elements are listed). The mass of the star is given after the reference.}
\label{table:oc}
\end{table}

During the helium-burning phase, the stellar envelope expands and cools, and the star becomes a \emph{red supergiant (RSG)}, with photospheric temperature on the order of a few thousand degrees.

\subsection{Carbon burning}
At $T\gtrsim 8\e{8}$ K, carbon nuclei fuse as
\begin{eqnarray}
2~\mbox{\iso{12}C}   &\rightarrow & \mbox{\iso{24}Mg}\\
                     &\rightarrow & \mbox{\iso{20}Ne} + \alpha\\
                     & \rightarrow & \mbox{\iso{23}Na} + \mbox{p}\\
                     & \rightarrow & \mbox{\iso{23}Mg} + \mbox{n}~.
\end{eqnarray}
The neutrons, protons, and $\alpha$-particles will cause further reactions, resulting in a large set of nuclides. The main products are neon and magnesium, complemented by some carbon, sodium, silicon, and aluminium. The zone is referred to as the \emph{O/Ne/Mg zone} later in the thesis, and we shall see in \textbf{paper III} how this zone is key for linking SNe to their progenitors. 

Table \ref{table:onemg} shows the composition of this zone from a few different models. A major down-revision in modern calculations is the amount of silicon produced in this zone (compare the H89 value with the other ones). Since silicon is an efficient coolant \citep[see e.g][]{Kozma1998I}, this has major impact on the thermal emission from this zone. One may also note that this is the main production site of neon and sodium in the universe, so one may compare the solar ratio of 33 \citep{Lodders2003} to the model outputs.  

From the carbon burning stage and on, the burning time-scales shorten significantly compared to the hydrogen and helium-burning time-scales. The reason is that internal energy is now efficiently converted to neutrinos that escape freely.

\begin{table}[htb]
\centering
\begin{tabular}{ c|c|c|c|c}
\hline
\emph{El} & \emph{H89-20}     &  \emph{WH07-12}    & \emph{H04-20} &  \emph{N97-18}\\
\hline
O  &  0.73         & 0.66          & 0.58       &   0.66\\ 
Ne & 0.12         & 0.25          & 0.31       &  0.21\\
Mg & 0.075        & 0.043         & n/a       &  0.037\\
Na & $7.0\e{-4}$ & $7.7\e{-3}$  & n/a       &  0.010 \\
C  & $5.7\e{-3}$  & $3.9\e{-3}$ & $4.5\e{-3}$      &  0.045\\
Al & $8.0\e{-3}$  & $4.0\e{-3}$ & n/a      & $5.0\e{-3}$\\
Si & 0.020  & $3.0\e{-3}$     & $4.0\e{-3}$    &   $2.5\e{-3}$\\
\hline
\end{tabular}
\caption{Composition (mass fractions) of the carbon-burning ashes, from a few stellar evolution models (only the three most common elements are listed). The mass of the star is given after the reference.}
\label{table:onemg}
\end{table}

\subsection{Neon burning}
Before the oxygen ignition temperature is reached, neon nuclei start photo-disintegrating to oxygen, as the photon energies are now in the MeV range. The released $\alpha$-particles are captured by other neon nuclei to make magnesium, silicon, and sulphur.
The reactions begin at $T\sim 1.7\e{9}$ K, and can be summarized as
\begin{eqnarray}
\mbox{\iso{20}Ne} + \gamma &\rightarrow& \mbox{\iso{16}O} + \alpha\\
\mbox{\iso{20}Ne} + \alpha &\rightarrow& \mbox{\iso{24}Mg}\\
\mbox{\iso{24}Mg} + \alpha &\rightarrow& \mbox{\iso{28}Si}\\
\mbox{\iso{28}Si} + \alpha &\rightarrow& \mbox{\iso{32}S}~.\\
\end{eqnarray}
The net effect is that two neon nuclei are converted to one oxygen nucleus and one magnesium nucleus, with some of the magnesium nuclei being further converted to silicon and sulphur. The composition after this burning stage is mainly oxygen, silicon, sulphur, and some magnesium, and is referred to as the \emph{O/Si/S zone} in the thesis. Table \ref{table:osis} shows the composition in this zone from a few models.


\begin{table}[htb]
\centering
\begin{tabular}{c|c|c|c}
\hline
\emph{El.} & \emph{N97-18} & \emph{H04-20} & \emph{WH07-15} \\
\hline
O  & 0.80 & 0.73 & 0.82 \\
Si & 0.10  & 0.08 & 0.10\\
S & 0.025  & 0.01 & 0.015 \\
Mg & 0.032 & n/a & 0.044\\
\hline
\end{tabular}
\caption{Composition (mass fractions) of the neon-burning ashes, from a few stellar evolution models (only the three most common elements are listed). The mass of the star is given after the reference.}
\label{table:osis}
\end{table}

\subsection{Oxygen burning}
\label{sec:oxburn}
At $T \sim 2.1\e{9}$ K, oxygen burns as
\begin{eqnarray}
2~\mbox{\iso{16}O} & \rightarrow &\mbox{\iso{32}S}\\
          &   \rightarrow     &\mbox{\iso{31}S} + \mbox{n}\\
          &   \rightarrow     & \mbox{\iso{31}P} + \mbox{p}\\
          &    \rightarrow         & \mbox{\iso{28}Si} + \alpha\\
          &    \rightarrow       & \mbox{\iso{30}P} + \mbox{d}~.      
\end{eqnarray}
The main products are silicon and sulphur (an \emph{Si/S zone}), with traces of chlorine, argon, potassium, calcium and phosphorus. From this stage on, the burning ashes will be further modified in the eventual SN explosion, so I do not list the chemical compositions. The nucleosynthesis abundances of these ashes are strongly non-solar, and the fact that they are not observed supports the previous statement that these ashes will be further modified before ejection \citep{Woosley1972}. 


\subsection{Silicon and sulphur burning}
\label{sec:ssfusion}
After oxygen burning is complete, all further reactions occur by photodisintegrations followed by $\alpha$-captures (similar to the neon burning). 
Above $T \sim 3.7\e{9}$ K, silicon and sulphur nuclei fuse with $\alpha$-particles to produce iron-group nuclei through steps of \iso{36}Ar, \iso{40}Ca, \iso{44}Ti, \iso{48}Cr, \iso{52}Fe and \iso{56}Ni. The burning occurs under conditions close to \emph{nuclear statistical equilibrium (NSE)}, where all electromagnetic and strong reactions are in detailed balance. The main isotope produced under NSE is the one that is most tightly bound for the given neutron excess\footnote{The neutron excess is the number of neutrons minus the number of protons, divided by the sum of them.}. Typically, the neutron excess is close to zero, and that nucleus is then \iso{56}Ni.


\subsection{Neutron and proton capture}
During the star's evolution, nucleosynthesis occurs partly via the \emph{slow} (s) process, in which nuclei capture free neutrons. By a series of such captures, elements more massive than \iso{56}Ni can also be made. The neutrons are mainly produced by successive $\alpha$-captures on \iso{14}N nuclei, and the s-process occurs mainly at the end of helium burning (WHW02). 

Protons are less likely to fuse with nuclei due to the electric repulsion. However, some important reactions still occur, one example being proton capture by \iso{21}Ne to produce \iso{22}Na. This isotope is radioactive with $\tau=3.75$ years, and serves as a source of radioactivity in the O/Ne/Mg zone (\textbf{paper II}). Another example is proton capture by \iso{25}Mg to produce \iso{26}Al. 

An important consequence of the neutron and proton captures is that the abundances of elements not participating in the main fusion cycles themselves may be altered from their natal values. For example, in the O/Ne/Mg zone the nickel and cobalt abundances are strongly enhanced, with significant amounts of the radioactive isotope \iso{60}Co produced (\textbf{paper II}).

\section{Mass loss}
\label{sec:massloss}
Observations of massive stars indicate that they are losing material from their surfaces \citep[e.g.,][]{Wright1975, Lamers1981}. Their optical spectra often show \emph{P-Cygni profiles} (blueshifted absorption and redshifted emission), which arise only from material in rapidly outflowing motion. From the line profiles, gas velocities higher than the escape velocity of the star can be inferred. Another method to detect the outflowing gas is by its radio free-free emission, or by far-infrared (FIR) emission from dust forming in the outflow. 

By studying a large sample of stars, the dependency of the mass-loss rate on stellar parameters such as luminosity, temperature, and metallicity can be determined. Such relations have been derived both for main sequence stars and for RSGs. A commonly used relation is the one by \citet{Nieu1990}. A high dust production efficiency, combined with low surface gravities, makes mass loss rates higher for RSGs than for main sequence stars.

Theoretically, mass loss on the main sequence is understood as radiation pressure acting on atomic lines. This idea was first proposed by \citet{Lucy1970}, and subsequently quantitative models have been developed that accurately reproduce the observed mass loss rates. The importance of metallicity is obvious for this type of mass loss.

In the RSG phase, understanding of the mass loss mechanism is poorer. Here, radiation pressure probably acts on dust grains forming in the cool envelope. It is also possible that large-scale convective motions and pulsations play a role, as recent imaging of Betelgeuse suggests \citep{Kervella2011}. Self-consistent modeling is challenging, and existing models still have to make a large number of assumptions.



More massive stars lose more mass, in total, than less massive ones \citep{Nieu1990}. Models suggest that the final envelope mass is therefore quite insensitive to the ZAMS mass of the star, being $6-10$ $M_\odot$ for non-rotating stars below 25 M$_\odot$, and steadily smaller for more massive ones (H04, WH07). Rotation increases the mass loss rate, and always leads to a smaller final mass (H04). 

For stars more massive than $\sim$35 \msun, also the helium core may be affected by mass loss (WH07). 
A 100 \msun~star ends up with a final mass of only $\sim$6 \msun, with no hydrogen or helium. 
In the models of WH07, the helium core reaches its maximum size (13 $M_\odot$) for a 45 \msun~star. We therefore find the important result that most types of (single) massive stars reach final masses of $5-15$ \msun~at the end of their evolution.

\section{Convection}
\label{sec:convection}
If stars were stable objects, it would be relatively easy to compute their evolution through the different burning stages. Unfortunately, they are usually not. \emph{Convection} refers to the turbulent mass motions that represent the second way for the star to transport energy besides radiation. A detailed theory is so far not available, because the solutions to the equations of hydrodynamics become chaotic for turbulent flows, preventing meaningful results to be obtained. However, phenomenological models such as the \emph{mixing-length theory} are available which are believed give reasonable results. 

Convection transports energy and alters the chemical composition by mixing gas between different layers. H-burning ashes can be convectively dredged up into the hydrogen envelope, altering the fractions of helium and CNO elements (Sect. \ref{sec:Hburn}). 
Similarly, He-burning ashes can be dredged up into the He zone, altering its content of oxygen, carbon, and neon (Sect. \ref{sec:heburn}). This mixing typically does not reach all the way through the He layer, but produces a \emph{He/C} zone in its lower parts. In \textbf{paper IV}, we see a strong [C I] 8727 \AA~line in the spectrum of SN 2008bk, that appears to arise from such a dredged-up layer.


Large uncertainties currently exist in the treatment of convection. Two different stability criteria for convection may be derived, the \emph{Schwarzschild} criterion and the \emph{Ledoux} criterion. These are identical for a perfect gas of constant composition. But if radiation pressure is important, or if there is a composition gradient, the Schwarzschild criterion may predict instability whereas the Ledoux criterion may not. In this situation, the region is said to be \emph{semi-convective}. It is not known exactly how the gas behaves in this situation. \citet{WW88} allow diffusion for the composition, but not for the thermal energy, a recipe followed in many models also today. 

Another source of uncertainty regards the so called \emph{convective over-shooting}. The stability criteria define exact radii in the star where the convection-driving buoyancy force disappears. But when the convective bubbles reach these points, they have non-zero velocities and will continue a bit passed the limit. The treatment of this over-shooting critically influences all aspects of the evolution of the star \citep{Schaller1992}. Strong over-shooting leads to a more massive core, a faster burning rate, and a higher luminosity \citep{Mowlavi1994}. Unfortunately, the predicted extent of over-shooting ranges from being neglegible to almost doubling the size of the convective regions.

\section{Rotation}
\label{sec:rotation}

It is well known that massive stars on the main sequence rotate rapidly (WHW02). \citet{Fukuda1982} find typical rotation velocities of 200 \kms. Rotation affects the star mainly during its hydrogen and helium burning phases, with the later burning stages transpiring too quickly for the rotational instabilities to have any impact (H04). Rotation leads to larger helium cores for a given stellar mass and metallicity, caused by an enhanced mixing of fuel into the nuclear burning core. For example, a 15 \msun~star has a helium core of 3.0 \msun~in the non-rotating models of WHW02, but of 4.9 \msun~in the rotating models of \citet{Heger2000}. For a 20 \msun~star, the numbers are 5.0 \msun~and 7.8 \msun. Rotation can also induce outward mixing of burning ashes to the surface, and increase the mass loss rate both by the centrifugal effect and by increasing the core luminosity (WHW02).

All models used in this thesis are non-rotating ones. When we connect the amount of metals observed in the SN to the progenitor mass of the star (\textbf{Paper III}), we therefore actually derive an upper limit. We do know, however, that rotation does \emph{not} have any strong influence on the chemical composition of the nuclear burning ashes (H04), so the metal line ratios will not change much by computing spectra from rotating progenitors. 

We may finally note that rotation influences the final point in the HR diagram. H04 finds the SN progenitor of a  20 \msun~star to be red, yellow, and blue, for rotation velocities of 100, 200, and 300 \kms, respectively.

\section{The final structure of the star}
An important property of the sequence of fusion reactions described in Sect. \ref{sec:fusion} is that each ash is only \emph{partly} burned in the next burning stage. The star is therefore develops an onion-like structure, with successively heavier elements further in.
The details of the final structure depends on the specific prescription used for rotation, convection, mass loss, and uncertain reaction rates. Currently, the uncertainties associated with convection (Sect. \ref{sec:convection}) is the largest sources of diversity in stellar evolution models (WHW02). 

\begin{figure}[h!]
\centering
\includegraphics[width=1\linewidth]{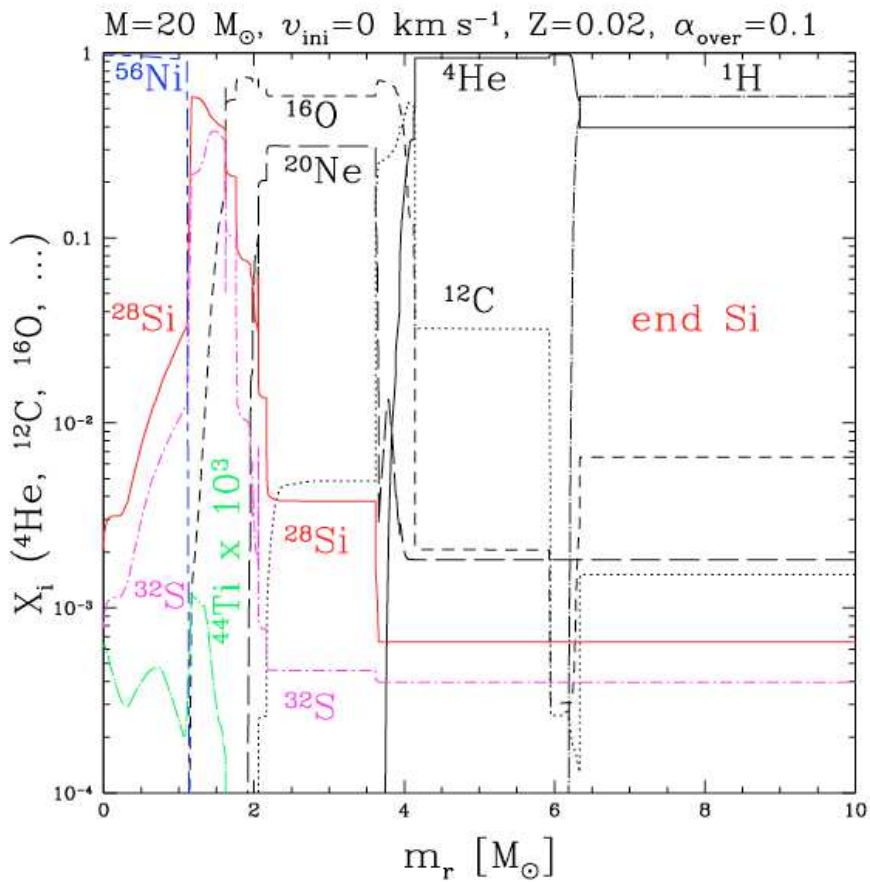}  
\caption{Final structure (mass fractions) of a 20 \msun~non-rotating star. From \citet{Hirschi2004}.}
\label{w02}
\end{figure}

Fig. \ref{w02} shows the final structure of a 20 \msun~non-rotating star as computed by H04. Note the stratification arising from the nuclear burning sequence H$\rightarrow$ He, He$\rightarrow$C/O, C$\rightarrow$O/Ne/Mg, Ne$\rightarrow$ O/Si/S, O$\rightarrow$ Si/S, and Si$\rightarrow$ \iso{56}Ni, modified by convective dredge-up of carbon into the He zone, and helium into the H envelope.


\section{Concluding remarks}
Today, we know that all atomic nuclei in the universe, apart from hydrogen, helium, lithium, beryllium, and boron, are created inside stars. Up until 1957, this was only one of four theories being considered for the origin of the elements. 
By the key role played by helium nuclei in nucleosynthesis (Sect. \ref{sec:fusion}), we have already come a long way to understand the large abundances of elements with even proton numbers in the universe.

In Chapter 2, I describe how the elements produced are ejected into the interstellar medium, to eventually become building blocks for new stars, planets, and humans.

\chapter{Supernovae}
\begin{centering}
\small{
\emph{'In the second year of the Chung-phing reign period, the 10th month, on a Keui-day (December 7, AD 185), a guest star appeared in the midst of the Southern Gate. It seemed to be as large as half a bamboo mat, and showed five colours in turn, now brightening, now dimming.  It diminished in brightness little by little and finally disappeared in the sixth month of the Hou-year (Hou-nian, 24 July to 23 August~AD~187).'}}\\
\end{centering}
\begin{flushright}
\small{\emph{Astrological Annals of the Hou Han Shu (China)}}\\
\end{flushright}
Before the invention of the telescope at the beginning of the 17th century, not much seemed to happen in the night sky. Only two types of objects provided the occasional drama; comets and \emph{guest stars}. 

Historical records of guest stars date almost two millenia back in time. The oldest one is from year 185, when Chinese astronomers recorded a bright new star that visited the night sky for a few weeks (the excerpt above). Similar events happened in years 386 and 393. 
The next records are from Chinese and Arabic astronomers in 1006 and 1054. These new stars were taken to convey important messages, surely altering the fates of many individuals and maybe even civilizations. 

Today, we know that these guest stars are \emph{novae} and \emph{supernovae (SN)}. The Swedish astronomer Knut Lundmark established, in 1920, that SNe are about a thousand times intrinsically brighter than the novae. Later it has been clarified that novae are surface eruptions on stars, while SNe are the complete explosive disruptions of them.

That SNe are explosions of stars was first suggested by Baade and Zwicky in 1939, who asked themselves what would happen to a massive star once it reached the degenerate iron core stage described in the previous chapter. Inspired by the discovery of the neutron by James Chadwick in 1932, their answer was that the core would collapse to a \emph{neutron star}, with the gravitational binding energy released in that process ejecting the outer layers in a violent explosion. 
This type of event is called a \emph{core-collapse supernova}. There is also another class, the \emph{thermonuclear supernovae}, which are the explosions of carbon-oxygen white dwarfs. 

\begin{figure}[htb]
\centering
\includegraphics[width=0.4\linewidth]{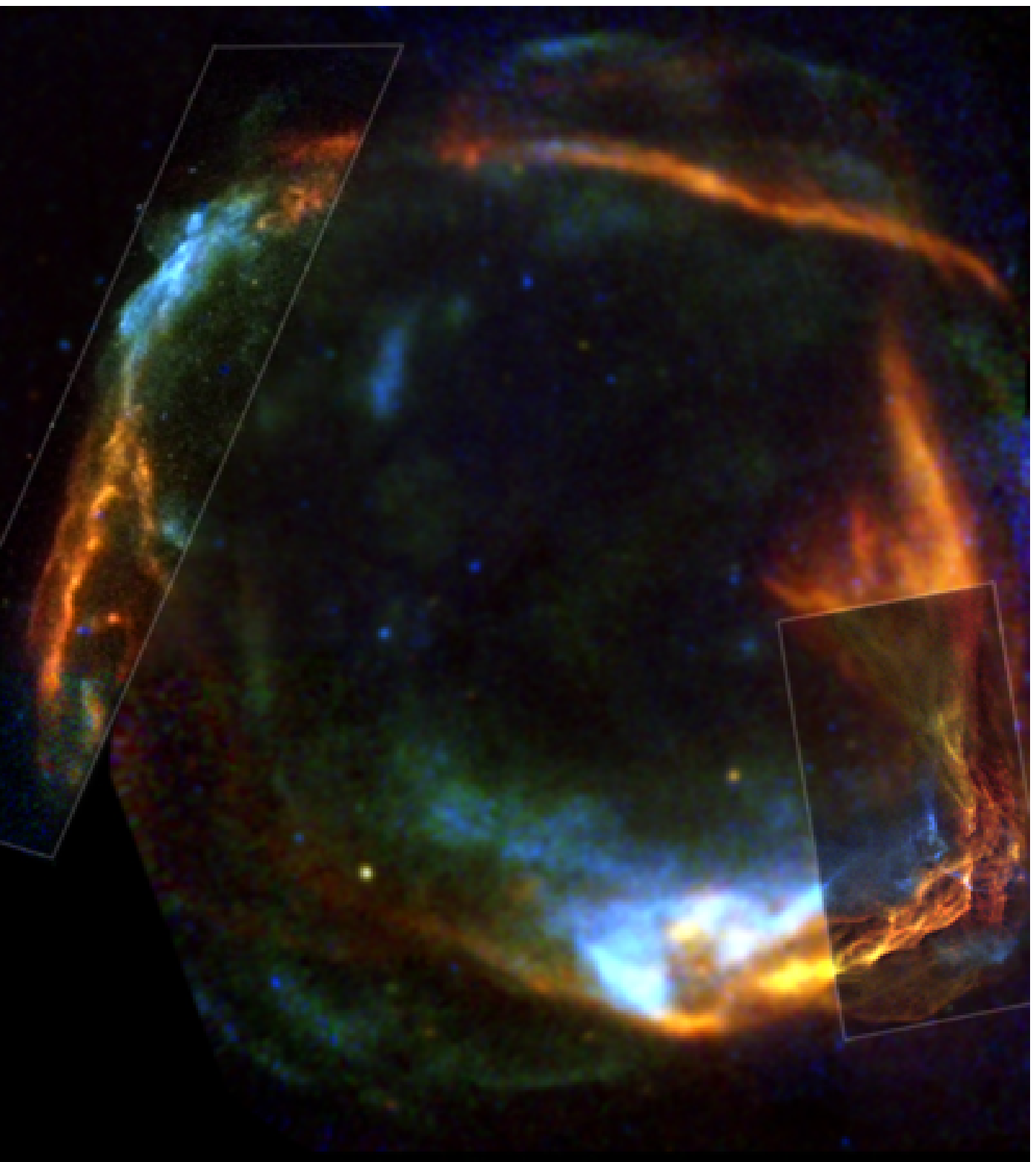}
\includegraphics[width=0.4\linewidth]{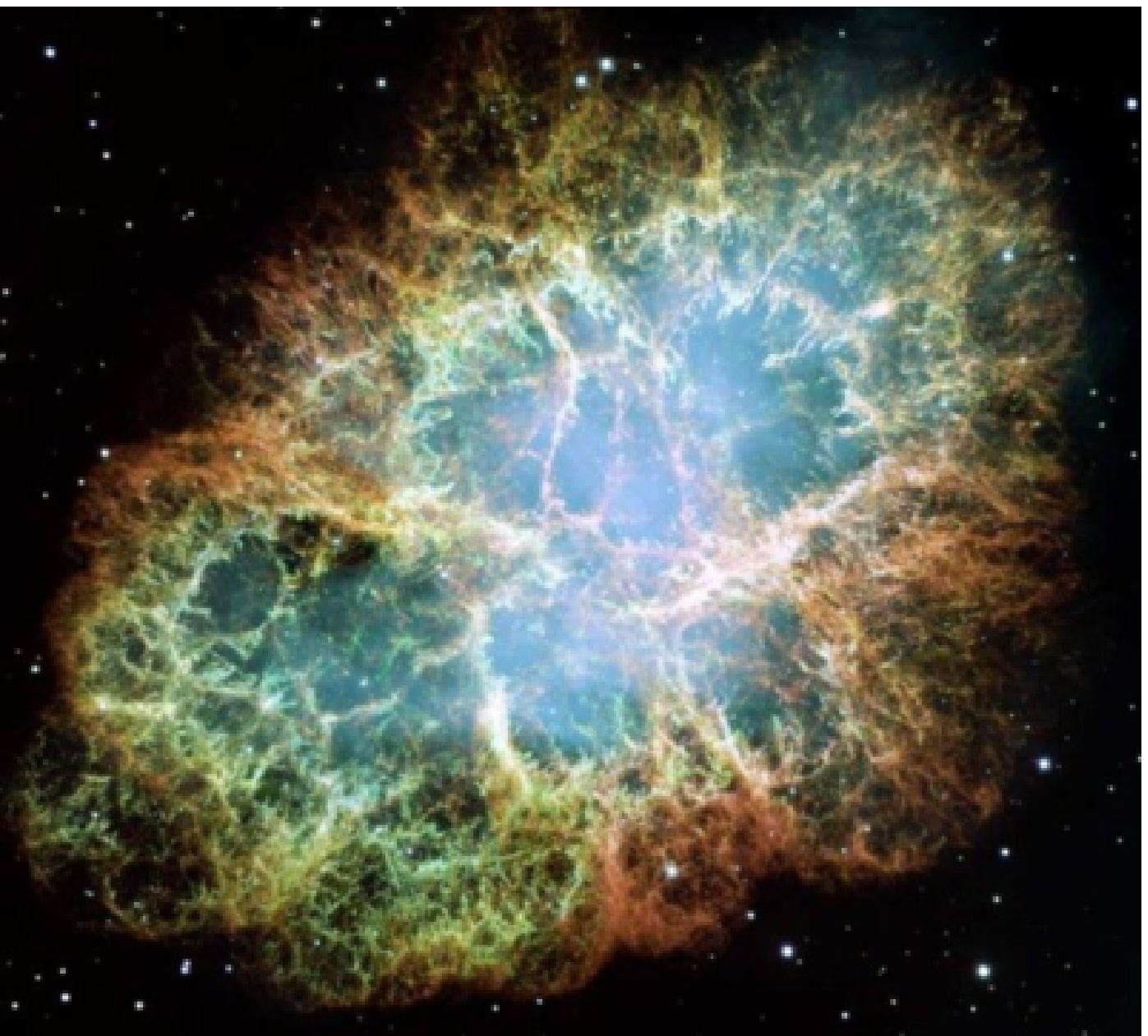}
\caption{The remnants of SN 185 and SN 1054 (the Crab nebula). \emph{Image credit : Chandra/NASA.}}
\label{fig:sn185}
\end{figure}

Today, we use our telescopes to zoom in on the locations of the historical guest stars mentioned above. Indeed, Baade and Zwicky were right; we find the expanding debris of violent explosions, sometimes with a pulsating neutron star at the center! Fig \ref{fig:sn185} shows what is seen in the direction of the year 185 and 1054 guest stars.

\section{Collapse and explosion}
\label{sec:explm}
If fusion reactions were to cease inside the Sun, it would slowly start contracting as thermal energy diffused out, lowering the pressure support inside. 
In the core of a massive star, thermal energy is not lost by such slow radiative diffusion, but by immediate neutrino losses. Once fusion stops, things therefore happen quickly. Apart from the neutrino losses, which follow pair annihilations and electron captures, internal energy is also lost by endoergic photodisintegrations of the iron nuclei. In less than one second, ten million years of fusion reactions are undone in the core as the iron nuclei are disintegrated into $\alpha$-particles. In its own immense gravitational field, the core collapses on a free-fall time-scale of $\sim$ 100 milliseconds. Matter crashes towards the center at about a quarter of the speed of light, 
compressing the core from the size of the earth to the size of a city. At that point, nuclear densities are reached, and the repulsive part of the nuclear force kicks in. The inner parts of the collapsing core come to a screeching halt, forming a proto-neutron star (PNS). The sudden deceleration launches a shock wave 20 km from the center that travels upstream through the infalling remainder of the iron core, which accretes onto the PNS after being shocked. The shock loses energy as it fights its way out through the inflow, but copious neutrino emission from the PNS pushes it on. For a few seconds, the PNS emits as much energy in neutrinos as the rest of the universe does in light! 
The shock wave runs through the rest of the star, exploding it. Approximately $10^{51}$ ergs of energy is endowed to the outer layers, which are expelled at thousands of kilometers per second. A supernova is born.

The details of the explosion mechanism outlined above are still being worked out. No spherically symmetric models succeed in achieving the required energy transfer from the neutrinos to the shock wave, instead showing the whole star collapsing to a black hole, simply disappearing from the night sky. It has therefore become clear that some sort of multidimensional effects must be crucial. Several types have been proposed, including rotationally-driven explosions along the star's rotation axis, convection, and acoustic oscillations. In the last few years, 2D and 3D models have emerged that show that so called \emph{Standing Accretion Shock Instabilities (SASI)} occur in the first few seconds after explosion \citep{Blondin2007, Scheck2008}. These pick out one or a few directions by spontaneous symmetry breaking, enhancing the neutrino deposition efficiency by convective motions along these modes. In \textbf{Paper I}, the 3D mapping of the ejecta of SN 1987A shows a morphology consistent with SASI-like instability. 

\subsection{Explosive nucleosynthesis}
\label{sec:explburn}
The shock wave running through the star is strongly supersonic and radiation-dominated. One may show that such a shock accelerates material to 6/7 of the shock velocity, and heats it to \citep{Sedov1959}
\begin{equation}
T = 4300~\rho^{1/4} V_s^{1/2}~~K~,
\label{eq:Tshock}
\end{equation}
where $V_s$ is the shock velocity, and $\rho$ is the density, which is of order $10^7$ g cm$^{-3}$  in the layers outside the collapsed core \citep[e.g.][]{Hirschi2004}. Combining this value with the typical shock velocity $V_s \sim 10^4$ \kms, we can estimate $T\sim 8\e{9}$ K. Since this is higher than the the silicon-fusion limit ($T_{\rm Si/S}\sim4\e{9}~$K), the shock burns the material to iron-group elements. The burning occurs close to NSE (Sect. \ref{sec:ssfusion}), producing \iso{56}Ni as the main isotope. However, the density soon falls to a regime where so called \emph{$\alpha$-rich freeze-out} produces large amounts of \iso{4}He nuclei, as well as radioactive isotopes such as \iso{44}Ti. The excess of $\alpha$-particles is built up as as the inefficient triple-alpha reaction falls out of equilibrium. In \textbf{paper II}, we discuss the signatures of this explosive burning in the late-time spectrum of SN 1987A.

It can be shown that the shock decelerates if the density falls off less steeply than a $r^{-3}$ power law, but accelerates otherwise. As the shock travels through the star, both decelerations and accelerations occur.
However, for all reasonable density profiles, the quantity $\rho(r)^{1/4} V_s(r)^{1/2}$ in Eq. \ref{eq:Tshock} steadily decreases, so the post-shock temperature decreases. Explosive burning occurs in reverse order compared to the hydrostatic burning. 
For a 20 \msun~star, the whole silicon/sulphur zone, and part of the oxygen zone, is burnt under NSE to \iso{56}Ni, followed by partial burning of the oxygen zone to Si/S (oxygen burning) and  O/Mg/Si/S (neon burning) \citep{Hashimoto1989}. The carbon burning regime occurs in regions where only trace carbon exists, and by the time the shock reaches the helium and hydrogen layers, the shock temperature is too low to burn them. 
In total, elements heavier than magnesium are mostly made explosively, and lighter ones mostly by hydrostatic burning. 

Since the shock is radiation-dominated, the energy density is given by $e=aT^4$, and we can estimate how much material will be burned by approximating the temperature as constant and equating
\begin{equation}
E_s = \frac{4\pi R^3}{3}a T^4~,
\end{equation}
where $E_s$ is the shock energy, $a$ is the radiation constant, and $R$ is the radius. For $E_s=10^{51}$ ergs and $T=T_{\rm Si/S}$, the solution is $R=3700$ km, which roughly corresponds to the radius within which all material will be burned to iron-group elements \citep{Thielemann1996}. For typical pre-SN models, the corresponding mass coordinate is $\sim$ 1.7 \msun. With the mass cut\footnote{The division line between material falling back to form the compact object, and material that is ejected in the SN.} for the neutron star being around 1.6 \msun, the estimate for the amount of \iso{56}Ni ejected is then 0.1 \msun, in good agreement with the observed values. Similarly, the explosive oxygen-burning and neon-burning occurs out to $\sim$6400 km and $\sim$12000 km, with corresponding mass coordinates $\sim$1.8 and $\sim$2.1 \msun, for a 20 \msun~star \citep{Thielemann1996}. The empirical correlation between explosion energy and \iso{56}Ni mass, shown in \textbf{paper IV}, can likely be understood from these simple considerations. If the mass cut stays roughly constant, a higher explosion energy simply keeps the shock hot enough for silicon-burning over a larger volume.  

\subsection{Reverse shock and mixing}
\label{sec:rsm}

Several types of hydrodynamical instabilities may occur in the expanding gas. The \emph{Rayleigh-Taylor (RT) instability} occurs when lighter parcels of gas are accelerated into denser ones. The interface between the parcels then fragment and the fluids get mixed, resulting in finger-like structures. The instability arises when the pressure gradient and the density gradient have opposite directions. \citet{Chevalier1976} and \citet{ChevalierKlein1978} showed that such conditions indeed arise in the flow of hydrogen-rich SNe. \citet{Herant1994} showed that the RT mixing is stronger in RSGs compared to blue supergiants. The mixing in SNe with RSG progenitors (\textbf{paper III, IV}), should therefore be at least as strong as observed for SN 1987A, which had a blue progenitor.

These early simulations used one-dimensional codes until the shock had reached the hydrogen envelope, and only then mapped the ejecta to a multi-dimensional grid, necessitating that perturbations be put in by hand. 
Recently, however, computational advancements have allowed the multi-dimensional computations to cover also the early phases.
As mentioned in Sect. \ref{sec:explm}, these simulations find strong instabilities arising from the earliest times. \citet{Kifonidis2003} and \citet{Kifonidis2006} compute the evolution in 2D, finding the SASI instability to arise already during the first second, leading to a strongly asymmetric shock. These asymmetries provide seeds for the RT instabilities at the Si/O and O/He interfaces, fragmenting and mixing them within the first few minutes after core bounce. The He/H interface is, however, found to be mixed mainly by \emph{Richtmyer-Meshkov instabilities}, which arise when a shock front impacts a composition interface at an oblique angle. These results are important in showing that also the core zone interfaces are subject to instabilities, which leads us to apply strong artificial mixing throughout the core region of the input models used in this thesis. 

One may show that the diffusion time scale for mixing on the \emph{microscopic} (i.e., atomic) scale, is long compared to the hydrodynamic timescale \citep{Fryxell1991}. In addition, models assuming that microscopic mixing does not occur generally agree better with observations than models that assume that it does \citep[e.g.][]{Fransson1989, Maeda2007}. Any significant microscopic mixing between the metal zones and the helium and hydrogen zones can be ruled out based on the empirically established production of CO and SiO, as helium ions destroy these molecules \citep{Lepp1990, Liu1996, Gearhart1999}. 
The mixing discussed above is therefore believed to occur only on \emph{macroscopic} scales. The difference is crucial for spectral modeling, since even small amounts of certain elements may completely change the spectrum formed in a particular zone.

As the ejecta expand, the internal energy steadily decreases through radiative and adiabatic losses. Internal energy is, however, also resupplied by radioactive decay of \iso{56}Ni. \citet{Herant1991} find that the \iso{56}Ni clumps expand on a time-scale of $\sim$1 week, inflating themselves to hot, low-density bubbles. This result, and others, leads us to assume a large filling factor for the iron clumps in all models in this thesis. 

After some time, pressure forces become neglegible and hydrodynamical interaction ceases. Each parcel of gas then continues coasting on whatever trajectory it is on. The SN is then said to have entered the \emph{homologous phase}.


\begin{figure}
\centering
\includegraphics[width=1\linewidth]{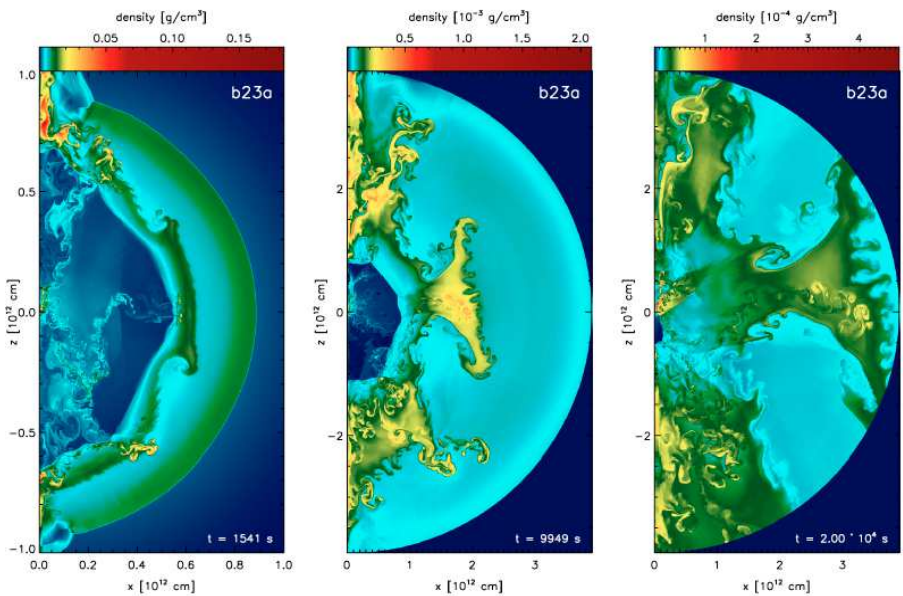}
\caption{The evolution of hydrodynamical instabilities in the 2D explosion models of \citet{Kifonidis2006}. The three density snapshots are at 1500, 10000, and 20000 seconds after core bounce.}
\end{figure}

\section{The diffusion phase}
As the shock breaks out through the surface of the star, the SN becomes visible. The initial burst will be hot and peak in X-rays, but soon the gas expands and cools to emit mostly ultraviolet and visible light. Only one SN has been caught in the act of the actual shock breakout, SN 2008D \citep{Soderberg2008}. The early UV/optical cooling phase has been observed in several objects \citep[e.g.][]{Stritzinger2002}. 

The highly ionized nebula is initially optically thick, and the energy inside can only slowly diffuse towards the surface. The spectrum is that of a blackbody scattered by lines in the atmosphere. The major difference to the spectrum of an ordinary (hot) star is that the absorption and emission lines are much broader due to the high expansion velocities, which completely dominate the thermal velocities. Many lines also form P-Cygni profiles, where the velocity field causes a blue-shifted absorption through combined with emission on the red side.

The early light-curve of the SN evolves in a way that depends on how much internal energy has been deposited by the shock passage, what fraction of this energy will be radiated rather than lost to adiabatic expansion, and on what time-scale that emission will occur. The major determinant for the latter two quantities is the compactness of the star, $M/R$ \citep[e.g.][]{Imshennik1992}. For a RSG, about 1\% of the internal energy will be radiated away, or $\sim10^{49}$ ergs. If the time-scale for emission is one month, the luminosity is then $\sim10^9$ L$_\odot$, still matching that of an entire galaxy. The radioactive elements produced may also influence the diffusion phase light-curve, especially for compact stars where the explosion energy is rapidly lost to adiabatic expansion.



\section{The steady-state nebular phase}
\label{sec:ssnp}

After a few months, the expanding nebula becomes optically thin in the continuum\footnote{Continuum absorption processes are those that occur for a wide range of photon energies, such as electron scattering, photoionization, and free-free absorption. See Chapter 5 for more details.}, and is said to have entered the \emph{nebular phase}. The remaining internal energy can then be efficiently radiated away, and if no other energy source is present the SN will rapidly fade away to undetectability. Fortunately, the radioactive elements produced in the explosion provide a new power source. 
As we saw in Sect. \ref{sec:explburn}, the initial shock burning produces \iso{56}Ni, which is radioactive with a life-time of 8.8 days. Its daughter nucleus, \iso{56}Co, is also radioactive with a life-time of 111.5 days. Most SNe enter the nebular phase when the \iso{56}Ni has decayed and \iso{56}Co is the dominant power source. It provides a powering of, for $t \gg 8.8$ days:
\begin{equation}
L(\mbox{\iso{56}Co}) = 3.4\e{8} L_\odot \left(\frac{M(\mbox{\iso{56}Ni})}{0.1~M_\odot}\right)e^{-t/111.5~\mbox{d}}~,
\label{eq:56co}
\end{equation}
where $M(\mbox{\iso{56}Ni})$ is the mass of \iso{56}Ni synthesized in the explosion, typically around 0.1~\msun. The decay products thermalize to produce UVOIR (UV-optical-near-infrared) radiation from the SN.


One can show that all atomic processes are fast in the early nebular phase (Chapter 4), which means that the SN almost instantaneously re-emits all the energy put in by radioactivity. We can therefore refer to this phase as the \emph{steady-state nebular phase}. As long as the decay particles are trapped by the ejecta, the light-curve then follows the time-evolution of the \iso{56}Co nuclide, $L(t)\propto e^{-t/111.5~\mbox{d}}$. In this phase, we do not need to know the history of the nebula to predict its output, just the instantaneous radioactive input. 

In general, the nebula will still be optically thick in various \emph{lines} even in the nebular phase. This fact has been one of the key drivers for this thesis project; to develop models that include transfer through these lines throughout the ejecta. In \textbf{paper II}, we show that this line transfer indeed has a strong influence on the emerging spectrum.

Can the line opacity trap radiation so that the output is actually lagging the radioactive input? The empirical fact that the bolometric light curves do not show any evidence of such time-delay suggests no. The velocity field makes transfer through individual lines a \emph{local} process (see Sect. \ref{sec:lt}), so a photon in an optically thick line does not have to diffuse through the whole nebula. In Sect. \ref{sec:bb}, we show that lines are active over a length scale $\sim 10^{-3}$ of the size of the nebula, so the mean-free-path $\lambda$ is
\begin{equation}
 \lambda \sim \frac{10^{-3}Vt}{\tau}~,
\end{equation}
where $V$ is the expansion velocity scale, and $\tau$ is the line optical depth. The average number of scatterings equals the optical depth $\tau$ in the Sobolev approximation (Sect. \ref{sec:lt}), so the time the photon is trapped in the line is 
\begin{equation}
t_{\rm scatter} = \frac{\tau \lambda}{c} = 10^{-3}\frac{Vt}{c} \lesssim 10^{-5}t~,
\end{equation}
where we have used $V/c \sim 10^{-2}$ in the last step. We see that the trapping time is independent of the optical depth of the line, and neglegible compared to the evolutionary time. \emph{Line-to-line} scattering will, however, introduce longer trapping times. A photon that scatters in $N$ lines will, in worst case, be trapped for a time 
\begin{equation}
t_{\rm scatter} \sim N\frac{Vt}{c}~\sim 10^{-2}N t~,
\end{equation}
which can potentially violate the steady-state assumption if $N$ is greater than a few. 
In practice, $N$ is kept small by the fact that fluorescence occurs with a relatively high probability for most lines, transferring the photons to longer wavelengths where the line opacity is lower \citep{Pinto2000b}.

\section{The time-dependent nebular phase}
After several years, the radiative output from the SN no longer matches the instantaneous input by radioactivity, because the reprocessing is going too slow \citep{Axelrod1980, Clayton1992, Fransson1993}. The SN is then said to have entered the \emph{time-dependent nebular phase}. From this time on, solutions of the \emph{time-dependent} equations for temperature, ionization, and excitation are required, rather than their steady-state variants (see Chapter 4). This means that the history of the nebula now influences its output at any given time.


Regions of low density and low ionization are the ones to be firstly affected by this time-dependence. 
The code developed in this thesis does not take time-dependence into account. In \textbf{paper I and II}, where the outer envelope of the SN has entered the time-dependent phase, we compute conditions in those layers with the time-dependent code by KF98.

\section{Supernova classes}

One can estimate that about one SN occurs per second somewhere in the universe. Of these, we currently discover about one in 100,000, or a few hundred per year. The discovery rate is expected to increase dramatically in the coming years thanks to robotic wide-field surveys such as Pan-Starrs, The Supernova SkyMapper, Large Synoptic Survey Telescope (LSST), and the Palomar Transient Factory (PTF).

The empirical classification of SNe are divided into an initial branch of \emph{Type I} (hydrogen lines present) and \emph{Type II} (hydrogen lines absent). The Type I class then divides into \emph{Type Ia} (strong Si II 615 nm line), \emph{Type Ib} (helium lines present), and \emph{Type Ic} (helium lines absent). The Type II branch subdivides into \emph{Type IIP} (a plateau in the light-curve), \emph{Type IIL} (a linear decline of the light-curve), and \emph{Type IIn} (narrow lines present). The Type Ia class corresponds to thermonuclear SNe, whereas all other classes correspond to core-collapse events. Some SNe not fitting into the standard classification scheme are given the \emph{peculiar (pec)} designation.  

\begin{figure}[tbp]
\includegraphics[width=1\linewidth]{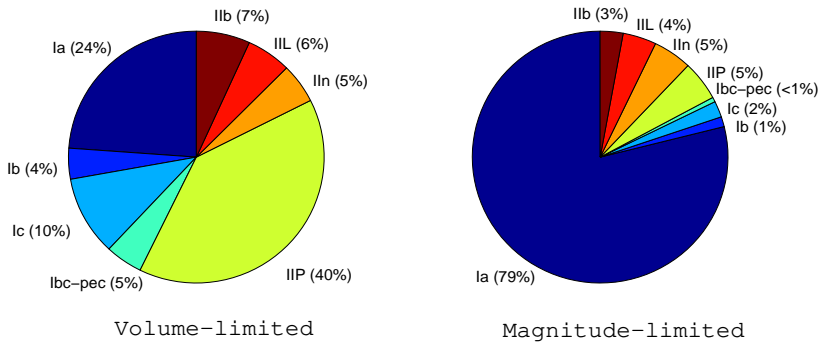}
\caption{Observed fractions of SN classes in a volume-limited sample (left) and a magnitude-limited sample (right). Data from \citet{Li2011}.}
\label{fig:class}
\end{figure}

Fig. \ref{fig:class} shows the observed fractions of the various SN classes. Since brighter SNe are more easily detected, the observed fractions of those (magnitude-limited) is higher than the actual fraction (volume-limited).

\subsection{Type II} 
The most common class in the universe is the Type II class, which makes up $\sim$57\% of all events \citep{Li2011}. Only in the last decade or two have decent nebular phase spectral data sets been obtained for Type IIs, with SN 1987A long being the only well-observed event. A listing of well-observed Type IIPs can be found in \textbf{paper IV}. Some of these appear to have similar \iso{56}Ni-masses and ejecta velocities as SN 1987A, but others seem to come from much weaker explosions, with low velocities and low \iso{56}Ni-masses. 




\subsection{Type Ib/c} Explosions of stars that have lost their hydrogen envelope, or both the hydrogen and helium envelopes, are called SN Type Ib and Ic, respectively. Due to their lower frequency of 19\% (volume-limited) and 4\% (magnitude-limited) \citep{Li2011}, Type Ib/c SNe have even fewer nebular data sets than the IIPs. However, with their small envelopes and large metal cores, they represent an important class to analyse with respect to nucleosynthesis. 

\subsection{Type Ia}
Type Ia SNe arise as a white dwarf accretes matter until is reaches the Chandrasekhar limit, at which time it ignites thermonuclear fusion of carbon and oxygen under degenerate conditions, leading to a run-away explosion. The Ias differ from core-collapse events in that they produce more \iso{56}Ni ($\sim$ 1 \msun~versus $\sim$0.1 \msun), and reach higher ejecta velocities ($\sim$0.1c versus $\sim$0.01c).

\begin{figure}[htb]
\includegraphics[width=1\linewidth]{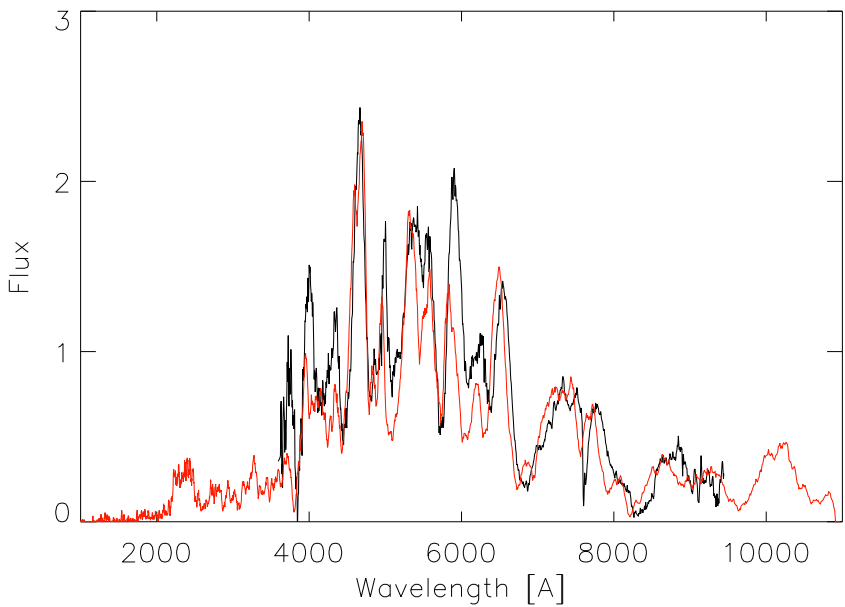}
\caption{The spectrum of the Type Ia SN 2005cf at day 94 post-explosion (black line), compared to the model spectrum of explosion model W7 \citep{Nomoto1984} (red line). From \citet{Maurer2011}.}
\label{fig:2005cf}
\end{figure}

This thesis is mainly focused on the core-collapse SN class. However, the code developed (described in Chapters 4 and 5, and in papers \textbf{II} and \textbf{III}) has been tested on a Type Ia explosion model and compared to the 'typical' Type Ia SN 2005cf. The spectrum is shown in Fig. \ref{fig:2005cf}. The general agreement is encouraging for the future application of the code for Type Ia modeling.

\section{SN 1987A}
On February 23 1987, the first SN to be seen by the naked eye since 1604 exploded in our satellite galaxy, the Large Magellanic Cloud. The proximity (50 kpc) gave an opportunity to study a stellar explosion in unprecedented detail, and SN 1987A continues to be carefully studied today. 

\begin{figure}[htb]
\centering
\includegraphics[width=0.75\linewidth]{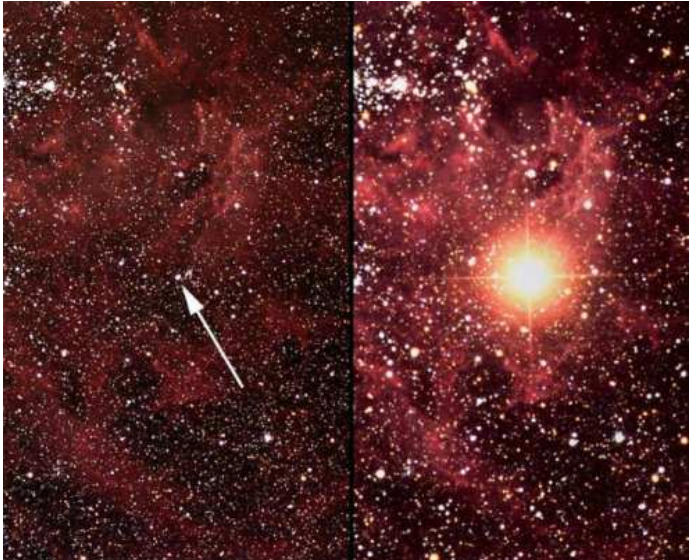}
\caption{The progenitor star (left) and the explosion (right) of SN 1987A. \emph{Image credit: Anglo-Australian Observatory.}}
\end{figure}

The early spectrum of the SN showed clear hydrogen lines, establishing it as a Type II event. However, the light-curve did not evolve as any other Type II previously observed, being unusually dim and rapidly evolving. It became clear that the exploded star must have been more compact than a RSG, which are the usual progenitors of Type II SNe.

\subsection{The progenitor}
\label{sec:87Aprog}
The progenitor star, Sk $-$69$^{\rm{o}}$ 202, had luckily been registered in pre-explosion surveys of the region \citep{Gilmozzi1987}, the first time a SN progenitor had been caught on a photograph. It was confirmed to be a compact and hot star, a blue supergiant of spectral class B3 Ia, with radius 50 R$_\odot$ and luminosity $10^5~L_\odot$ \citep{Arnett1989}. It did not have any particularities associated with it (the region was actually observed regularly for over a century prior to 1987), confirming the theoretical expectations that the late burning stages would transpire too quickly to affect the appearance of the star.

Stellar evolution codes were scrambled to see what models could explain Sk $-$69$^{\rm{o}}$ 202. 
\citet{Woosley1988} found a 20 \msun~star with reduced semi-convection to end as a blue supergiant with the correct luminosity. Other models invoked a binary system where a merger between a $15-20$ \msun~star and a $\sim$3 \msun~main-sequence star took place $\sim 2\e{4}$ years prior to explosion \citep{Hillebrandth1989, Podsiadlowski1990}. The exact nature of the progenitor system still remains unknown. For our modeling of SN 1987A in \textbf{paper I} and \textbf{paper II}, we assume that the progenitor had a mass of 19 \msun.

\subsection{Spectral analysis}
\label{sec:specan87A}
The analysis of SN 1987A had as of late 2011 spawned some 1100 refereed scientific articles, which comprises a body of information almost impossible to adequately overview. I mention here a few of the most important results. A good review of its first five years of evolution can be found in \citet{McCray1993}. Some overview of modeling results can also be found in \textbf{paper II}.

\paragraph{Light curve}
Hydrodynamical modeling of the light-curve indicated a hydrogen envelope of size $7-10$ M$_\odot$ and an explosion energy of $0.8-1.5\e{51}$ ergs \citep{Woosley1988, Shigeyama1990, Bethe1990}. The shock-deposited energy dominated the light-curve for the first four weeks, but due to the rapid adiabatic cooling, radioactivity soon took over \citep[e.g.][]{Arnett1989}. From multi-band observations in the steady-state nebular phase, the \iso{56}Ni mass was determined to 0.069 \msun \footnote{For D=50.1 kpc and color excess E(B-V) = 0.15 mag.} \citep{Bouchet1991a}.

\paragraph{Spectra}
Most emission lines in the nebular spectra showed similar line profiles, suggesting a similar velocity distribution \citep[e.g.][]{Meikle1993}. Some of the clearly identifiable optical lines were H$\alpha$, [O I] \wll6300, 6364, [Ca II] \wll 7291, 7323, [Ca II] IR-triplet, [Fe II] \wl7155, 7172, and Mg I] \wl4571. The absence of H$\beta$ in the spectrum could be explained by models showing that the optical depths for the Balmer lines were high, resulting in conversion of H$\beta$ to Pa$\alpha$ + H$\alpha$ \citep{Xu1992}. The hydrogen recombination line ratios in Type IIP SNe therefore follow the unusual 'Case C' scenario, with thick Lyman and Balmer series.

Not even the hydrogen lines showed any flat tops, indicating that hydrogen must have been mixed down to almost zero velocities. At the same time, the iron lines showed broad wings suggesting that some of the \iso{56}Ni had been mixed far out into the envelope. Recently, the first 3D simulation of a realistic progenitor \citep{Hammer2010} succeeded in reproducing the high-speed \iso{56}Ni bullets inferred to be present in the ejecta. 


\paragraph{Molecules}
For the first time, molecules were detected in a SN. Carbon monoxide (CO) was detected in its fundamental band (4.6 $\mu$m) and first overtone band (2.3 $\mu$m) a few months after explosion \citep{Spyromilio1988}. The flux peaked at 200 days and persisted to at least day 775 \citep{Wooden1993}. 
\citet{Liu1992} estimated a total CO mass of $1\e{-3}$ \msun. This is in agreement with models showing that the non-thermal electron population limits the condensation efficiency to a few percent \citep{Clayton2001}. This in turn means that CO cannot be important in zones with low carbon abundance, and that the carbon in the C/O zone is mostly available for dust formation. 

Silicon monoxide (SiO) was detected in its fundamental band (8.1 $\mu$m) at day 160 \citep{Aitken1988}, persisting to day 519. \citet{Liu1996} estimate a mass of $5\e{-4}$ \msun, implying that neither SiO locks up much material.

Molecules have since been detected also in other core-collapse SNe, carbon monoxide in 1995ad, 1998S, 1998dl, 1999em, 2002hh, 2004et, 2004fj, and 2005af, and silicon monoxide in SN 2005af and 2004et. In \textbf{paper III}, we use these observations to motivate an assumption of CO and SiO cooling in the C/O and O/Si/S zones.
\paragraph{Dust}
Several indicators showed that dust formed in the supernova during its second year, see \textbf{paper II} (Sect. 2.6) for a review.
Dust formation has since been observed in several other SNe (1998S, 1999em, 2003gd, 2004dj, 2004et, 2006jc, and 2010jl), as well as in SN remnants Cas A, the Crab, Kepler and 1E0102.2-7219.

\subsection{Circumstellar interaction}
The presence of narrow emission lines in the spectra quickly made it clear that the star was surrounded by circumstellar material that had been flash-ionized in the shock breakout \citep{Fransson1989UV}. A few years after explosion, this material could be resolved as a triple-ring nebula consisting of an inner ring, located at a distance of 0.67 light-years and moving out with a velocity of 10 \kms, and two outer rings, located about three times further out and moving three times faster. The kinematics suggest a common origin in an ejection event some 20,000 years ago, the cause of which is still unclear.

The shock, travelling at $\sim$25,000 \kms, reached the inner ring in 1995. A dramatic display of radiation, from X-rays to the radio regime, has since followed, caused both by the shock transmitted into the ring, and a reverse shock created by the impact. The emission has steadily increased over the years as the shock has penetrated into denser parts of the ring. In 2010, the X-ray curve flattened, possibly suggesting that the densest parts of the ring have now been passed \citep{Park2011}. 

Recent calculations show that the X-ray emission from the ring started to influence the ejecta in 2001 \citep{Larsson2011}. Most of the X-rays hitting the ejecta are estimated to be absorbed in the hydrogen envelope
, increasing the flux in hydrogen and helium lines.

The ejecta itself, mostly confined to within 5000 \kms, will start impacting the ring around 2015. A significant fraction of the $10^{51}$ ergs of kinetic energy will then be converted to radiation (more than what has been emitted so far), however over a time-span of several decades. 


\subsection{Missing compact object}
\label{sec:mco}
A total of 24 neutrinos emitted in the collapse were detected by the Kamiokande II, IMB and Baksan neutrino detectors, representing the only neutrinos detected from outside the solar system. The neutrino detection gave spectacular confirmation of the core-collapse model, with the number of neutrinos and their energy distribution being in excellent agreement with quantitative models for neutron star formation \citep{McCray2007}. That a neutron star was formed rather than a black hole is also suggested by the inferred explosion energy and the \iso{56}Ni mass, which correspond to a mass-cut at $\sim$ 1.6 \msun~for any reasonable progenitor model \citep{Hashimoto1989}, well below the upper mass limit of $2-3$ \msun~for neutron stars.

However, as of 2011 no sign of the neutron star has been seen. It is possible that an optically thick dust cloud happens to obscure it. If the neutron star would be a \emph{pulsar}, i.e. magnetized and spinning, it probably would have been detected by now \citep{McCray2007}. 

Could a (non-spinning) neutron star influence the SN spectra? Neutron star cooling models by \citet{Tsuruta2002} show a total thermal luminosity of about 10 $L_\odot$ for a decade-old neutron star. But the energy input by radioactivity is never below 500 $L_\odot$ for the first 50 years, assuming a \iso{44}Ti mass of $1\e{-4}$ \msun. Thus, the SN spectrum will not be noticeably influenced by the thermal emission of the neutron star. This is an important check for the validity of the \iso{44}Ti-mass derived in \textbf{paper II}.


\chapter{Physical processes}

\begin{figure}[htb]
\centering
\includegraphics[width=1\linewidth]{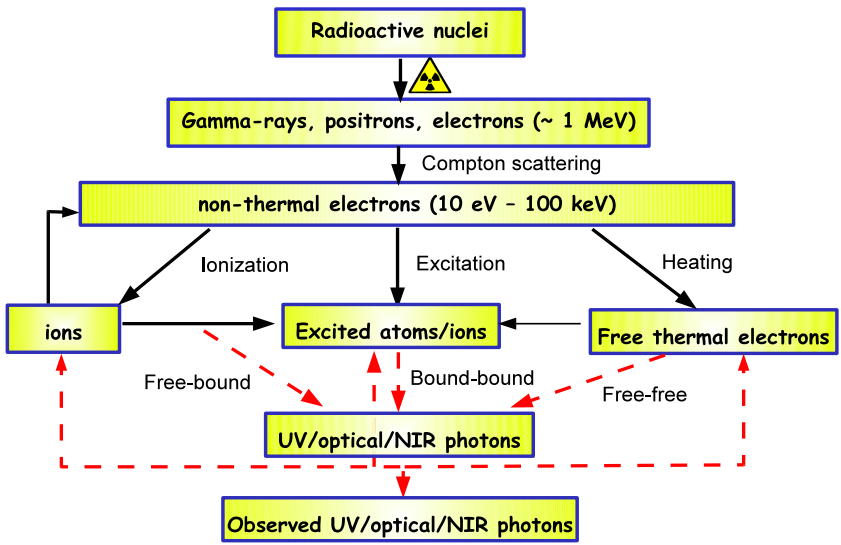}
\caption{Schematic overview of the physical processes occurring in SNe. Radiative processes involving the internal UVOIR field are drawn as red, dashed arrows.}
\label{fig:block}
\end{figure}

I now proceed to describe the main physical processes occurring in SNe, and how the state of the gas (temperature, ionization and excitation) can be calculated. The power source in the nebular phase is radioactivity, which heats, ionizes, and excites the gas, which then emits radiation. This radiation is partly reabsorbed, and partly escapes for us to observe. A schematic diagram showing the relevant processes is presented in Fig. \ref{fig:block}. 
 
\section{Radioactive decay}
In the first step of a radioactive decay, a nuclear transmutation changes the number of protons in the nucleus. 
This occurs either by emission of an $\alpha$-particle ($\alpha$-decay), an electron plus an electron anti-neutrino ($\beta^-$-decay), a positron plus an electron neutrino ($\beta^+$-decay), or by electron capture (EC), where one of the orbiting electrons (usually from the $K$ or $L$ shell) is captured to convert one of the protons to a neutron, and the energy is usually emitted as an electron neutrino. The re-filling of the inner electron orbit occurs by emission of an X-ray or by the ejection of an electron (Auger electron). Apart from the decay processes mentioned above, some rare decays also involve the emission of a neutron or a proton. 

The new nucleus is usually formed in an excited configuration. It decays to the ground state, usually in several steps, by emission of $\gamma$-rays ($\gamma$-decay) and/or by internal conversions (IC), where the excitation energy is transferred to an orbiting electron, which is ejected. Following internal conversions, an X-ray or an Auger electron is also produced, just as after electron captures. 


A radioactive decay channel is thus associated with emission of $\gamma$-rays, X-rays, electrons, and positrons, all with individual energy distributions. Some channels may emit all of these particles, while others may emit only one or two types. Table \ref{table:rd} lists the fraction of the decay energy going to different particles for the most common radioactive isotopes in SNe. Note that \iso{56}Co and \iso{44}Sc are the decay products of \iso{56}Ni and \iso{44}Ti, respectively. One should observe that channels carrying only a small part of the total energy can still be important if the \emph{trapping} of the decay product is more efficient than for the other ones.


\begin{table}[htb]
\centering
\begin{tabular}{ c|c|c|c|c}
\hline
Isotope & Gammas & Positrons & Electrons & X-rays  \\
\hline
\iso{56}Ni & 100\% & 0 & $<$0.1\% & $<$0.1\%\\
\iso{56}Co & 96.5\% & 3.4\% & $<$0.1\% & $<$0.1\% \\
\iso{57}Co & 86.4\% & 0 & 11.3\% & 1.8\% \\
\iso{44}Ti & 92.7\% & 0 & 6.7\% & 0.49\% \\
\iso{44}Sc & 66.3\% & 33.7\% & $<$0.1\% & $<$0.1\%\\
\hline
\end{tabular}
\caption{Fraction of the decay energy (excluding the neutrinos) carried by various particles, for the most important radionuclides in SNe. The electron rest mass is not counted in the energy budget, whereas the positron rest mass is, being included in the gamma column. Data from \citet{Endt1990, Junde1992, Seitenzahl2011}.}
\label{table:rd}
\end{table}

\section{Deposition of decay products}
\label{sec:deposition}

\subsection{Gamma-rays}

Gamma-rays see an opacity dominated by Compton scattering by free and bound electrons. Because the $\gamma$-ray energy ($\sim 1$ MeV) is much higher than the binding energy of the electrons ($<$ 1 keV), the $\gamma$-rays do not distinguish between bound and free electrons, and the deposition is therefore independent of the ionization/excitation state of the gas. The total cross section is given by the Klein-Nishina formula, which gives the Thomson value ($\sigma_T = 6.65\e{-25}\mbox{cm}^2$) at low energies,  but decreases monotonically with increasing energy. At 1 MeV the cross section is about $1/3~\sigma_T$, corresponding to an absorption coefficient
\begin{equation}
\kappa_\gamma(1~\mbox{MeV}) \sim \frac{1}{3}\sigma_Tm_p^{-1} Y_e=0.066\frac{Y_e}{0.5}~~\mbox{cm}^2\mbox{g}^{-1}~,
\label{eq:kn1mev}
\end{equation}
where $Y_e = Z/A$ is the number of electrons per nucleon.
Most zones have $Y_e \sim 0.5$, but the hydrogen zone has $Y_e = 0.86$, or lower if helium has been dredged up (Sect \ref{sec:convection}).

For $\gamma$-rays over 1.022 MeV, pair production can also occur in the vicinity of a nucleus. As the $\gamma$-rays are down-scattered below $\sim$100 keV, photoelectric absorption adds to the opacity as well. 

\paragraph{Effective opacity}
Several authors (see below) have found that transport with a purely absorptive, effective (gray) opacity 
\begin{equation}
\kappa^{eff}_\gamma \sim 0.03 \frac{Y_e}{0.5}~~\mbox{cm}^2\mbox{g}^{-1}~,
\label{eq:kappaeff}
\end{equation}
gives an accurate approximation to the exact solution for the $\gamma$-ray transfer. The value is lower than the Klein-Nishina coefficient (Eq. \ref{eq:kn1mev}) because this opacity represents pure absorption with no scatterings. The optimal value of $\kappa^{eff}_\gamma$ has some dependency on the geometry of the problem, but for typical geometries \citet{Colgate1980}, \citet{Axelrod1980}, and \citet{Woosley1980} found values in the range $\kappa_\gamma^{eff}=0.028-0.033(Y_e/0.5)$. \citet{Swartz1995} made the important realization that the optimal value is time-dependent, finding $\kappa^{eff}_\gamma=0.033(Y_e/0.5)$ as long as multiple scatterings occur, falling to  $\kappa_\gamma^{eff}=0.025(Y_e/0.5)$ when zero or one scatterings occur. \citet{Swartz1995} also provided a nice explanation \emph{why} the effective opacity works so well for approximating the Compton scattering process; the interactions are such that forward scatterings remove little energy from the $\gamma$-ray, while strongly direction-changing ones remove a lot. Thus, the $\gamma$-ray travels along approximately straight paths with little energy loss until a strong interaction occurs, at which point it loses most of its energy. This situation is obviously well approximated by a purely absorptive transfer.

In this thesis, gray transport with Eq. \ref{eq:kappaeff} is used. One should be aware that this treatment is only accurate to $10-20$\%. However, as we in general do not know the detailed distribution of the ejecta zones within the SN, performing exact Compton calculations is not very meaningful.

\begin{figure}[htb]
\centering
\includegraphics[width=0.8\linewidth]{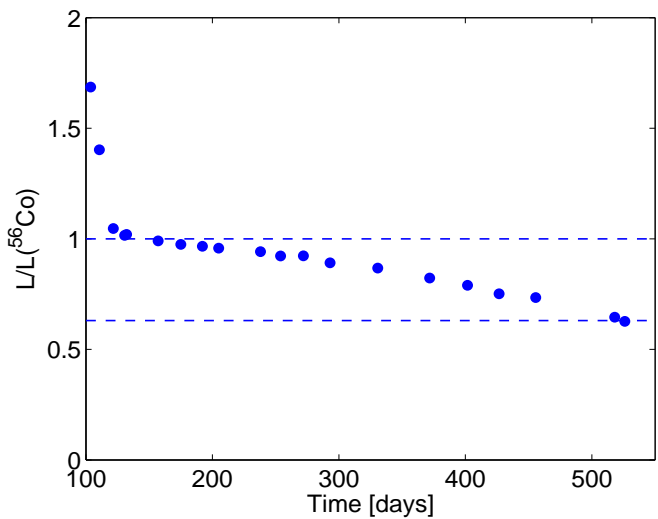}
\caption{The bolometric light-curve of SN 1987A normalized to the powering by 0.07 \msun~of \iso{56}Co. Data from \citet{Suntzeff1990}. Time is in days since explosion.}
\label{fig:87alightcurve}
\end{figure}

A homogeneous sphere of mass $M$ and expansion velocity $V$ reaches optical depth unity to the gamma-rays at a time
\begin{equation}
t_{\gamma}^{trap} = 460~\mbox{days}~\left(\frac{M}{10~M_\odot}\right)^{1/2}\left(\frac{V}{3000~\mbox{km/s}}\right)^{-1}~.
\label{eq:trapping}
\end{equation}
Since the expansion velocity $V$ can be estimated from the line widths, the epoch when the bolometric light-curve starts decreasing faster than $e^{-t/\tau_{\rm 56Co}}$ can therefore give us an estimate of the ejecta mass $M$. Fig. \ref{fig:87alightcurve} shows the bolometric light-curve of SN 1987A, normalized to the decay by 0.07 \msun~of \iso{56}Co. We see that the trapping reaches $1-e^{-1}=0.63$ at about 530 days after explosion. From Eq. \ref{eq:trapping}, with $V=3000$ \kms,  the ejecta mass can then be estimated as 13 \msun, in good agreement with estimates from more detailed modeling. The assumption of constant density, as well as which $V$ to use, are the limitations of the method, but Eq. \ref{eq:trapping} can still be used to tell us the ejecta mass to within a factor two or so. 

\begin{figure}[htb]
\centering
\includegraphics[width=0.8\linewidth]{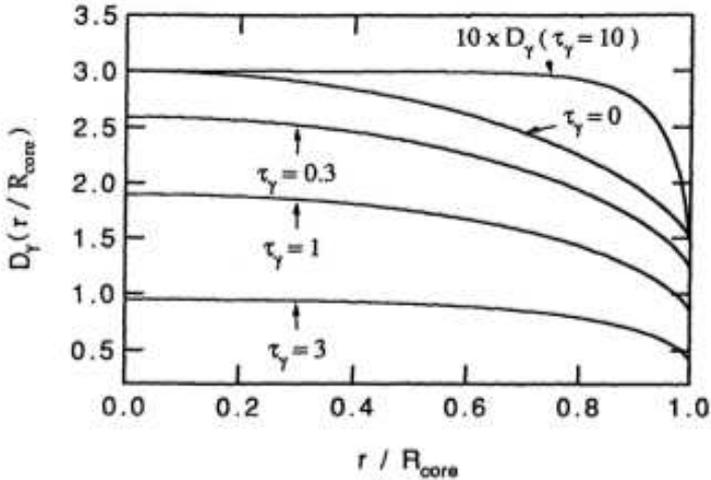}
\caption{Gamma-ray deposition (arbitrary units) as function of radius in a uniform sphere for different optical depths. From \citet[][KF92 hereafter]{Kozma1992}.}
\label{fig:kf92}
\end{figure}

The $\gamma$-ray deposition as function of radius in a uniform sphere is plotted in Fig. \ref{fig:kf92} for a few optical depths. 
The difference between the outer edge and the center never exceeds a factor of two. From the perspective of $\gamma$-ray deposition, it is therefore reasonable to use single-component zones in the core where the \iso{56}Ni is distributed. Such models are used throughout this thesis.

\subsection{Electrons and positrons}
The electrons and positrons emitted in the radioactive decays lose their energy by ionizing and exciting the gas. Since their energy is much higher than the excitation and ionization potentials, one may use the \emph{Bethe approximation} for both ionization and excitation cross sections \citep{Bethe1930}. The cross section for ion $i$, transition $j$, is
\begin{equation}
\sigma_{coll,ij}(E)= \frac{\pi a_0^2}{E}\left(A_{ij}\ln E +B_{ij}\right)~~~E \gg I~,
\end{equation}
where $E$ is the energy of the colliding electron, $a_0$ is the Bohr radius, 
and $A_{ij}$ and $B_{ij}$ are constants, depending only on level energies and A-values. For lower energies, experimental data for the collision cross sections are needed. KF92 provides a summary of the atomic data we use. $A_{ij}$ is, for bound-bound transitions, proportional to the oscillator strength $f_{ij}$. 
Allowed transitions ($f_{ij}\sim 1$) will therefore in general be more important than forbidden ones ($f_{ij} \ll 1$). Most elements have a continuum oscillator strength that is similar to the sum of excitation oscillator strengths, so about as much energy goes into ionizing as exciting the gas by these primary particles. However, the total ionization and excitation rates are influenced mainly by the \emph{secondary} electrons ejected in each ionization, described more in Sect. \ref{sec:ieh}. 

Since 1 MeV $\sim 10^5$ times the ionization/excitation potentials ($\sim$ 10 eV), some $10^5$ scatterings are involved before the kinetic energy has been converted to ionization and excitation energy. The time-scale for the down-scattering process can be shown to be, for a pure hydrogen nebula, \citep{Axelrod1980}
\begin{equation}
t_{slowdown} \sim 10^{-4}~\mbox{days}~\left(\frac{M}{10~M_\odot}\right)^{-1}\left(\frac{V}{3000~\mbox{\kms}}\right)^{3}\left(\frac{t}{100~\mbox{days}}\right)^{3}~,
\end{equation}
which stays short for many decades.

An effective, gray opacity can be used for the lepton transport just as for the $\gamma$-ray transport. \citet{Colgate1980} and \citet{Axelrod1980} found good values for the absorption coefficient to be 
\begin{equation}
\kappa^{eff}_{e+}=(7-10)\frac{Y_e}{0.5}~~\mbox{cm}^2\mbox{g}^{-1}~,
\end{equation}
from which it is clear that the electrons and positrons can travel only $\sim 1/300$ the distance of the $\gamma$-rays. 
A SN (without a magnetic field) becomes optically thin to the leptons at a time
\begin{equation}
t_{e-,e+}^{trap} \sim 23~\mbox{years} \left(\frac{M}{10~M_\odot}\right)^{1/2}\left(\frac{V}{3000~\mbox{km~s}^{-1}}\right)^{-1}~,
\label{eq:eltrap}
\end{equation}
which shows that they remain fully trapped for several decades, and furthermore we can assume on-the-spot absorption for the first year or so. A disordered magnetic field will further increase the trapping by locking the particles in Larmor orbits. The cyclotron radius $R_c$, relative to the radius of the nebula $R$, is \citep{Axelrod1980}
\begin{equation}
\frac{R_c}{R} = 1.8\e{-6}\left(\frac{B}{10^{-6}~\mbox{G}}\right)^{-1}\left(\frac{V}{3000~\mbox{\kms}}\right)^{-1}\left(\frac{t}{100~\mbox{d}}\right)^{-1}~.
\end{equation}
A magnetic field even at the interstellar level ($B \sim 10^{-6}$ G) will therefore produce local trapping of the positrons, and the trapping will \emph{increase} with time. In \textbf{paper II}, we find that such a field appears to be present in the ejecta of SN 1987A.




\subsection{X-rays}
\label{sec:xrays}
The X-ray opacity is dominated by photoelectric absorption on inner-shell electrons of metals such as C, O, Si, and Fe. The total photoionization cross section at 1 keV, for solar abundances, is $3\e{-22}$ cm$^2$ \citep[e.g.][]{Morrison1983}. The trapping time-scale can then be computed as 
\begin{equation}
t_{X-rays}^{trap} \sim 540~\mbox{years} \left(\frac{M}{10~M_\odot}\right)^{1/2}\left(\frac{V}{3000~\mbox{km~s}^{-1}}\right)^{-1}\left(\frac{E_X}{1~\mbox{keV}}\right)^{-3/2}~.
\end{equation}
Apart from radioactivity, the ejecta is exposed to (at least) two other sources of X-rays; the new-born neutron star and the shock-heated circumstellar gas. The neutron star will have a temperature of a few million degrees \citep[e.g.][]{Tsuruta2002}, emitting photons with an average energy of $3.82~k_BT\sim$1 keV. However, as we saw in Sect. \ref{sec:mco}, this flux is too weak to influence the SN spectrum. X-ray input from circumstellar gas may be more important. Recently, \citet{Larsson2011} showed that the X-rays from the forward shock in SN 1987A are influencing the ejecta conditions at late times.


\subsection{Deposition time-evolution}
\begin{figure}[htb]
\includegraphics[width=1\linewidth]{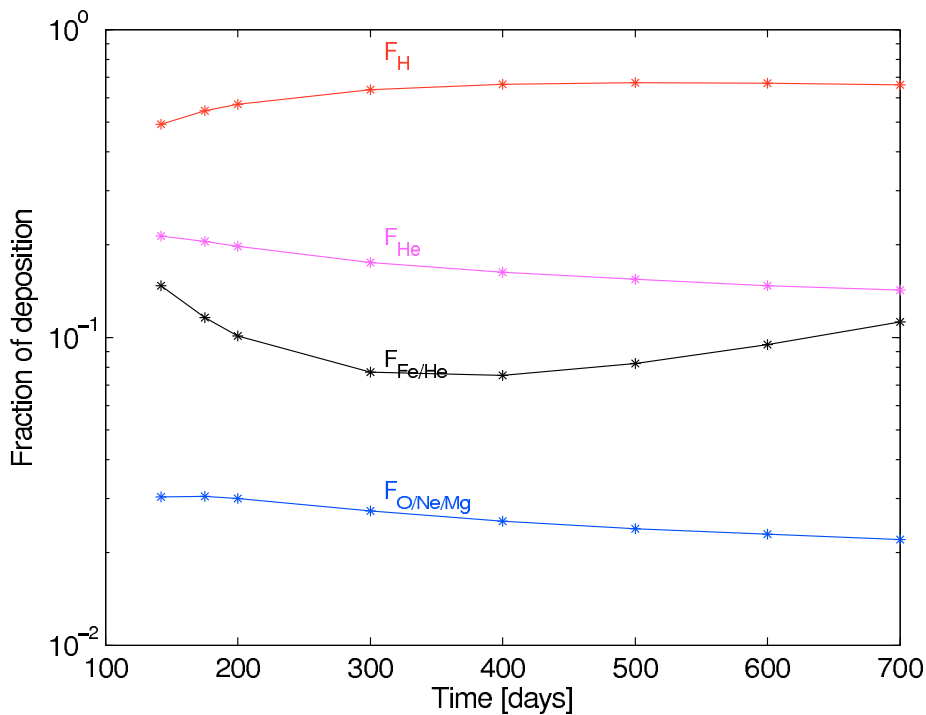}
\caption{Fraction of the total deposition going to the H, He, Fe/He, and O/Ne/Mg zones as function of time, for the 12 \msun~model in paper IV.}
\label{fig:depfracs}
\end{figure}

Fig. \ref{fig:depfracs} shows the fraction of the total energy deposited in the H, He, Fe/He, and O/Ne/Mg zones as function of time for the 12 \msun~model in \textbf{paper IV}. The optical depth to the $\gamma$-rays is initially quite high in the Fe/He-clumps, but as the trapping in these clumps decreases, so does the fraction deposited here. After 300-400 days, the deposition in the Fe/He clumps becomes dominated by positrons. Since the $\gamma$-ray trapping in the other core zones decreases as $t^{-2}$, their curves continue decreasing while the Fe/He curve turns up. The deposition in the H-zone is a monotonically increasing function with time as long as the total $\gamma$-ray deposition exceeds the total positron deposition. This is because the deposition in the H-envelope falls off more slowly than $t^{-2}$, as a larger and larger fraction of the $\gamma$-rays escape from the core. Clearly, the best time to study the Fe/He (and Si/S) zone is for times later than $\sim$ 1000 days, something exploited in \textbf{Papers I} and \textbf{II}. The best time to study the oxygen zones is instead as early as possible in the nebular phase, the basis of papers \textbf{III} and \textbf{IV}.

\section{Ionizations, excitations, and heating produced by radioactivity}
\label{sec:ieh}
\begin{figure}[htb]
\centering
\includegraphics[width=0.49\linewidth]{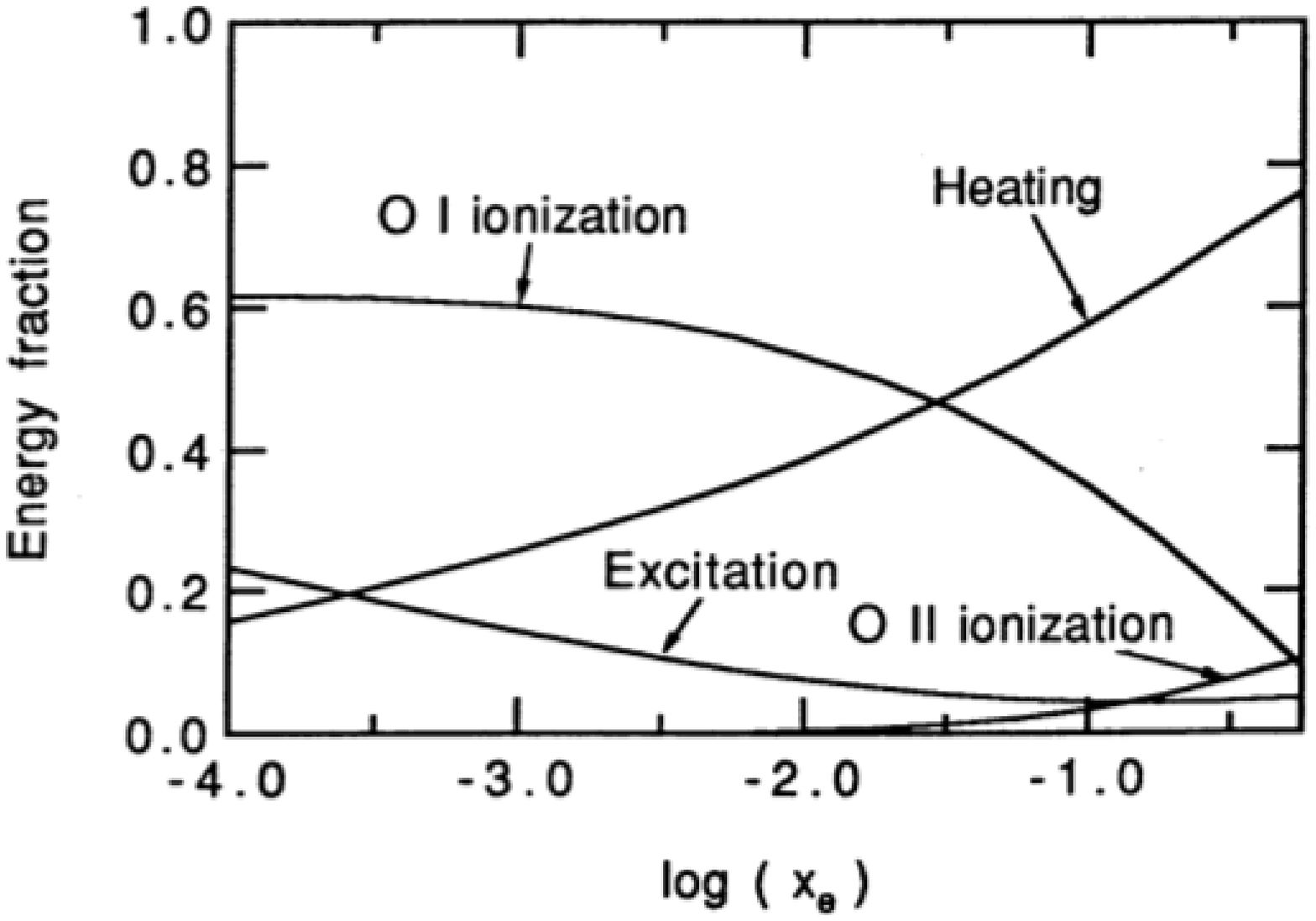}
\includegraphics[width=0.49\linewidth]{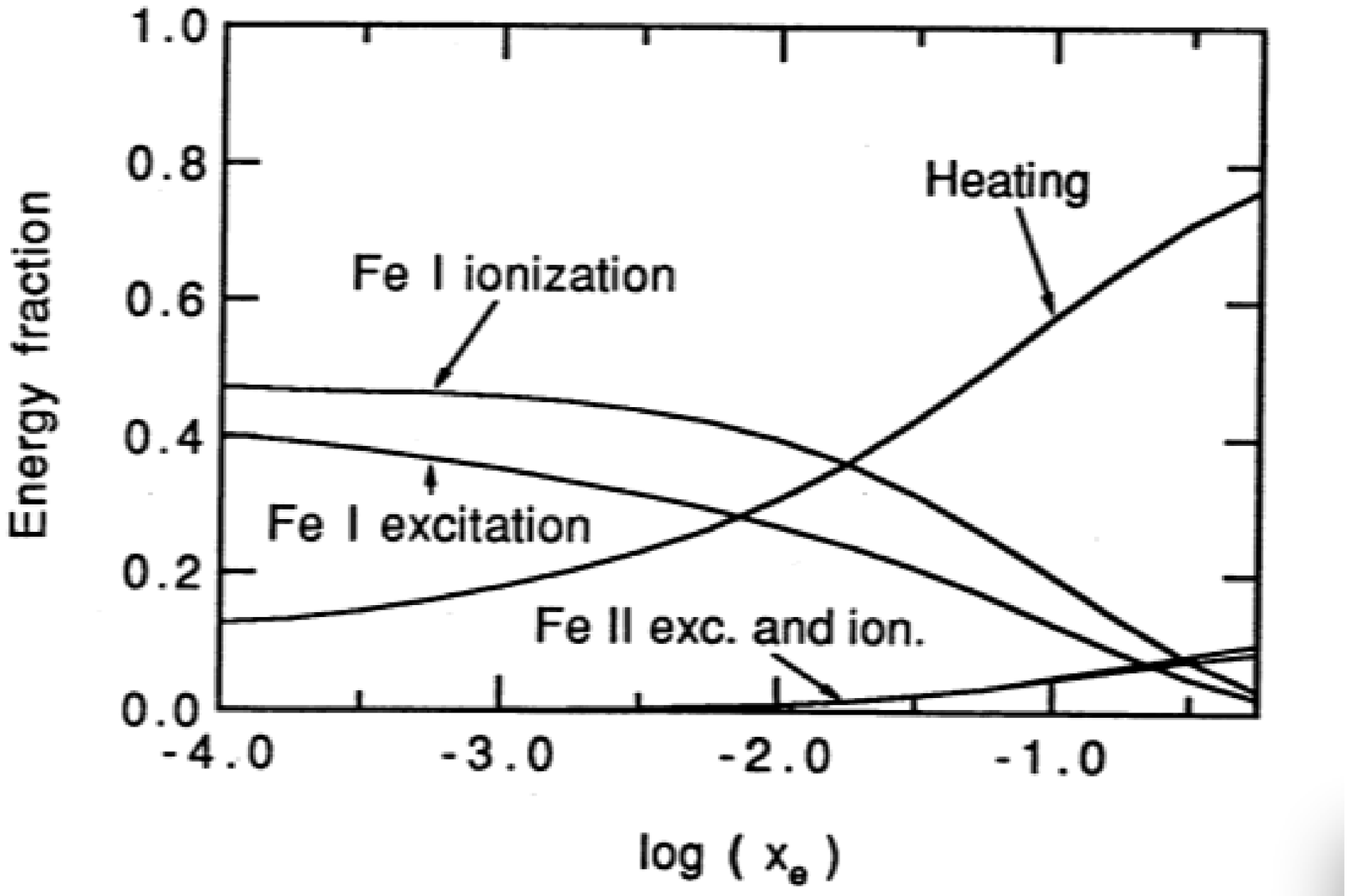}
\caption{The fraction of radioactive powering going into heating, ionization, and excitation in a pure oxygen plasma (left) and a pure iron plasma (right), as function of the electron fraction $x_e$. From KF92.}
\label{fig:kf92ox}
\end{figure}

Each ionization by a $\gamma$-ray or high-energy lepton leads to the creation of a new high-energy electron. These secondary electrons will excite and ionize the gas just as the primary ones. Furthermore, once the electrons are slowed down to keV energies, interaction with the thermal electron pool becomes important, leading to heating. In this way, the radioactive decays lead to ionizations, excitations, and heating of the gas. At each point in time and space we will have a distribution describing the number of non-thermal electrons per energy interval. If this distribution can be determined, the rates for all these processes can be calculated by integration over the relevant cross sections. Of course, these rates are precisely what determines the distribution, so the system has to be solved self-consistently. It was first solved in the SN context by \citet{Axelrod1980}, using the Continuous Slowing Down approximation (in which the electrons are assumed to lose only a small part of their energy in each interaction), and later by \citet{Lucy1991}, \citet{Xu1991} and KF92, using exact solutions. These authors found that the solution is insensitive to the exact properties (type and energy) of the original input particles, and to the density. The only major determinants are the chemical composition (the ion abundances) and the electron fraction. Fig. \ref{fig:kf92ox} shows the solutions for pure oxygen and iron plasmas.

Throughout this thesis, the KF92 code (with some updated atomic data) has been used for computing the non-thermal ionization, excitation, and heating rates. To save computation time, only ions with number fractions over $10^{-3}$ were included in the solutions. For the others, the average ionization rate for the corresponding ionization degree was used, whereas the non-thermal excitation rates were put to zero.

\section{Temperature}
\label{sec:temp}
Many of the processes important for the thermal structure of SNe are the same as in H I regions, discussed by \citet{Dalgarno1972}. Both environments are low-ionization nebulae heated mainly by non-thermal processes. 

The first law of thermodynamics states that the energy content $U$ of a parcel of gas changes as
\begin{equation}
\frac{dU(t)}{dt} = Q(t) - p(t)\frac{dV(t)}{dt}~,
\label{eq:td1}
\end{equation}
where $Q(t)$ is the net heating rate of the parcel, $Q(t)=H(t)-C(t)$, where $H(t)$ is the heating rate and $C(t)$ is the cooling rate, $p(t)$ is the pressure, and $V(t)$ is the volume. Cooling may, in general, occur by conduction, convection, or radiation, but in SNe only radiative cooling is important. For an ideal mono-atomic gas we have $U(t) = 3/2N(t)k_BT(t)$, where $N(t)$ is the total number of particles (atoms and electrons) at time $t$. If $N_a$ is the total number of atoms, then $N(t) = N_a(1+x_e(t))$, where $x_e(t)$ is the electron fraction. Further, $p(t) = n(t)k_BT(t)$, where $n(t)=N(t)/V(t)$. We also have, for a homologously expanding gas, $V(t)=V_0(t/t_0)^{-3}$, so $dV(t)/dt=3V(t)/t$. Eq. \ref{eq:td1} can then be written as
\begin{equation}
 \frac{dT(t)}{dt} = \frac{h(t) - c(t)}{\frac{3}{2}k_B n(t)} - \frac{2T(t)}{t} - \frac{T(t)}{1+x_e(t)}\frac{d x_e(t)}{dt}~,
\label{eq:dTdt}
\end{equation}
where $h(t)$ and $c(t)$ are now the heating and cooling rates per unit volume. The last term is usually much smaller than the other ones and may be neglected.

The cooling $c(t)$ is usually dominated by collisional excitations giving atomic line radiation, just as in H I and H II regions. This line cooling increases with temperature, since a higher temperature means that a larger fraction of the electrons have high enough energy to excite any given transition. If collisional deexcitations are unimportant, $c(t)$ also increases with density as $n_c(t) n_e(t)$, where $n_c(t)$ is the number density of cooling atoms and $n_e(t)$ is the electron number density. 

Since also the adiabatic cooling term, $-2T(t)/t$, increases with temperature, a cold gas has weak cooling compared to the (presumed temperature-independent) heating. If heating of this gas is initiated, only the $h(t)$ term in Eq. \ref{eq:dTdt} will be important and the temperature will increase. For each increase, $c(t)$ will increase, and so $h(t)-c(t)$ will become smaller. The adiabatic cooling increases as well if $T$ grows more quickly than $\propto t$. Eventually, heating and cooling will match, making $dT/dt$ small, and the temperature stabilizes.



The \emph{thermal equilibrium approximation} corresponds to setting 
\begin{equation}
h(t) = c(t)~, 
\label{eq:tea}
\end{equation}
turning the differential equation \ref{eq:dTdt} into an algebraic equation. Since both heating and cooling are radiative (if we count radioactive decay particles as radiation), this condition corresponds to \emph{radiative equilibrium} (in the co-moving frame). 
From Eq. \ref{eq:dTdt}, the approximation is good if (neglecting the last term)
\begin{equation}
\frac{h(t)}{\frac{3}{2}k_Bn(t)}, \frac{c(t)}{\frac{3}{2}k_Bn(t)} \gg ~\dot{T}(t) + \frac{2T(t)}{t},\\
\end{equation}
which can be written as
\begin{equation}
t_{\rm heat}(t), t_{\rm cool}(t) \ll t_{\rm change}(t)~,
\label{eq:ssc}
\end{equation}
where $t_{\rm change}(t)$ is the time-scale for change in physical conditions (density, temperature)
\begin{equation}
t_{\rm change}(t) = \left(\frac{\dot{T}(t)}{T(t)} + \frac{2}{t}\right)^{-1}~,
\end{equation}
and $t_{\rm heat}(t)$ and $t_{\rm cool}(t)$ are the time-scales for adding and removing the thermal energy content, respectively, at a given time:
\begin{eqnarray}
t_{\rm heat}(t) = \frac{\frac{3}{2}k_Bn(t)T(t)}{h(t)} = \frac{\frac{3}{2}k_BT(t)}{H_{\rm pp}(t)}\\
t_{\rm cool}(t) =   \frac{\frac{3}{2}k_Bn(t)T(t)}{c(t)} = \frac{\frac{3}{2}k_BT(t)}{C_{\rm pp}(t)}~,
\end{eqnarray}
where $H_{\rm pp}(t)$ and $C_{\rm pp}(t)$ are the heating and cooling rates per particle. By inspection of Eq. \ref{eq:dTdt}, it is is clear that either both conditions in Eq. \ref{eq:ssc} will be fulfilled, or none of them will. It is therefore enough to compare one of the time-scales, usually chosen as $t_{\rm cool}$, to $t_{\rm change}$.

The temperature will be able to significantly change no faster than on the shorter of the radioactive decay time-scale and the expansion time-scale. If the latter is the shorter one, we can approximate $\dot{T}/T \sim T(t)/t$, which gives $t_{change}(t) = t/3$. If the radioactive time-scale $\tau_{decay}$ is shorter, we can write $\dot{T(t)}/T(t) \sim 1/\tau_{decay}$, and then $t_{change} \sim \tau_{decay}$. Thermal equilibrium corresponds to to having the cooling time-scale (or equivalently, the heating time-scale)  shorter than the smallest value of $t/3$ and $\tau_{decay}$.



\begin{figure}[htb]
\centering
\includegraphics[width=0.7\linewidth]{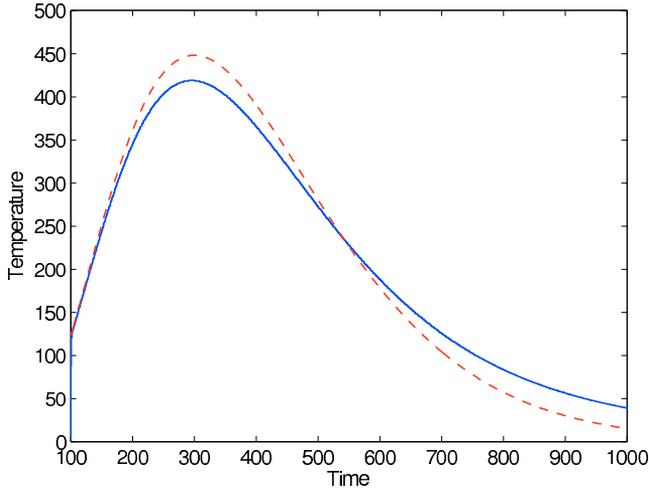}
\caption{Example calculation of the temperature evolution $T(t)$ for an exact solution (solid line), and a thermal equilibrium approximation (dashed line). The approximate solution starts breaking down at $\sim$ 200 days, giving an \emph{overestimate} of the temperature (and thereby the thermal emission) from $\sim$ 200-500 days. The loss of cooling efficiency (in the exact solution) means that deposited energy is stored up in the gas. This causes the approximate solution to \emph{underestimate} the temperature (and thereby the thermal emission) at late times.}
\label{fig:Tex}
\end{figure}

The code uses the thermal equilibrium approximation to compute the temperature at any given point and  time. As Fig. \ref{fig:Tex} illustrates with an example calculation, the steady-state approximation can cause both under and overestimates of the actual temperature.
 
\subsection{Heating rates}
\label{sec:heatingrates}
Heating may in general occur by radioactivity (non-thermal electron collisions), photoionizations, line absorptions followed by collisional deexcitations, free-free absorptions, exothermic charge transfer reactions, and Penning ionizations:
\begin{equation}
h = h_{nt} + h_{pi} + h_{cd} + h_{ff} + h_{ct} + h_{pe}~.
\end{equation}
These calculation of these are described in \textbf{paper II} (Sect. 2.3.3). 


\subsection{Cooling rates}
Cooling may in general occur by atomic line cooling, recombinations, free-free emission, endothermic charge transfer reactions, molecular line cooling, and dust cooling
\begin{equation}
c = c_{line} + c_{rec} + c_{ff} + c_{ct} + c_{mol} + c_{dust}~.
\end{equation}
These are also described in paper II, but I make some additional comments on them here.
\paragraph{Line cooling $c_{line}$}

The cooling in a particular line is in general not equivalent to the emission in the line, as population of the upper level may occur also by other processes than by energy transfer from the thermal electron pool. Instead we have to consider how the thermal electrons lose and gain energy by exciting and deexciting the transition. Letting $l$ and $u$ denote the lower and upper level:
\begin{equation}
c_{line}^{lu} = \Delta E_{lu}n_e \left[n_l q_{lu}(T)-n_u q_{ul}(T)\right]~,
\end{equation}
where $\Delta E_{lu}$ is the transition energy, and $q_{lu}(T)$ and $q_{ul}(T)$ are the rates for upward and downward collisions, related by
\begin{equation}
q_{lu}(T) = \frac{g_u}{g_l}q_{ul}(T)e^{-\Delta E_{lu}/k_BT}~,
\end{equation}
where $g_u$ and $g_l$ are the statistical weights. 
We can therefore write
\begin{equation}
c_{line}^{lu} = \Delta E_{lu} n_e q_{ul}(T)\left[ \frac{g_u}{g_l}n_l e^{-\Delta E_{lu}/k_BT} - n_u\right]~.
\label{eq:linecooling}
\end{equation}

If the upper level population $n_u$ exceeds $\frac{g_u}{g_l}n_l e^{-\Delta E_{lu}/k_BT}$ (its LTE value $n_u^*$ in the two-level approximation), the transition therefore induces net heating instead of cooling. This is often the case for low-lying H I and He I transitions, where non-thermal excitations and recombinations over-populate the excited states \citep{deKool1998}. 

The collisional cross sections typically have a $1/V_e^2$ dependence \citep{OB}, where $V_e$ is the velocity of the thermal electrons. The effect is that, even though more encounters occur per second in a hotter gas (in proportion to $V_e$), there are actually fewer collisions occurring (for a given electron energy) by a factor $V_e \propto T^{1/2}$. The $q_{lu}$ coefficients therefore have a temperature-dependence
\begin{equation}
q_{lu}(T) = T^{-1/2}C_{lu}(T)~,
\end{equation}
where $C_{lu}(T)$ are transitions-specific functions, weakly dependent on $T$.

Now, consider a two-level atom with only collisions and radiative decays occurring. Clearly, the cooling in this particular situation \emph{is} equal to the emissivity in the line (since no other process is allowed to populate the upper level):
\begin{equation}
c_{line}^{lu} = n_u \Delta E_{lu} A_{ul} \beta_{ul}~.
\end{equation}
where $\beta_{ul}$ is the escape probability. The upper level population $n_u$ has its maximum value in LTE: 
\begin{equation}
n_u* = \frac{g_u}{g_l}n_l e^{-\Delta E_{lu}/k_BT}~.
\end{equation}
But if we insert this into Eq. \ref{eq:linecooling}, we get $c_{line}^{lu}=0$! The resolution to this paradox lies in the realization that LTE and any net radiative losses are not compatible; strict LTE requires that all deexcitations are collisional. The radiative cooling corresponds, equation-wise, to the potentially small deviation of $n_u$ from its LTE value ($n_u*$)
\begin{eqnarray}
n_u = n_u^*\left(1 - \frac{A_{ul}\beta_{ul}}{q_{ul}n_e}\right) \rightarrow \\
c_{line}^{lu} = \Delta E_{lu} n_e q_{ul}(t)\left(n_u^*\frac{A_{ul}\beta_{ul}}{q_{ul}(T)n_e}\right) = n_u^* \Delta E_{lu} A_{ul} \beta_{ul}~.
\end{eqnarray}
The ratio $A_{ul}\beta_{ul}/q_{ul}n_e$ may become very small, and the actual computation of the cooling rate then constitutes taking the difference between two much larger numbers. This may potentially cause problems by numerical cancellation \citep{Heath}. The machine precision at double precision is $\sim 10^{15}$. The minimum value of $A_{ul}\beta_{ul}/q_{ul}n_e$ is of order $1/n_e$, so for $n_e \gtrsim 10^{15}$ cm$^3$, numerical cooling calculations by this method break down, and one instead has to use the emissivity combined with some method to estimate the non-thermal contribution. This situation, fortunately, occurs well above the electron number densities in nebular phase SN environments. Potential problems may instead arise by insufficient accuracy in the NLTE levels computations. If we compute these to an accuracy of 1\%, but the cooling corresponds to a difference of 0.1\% between upward and downward collision rates, our cooling rates become wrong! This situation is usually salvaged by the fact that the low-lying levels that are important for the cooling converge more rapidly than the high-lying levels. When global convergence is reached at the 1\% level, the accuracy in the important levels is therefore much higher.


An important point is that line cooling cannot be quenched by increasing the density so that the collisional deexcitations increase; this effect is always superseded by an increase in the number of upward collisions. The maximum cooling rates are achieved in LTE. 

\paragraph{Recombination cooling $c_{rec}$}

The recombination cooling rate per unit volume equals the recombination rate times the average energy of the recombining electrons. This value is $\sim 0.8 k_BT$ for hydrogen at 5000 K \citep{OB}, which we use for all atoms. Thus,
\begin{equation}
c_{rec} = n_e \sum_k \alpha_k(T) n_k 0.8 k_BT~,
\end{equation}
where $\alpha_k(T)$ is the total recombination rate for ion $k$.

\paragraph{Free-free cooling $c_{ff}$}
The free-free cooling is given by \citep[e.g.][]{OB}
\begin{equation}
c_{ff} = 1.42\e{-27} n_{\rm e} T^{1/2} \sum_{\rm k} Z_{\rm k}^2 g_{\rm ff}^k n_{\rm k}~,
\end{equation}
where summation occurs over all ions $k$. We follow \citet{deKool1998} and use a constant Gaunt factor $g_{\rm ff}^k=1.3$.

\paragraph{Charge transfer cooling $c_{ct}$}
Charge transfer cooling may occur by endothermic reactions, which are included as negative terms in the net charge transfer heating rates (Sec. \ref{sec:heatingrates}).

\paragraph{Molecular cooling $c_{mol}$}
Molecules are efficient coolers due to their rich level structure. The models in this thesis do not include calculation of the molecular abundances, and so we cannot compute the molecular cooling rates explicitly. In papers I and II, where the spectra anyway are rather insensitive to the temperature, we set $c_{mol}=0$. In papers III and IV,we attempt a more realistic treatment be setting $c_{\rm mol}=h$ in the O/Si/S and O/C zones, where SiO and CO are expected to form.

Significant work has already been done for the inclusion of molecules in the code, and will hopefully be achieved in the near future. 

\paragraph{Dust cooling $c_{dust}$}
Dust can be a coolant by inelastic collisions of the electrons with the grain surfaces. We do not include this process in the code, so $c_{dust}=0$. 


\section{Ionization}
\label{sec:ionization}

In nebular phase SNe, the gas is far from LTE, so we have to compute the ionization balance from the rate equations. 
Letting $x_{j,i}$ denote the fraction of element $j$ in ionization state $i$, the time-derivative of $x_{j,i}$ is
\begin{equation}
\frac{dx_{j,i}}{dt}  = \Gamma_{j,i-1}x_{j,i-1} - \left(\psi_{j,i} + \Gamma_{j,i}\right) x_{j,i} + \psi_{j,i+1} x_{j,i+1}~,
\label{eq:exaction}
\end{equation}
where $\Gamma_{j,i}$ is the ionization rate per particle and $\psi_{j,i}$ is the total recombination rate per particle.
The \emph{ionization equilibrium approximation} corresponds to setting the time-derivative to zero, which gives equations of the form

\begin{equation}
\Gamma_{j,i-1}x_{j,i-1} = \psi_{j,i} x_{j,i}~.
\label{eq:approxion}
\end{equation}
Together with the number conservation equations (one for each element)
\begin{equation}
\sum_{i=0}^{Z(j)} x_{j,i} = 1~,
\end{equation}
and the charge conservation equation 
\begin{equation}
n_e = \sum_j n_j \sum_{i=0}^{Z(j)} i x_{j,i}~,
\end{equation}
the ionization balance, i.e. the solutions for all $x_{j,i}$, can be computed. The system of equations is non-linear because the ionization and recombination rates generally depend on the ion abundances themselves (through dependence on ion densities for charge transfer reactions and on the electron number density for recombination rates), and is solved by the Newton-Raphson method. 

The ionization equilibrium approximation is valid for
\begin{eqnarray}
\frac{dx_{j,i}}{dt}&\ll& \Gamma_{j,i-1}x_{j,i-1} + \psi_{j,i+1}x_{j,i+1} \nonumber \\
          & \ll &  \left(\Gamma_{j,i} + \psi_{j,i}\right) x_{j,i}~,\nonumber \\
\end{eqnarray}
with the same reasoning as in Sect. \ref{sec:temp}, either both conditions will be fulfilled, or none of them will. We can therefore choose the one easier to test. This is, usually, the second one. We can then write
\begin{equation}
t_{\rm out}^{j,i}(t) \ll \left(\frac{dx_{j,i}/dt}{x_{j,i}}\right)^{-1} \sim \mbox{min}[t, \tau_{decay}]~,
\end{equation}
where
\begin{eqnarray}
t_{\rm out}^{j,i}(t) &=& \frac{1}{\Gamma_{j,i} + \psi_{j,i}}
\end{eqnarray}
Since $t_{\rm out}^{j,i} \leq t_{\rm rec}^{j,i}$, where
\begin{eqnarray}
t_{\rm rec}^{j,i}(t) &=& \frac{1}{\psi_{j,i}}\\
\end{eqnarray}
ionization equilibrium is fulfilled if $t_{rec}^{j,i} \ll \mbox{min}[t,\tau_{decay}]$. Assuming the recombination rate to be dominated by electron recombinations, we have
\begin{equation}
t_{\rm rec}^{j,i} = \frac{1}{\alpha_{j,i} n_e}~.
\end{equation}
The recombination rate is of order $\alpha_{j,i} \sim 10^{-12}$ cm$^3$s$^{-1}$ at a few thousand degrees, so the recombination time-scale is, for $x_e=10^{-2}$
\begin{equation}
t_{\rm rec}^{j,i}\sim 0.3~\mbox{days} \left(\frac{M}{10~\msun}\right)^{-1}\left(\frac{V}{3000~\kms}\right)^3\left(\frac{x_e}{0.01}\right)^{-1}\left(\frac{t}{1~\mbox{year}}\right)^{3}
\end{equation}
The core region of SNe will therefore have a recombination time scale shorter than $t$ and $\tau_{decay}$ for a decade or so. The density gradient in the outer envelope is so steep, however, that the recombination time-scale there becomes long already after $1-2$ years \citep{Fransson1993}. 

The ionization rates will decline exponentially with the radioactivity, whereas the recombination rates will evolve roughly with density as $t^{-3}$. For small $t$, the $t^{-3}$ decline is more rapid, but at later times the exponential decay is the more rapid. The result is that the ionization degree reaches a maximum after some time, and then declines. If we approximate $\Gamma_{j,i}x_{j,0} = C e^{-t/\tau_{decay}}$ and $\alpha_{j,1}=constant$, the solution for the ionization fraction in a single-species gas is
\begin{equation}
x_e(t) = \tilde{C} e^{-t/2\tau_{decay}}t^{3/2}~,
\label{eq:xe}
\end{equation}
which reaches its maximum at $t=3\tau_{decay}$, so we can expect the ionization degree to peak around $\sim$ 1 year ($3\cdot 111$ days), independent of the \iso{56}Ni mass as well as the ejecta mass and velocity.

\begin{figure}[htb]
\centering
\includegraphics[width=0.7\linewidth]{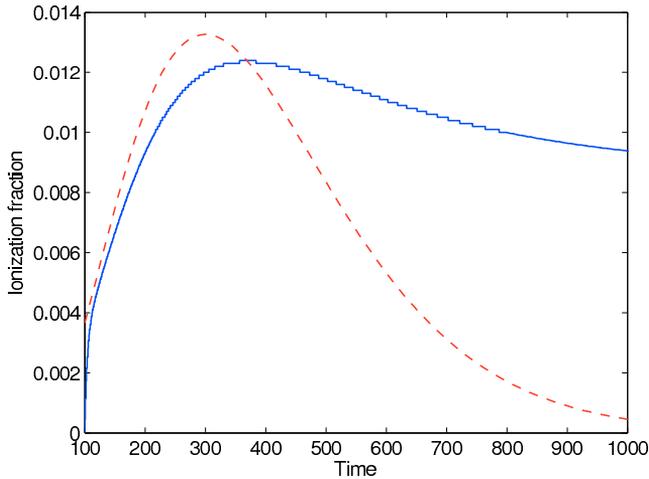}
\caption{Example calculation of the ionization fraction of a two-level system using the exact solution (solid blue), and the ionization equilibrium approximation (dashed red).}
\label{fig:freeze}
\end{figure}

Fig. \ref{fig:freeze} shows an example calculation using the exact equation (Eq. \ref{eq:exaction}) and the ionization equilibrium approximation (Eq. \ref{eq:approxion}). Just as for the temperature, the ionization fraction is initially overestimated, but then underestimated.

\subsection{Ionization rates}

Ionizations can occur by radioactivity (non-thermal electron collisions), photoionizations, or by charge transfer:
\begin{equation}
\Gamma_{j,i} = \Gamma_{j,i}^{\rm nt} + \Gamma_{j,i}^{\rm pi} + \Gamma_{j,i}^{\rm ct}~.
\end{equation}
In SN nebular phase environments, the temperature and density are too low for collisional ionizations by thermal electrons to be important.

\paragraph{Non-thermal ionization rates ($\Gamma_{j,i}^{\rm nt}$)} These are obtained from the solution to the degradation of radioactive products described in Sect. \ref{sec:ieh}.  
In this model, ionizations may occur for any of the outer shell electrons, as well as for electrons in the shell inside that. The total ionization rate should be the sum over the individual contribution from these. Letting $E_{j,i}$ denote the energy per second and unit volume going into ionization of ion $j,i$, and letting $n_{j,i}^{gm}$ denote the ground-multiplet number density, and $\chi_{j,i}^k$ the ionization potential of bound electron $k$, the ionization rate for the ground state as a whole is 
\begin{equation}
\Gamma_{j,i}^{\rm nt,exact} = \frac{E_{j,i}^1}{n_{j,i}^{gm} \chi_{j,i}^1} + \frac{E_{j,i}^2}{n_{j,i}^{gm} \chi_{j,i}^2} + ... = \frac{E_{j,i}}{n_{j,i}^{gm} \bar{\chi}_{j,i}}
\end{equation}
where
\begin{equation}
\bar{\chi}_{j,i} = \left(\frac{E_{j,i}^1}{E_{j,i}}\frac{1}{\chi_{j,i}^1} + \frac{E_{j,i}^2}{E_{j,i}}\frac{1}{\chi_{j,i}^2} +...\right)^{-1}~.
\label{eq:exactpotential}
\end{equation}
We currently approximate the averaged ionization potential $\bar{\chi}_{j,i}$ with the outermost electron ground state potential, $\bar{\chi}_{j,i} = \chi_{j,i}^{valence}$ and leave the ionized atoms in their ground states. The assumption of only ground multiplet ionizations is a good approximation in the nebular phase when few atoms are in excited states. We thus put
\begin{equation}
\Gamma_{j,i}^{\rm nt} = \frac{E_{j,i}}{n_{j,i}^{gm}\chi_{j,i}^{valence}}~.
\end{equation}
An exact treatment would involve using the exact expression (Eq. \ref{eq:exactpotential}), but then also creating some ions with electron gaps, and following the subsequent fluorescence and Auger events.

\paragraph{Photoionization rates  ($\Gamma_{j,i}^{\rm pi}$)} The photoionization rates are, in general, related to the mean intensity of the radiation field $J_\nu$ (see Chapter 5). If a fraction $y_l$ of a particular ion are in excitation state $l$, the contribution by level $l$ to the photoionization rate of the ion is
\begin{equation}
\Gamma_{j,i,l}^{\rm pi} = y_l \int_{\nu0(l)}^\infty \sigma_{l,\nu} \frac{4\pi J_\nu}{h\nu}d\nu~,
\end{equation}
where $\sigma_{l,\nu}$ is the photoionization cross section, and $\nu_0(l)$ is the threshold. The total ionization rate is thus
\begin{equation}
\Gamma_{j,i}^{\rm pi} = \sum_l y_l \int_{\nu0(l)}^\infty \sigma_{\nu,l} \frac{4\pi J_\nu}{h\nu}d\nu~.
\end{equation}
 One major advancement in this thesis is the
development of a radiative transfer calculation to be able to compute these rates, which is done by direct counting of the photoionization events in the Monte Carlo simulations. Details of these computations can be found in \textbf{paper II} (Sect. 2.3.1) and \textbf{paper III} (Appendix A). The photoionization cross sections we use are the ones for ionization to the ground state in the ionized atom. One exception is for O I ionizations, where ionizations to the first few excited states are included. 

\paragraph{Charge transfer rates  ($\Gamma_{j,i}^{\rm ct}$)} 
Charge transfer (CT) refers to reactions where an electron is transferred from one atom/ion to another. Letting superscripts denote the level of ionization, ionization of $X^i$ by $Y^{i'}$ can be written
\begin{equation}
X^i + Y^{i'} \rightarrow X^{i+1} + Y^{i'-1}
\end{equation}

The total ionization rate is the sum of the rates with each individual atom/ion.
\begin{equation}
\Gamma_{j,i}^{\rm ct} = \sum_{j',i'} \xi_{j,i,j',i'}^{ct,ion} n_{j',i'}~.
\end{equation}
The reaction rates can be large, $\xi_{j,i,j',i'}^{ct,ion} \sim 10^{-9}$ cm$^3$s$^{-1}$, especially if, in addition to fulfillment of selection rules, the energy defect of the reaction is small, $\lesssim~0.6$ eV \citep{Melius1974}. Often, this is fulfilled by reactions occurring to some excited state in one of the reaction products.

Both the non-thermal and photoelectric ionization rates can be expected to decline exponentially with the
radioactive decay. Charge transfer ionization rates, however, are proportional to the ion density, which Eq. \ref{eq:xe} combined with $n\propto t^{-3}$ shows to decline as $\sim e^{-t/2\tau}t^{-3/2}$. Thus, we have roughly

\begin{equation}
\frac{\Gamma_{j,i}^{\rm ct}}{\Gamma_{j,i}^{\rm nt} + \Gamma_{j,i}^{\rm pi}} \propto e^{t/2\tau}t^{-3/2}~,
\end{equation}
which has a \emph{minimum} at $t=3\tau$. The implication is that charge transfer has its smallest impact after about one year, but plays an increasingly important role for the ionization balance at late times. 

In highly ionized plasmas, charge transfer is less important because of the Coulomb repulsion between ions. But in the low-ionization environment of nebular phase SNe, they are critical, and lack of atomic data is a major obstacle.

\subsection{Recombination rates}

Recombinations can occur by radiative recombinations, dielectronic recombinations, or charge transfer. 
\begin{equation}
\psi_{j,i} = \psi_{j,i}^{\rm rr} + \psi_{j,i}^{\rm dr} + \psi_{j,i}^{\rm ct}
\end{equation}
Collisional recombinations are unimportant. Stimulated recombinations can be treated as negative photoionizations, but have not been considered in this thesis.

\paragraph{Radiative recombinations $\psi_{j,i}^{\rm rr}$}
The radiative recombination rate is
\begin{equation}
\psi_{j,i}^{\rm rr, exact} = \bar{\alpha}_{j,i}(T) n_e~,
\end{equation}
where $\bar{\alpha}_{j,i}(T)$ is the radiative recombination coefficient for the relevant ion, averaged over the internal states of the ion:
\begin{equation}
\bar{\alpha}_{j,i}(T) = \sum_l y_l \alpha_{j,i,l}(T)
\end{equation}
Since most atoms will be in the ground multiplet, we approximate $\bar{\alpha}_{j,i}(T) = \alpha_{j,i,gm}(T)$, so
\begin{equation}
\psi_{j,i}^{\rm rr} = \alpha_{j,i,gm}(T) n_e
\end{equation}
The temperature dependence for the recombination coefficient is typically $T^{-1/2}$. 

\paragraph{Dielectronic recombinations $\psi_{j,i}^{\rm dr}$} Dielectronic recombinations involve energy transfer to one of the bound electrons rather than emission of a free-bound photon; it is thus radiation-less. The atom is left in a state with two excited electrons. One of them may be ejected again in a so called autoionization, or they may both cascade to the ground state. Dielectronic recombination rates are usually significant only at very high temperatures ($\sim 10^6$ K), but can be comparable to the radiative recombination rates for some ions at low temperature \citep{Nussbaumer1983}. They are currently not included in the code:
\begin{equation}
\psi_{j,i}^{\rm dr} = 0~.
\end{equation}

\paragraph{Charge transfer recombinations $\psi_{j,i}^{\rm ct}$}
The charge transfer recombination rate is
\begin{equation}
\psi_{j,i}^{\rm ct} = \sum_{j',i'} \xi_{j,i,j',i'}^{ct, rec} n_{j',i'}~.
\end{equation}
As mentioned above, the reaction rates may be large, $\xi_{j,i,j',i'}^{ct, rec} \sim 10^{-9}$ cm$^3$s$^{-1}$, compared to $\alpha_{j,i} \sim 10^{-12}$ cm$^3$ s$^{-1}$ for recombinations with free electrons. In a neutral gas, where the number density of neutral species is much larger than $n_e$, it is clear that $\psi^{\rm ct}$ may be orders of magnitudes larger than $\psi^{\rm rr}$. Charge transfer recombination therefore dominates the recombination rate for many ions, especially the ones with high ionization potential that can ionize many other elements. One example of this is oxygen ions in SN cores; large abundances of C I, Na I, Mg I, Si I, and others lead to charge transfer recombinations dominating radiative ones, quenching oxygen recombination lines (\textbf{paper II}).

\section{Excitation}



Just as for the ions, we can set up equations for the population of excited states within each ion. Letting $y_{i,l}$ denote the fraction of some element in ionization state $i$ and excitation state $l$, we have (if someone knows how to write fun equations, let me know!)
\begin{eqnarray}
\frac{dy_{i,l}}{dt} &=& \sum_{l'} y_{i-1,l'}\Gamma_{i-1,l',i,l} + \sum_{l'} y_{i+1,l'}\psi_{i+1,l',i,l} + \nonumber \\ 
& & \sum_{l' \ne l} y_{i,l'} P_{l',l} - y_{i,l} \left(\sum_{l' \ne l} P_{l,l'} + \sum_{l'} \Gamma_{i,l,i+1,l'} + \sum_{l'}\psi_{i,l,i-1,l'}\right), 
\label{eq:excitation}
\end{eqnarray} 
where the terms represent ionizations into the level ($\Gamma_{i-1,l',i,l}$), recombinations into the level ($\psi_{i+1,l',i,l}$), radiative and collisional transitions from all other (internal) levels into the level ($P_{l',l}$), radiative and collisional rates out of the level to all other (internal) levels ($P_{l,l'}$), ionizations from the level ($P_{i,l,i+1,l'}$), and recombinations from the level ($\psi_{i,l,i-1,l'}$).

The \emph{excitation equilibrium approximation} corresponds to setting the time-derivative to zero. Analogously with the thermal and ionization equilibrium, the approximation is valid if the deexcitation time-scale is short compared to $t$ and $\tau_{decay}$. Since deexcitation rates are never smaller than $\sim 10^{-3}$ s$^{-1}$ for \emph{excited} states, this condition is always fulfilled for them. Thus, for \emph{given} ion abundances, we can always compute the excitation structure in steady-state.

The resulting system of equations would be linear (for a given $n_e$) if all lines would be optically thin. In general, the system is non-linear because the deexcitation rates $P_{l,l'}$ depend on the lower level populations for optically thick lines. We solve the system of equations by straight iteration; guess the solutions $y_{i,l}$, compute the deexcitation rates $P_{l,l'}$, make a new solution for $y_{i,l}$, and iterative until convergence.

To enhance the convergence stability, the model atoms should be kept as small as possible. 
Also reasonable execution times is a strong reason to keep the model atoms limited in size. All in all, the construction of the model atoms is a complex optimisation problem, where a compromise between completeness, accuracy, stability, and efficiency has to be found. 

\subsection{Transitions rates}
\label{sec:transrates}
Photoionization and recombination rates ($\Gamma$,$\psi$) have been described earlier in the chapter. Here, I describe the internal transition rates ($P_{l,l'}$). Letting $l$ and $u$ denote the lower and upper levels, the internal transition processes are spontaneous emission ($R_{u,l}^{spont}$), stimulated emission ($R_{u,l}^{stim}$), photoabsorption ($R_{l,u}^{abs}$), thermal collisions ($C_{u,l}^{thermal}, C_{l,u}^{thermal}$), and non-thermal collisions ($C_{u,l}^{non-thermal}, C_{l,u}^{non-thermal}$):
\begin{eqnarray}
P_{u,l}& =& R_{u,l}^{spont} + R_{u,l}^{stim} + C_{u,l}^{thermal} + C_{u,l}^{non-thermal}\\
P_{l,u} &=& R_{l,u}^{abs} + C_{l,u}^{thermal} + C_{l,u}^{non-thermal}~.
\end{eqnarray}

\paragraph{Spontaneous emission rate $R_{u,l}^{spont}$}
The spontaneous emission rate is given by the Einstein A-coefficient. However, in the Sobolev approximation, one cancels the contribution by this spontaneous emission to the $R_{l,u}^{abs}$ and $R_{u,l}^{stim}$ rates (see Chapter 5, Sect. \ref{sec:lt}), and instead uses
\begin{equation}
R_{u,l}^{spont} = A_{ul}\beta_{ul}^S~,
\end{equation}
where $\beta_{ul}^S$ is the Sobolev escape probability (see Chapter 5). For some lines, $\beta_{ul}^S$ is replaced by an effective escape probability $\beta_{ul}^{eff}$, which includes absorption in the continuum and/or in other lines. See Sect. \ref{sec:lt} for more details.

\paragraph{Radiative absorption rates $R_{l,u}^{abs}$}
The general expression for the radiative absorption rate is 
\begin{equation}
R_{l,u}^{abs,exact} = B_{lu}\int_0^\infty J_\nu \phi_\nu d\nu = B_{lu}\bar{J}_{lu}
\end{equation}
As mentioned above, contributions by spontaneous emission in the line itself should be excluded in $\bar{J}_{lu}$ in the Sobolev approximation. One may then show (see Sect. \ref{sec:lt}) that the expression to use is

\begin{equation}
R_{l,u}^{abs} = B_{lu}J_\nu^b\beta_{ul}^S
\end{equation}
where $J_\nu^b$ represent the radiation field in the blue wing of the line (where no line photons exist). Our method for computing $R_{l,u}^{abs}$ is described in \textbf{paper III} (Appendix A).

\paragraph{Stimulated emission rate $R_{u,l}^{stim}$}

We obtain the stimulated rates from $R_{l,u}^{abs}$ according to
\begin{equation}
R_{u,l}^{stim} = \frac{g_l}{g_u}R_{l,u}^{abs}
\end{equation}

\paragraph{Thermal collision rates $C_{u,l}^{thermal}$}
No selection rules apply for collisional transitions, so they may be important for all lines. 
Typically only electron collisions are considered, as the thermal velocities, and hence the collision frequencies, are much lower for atoms and ions 
\begin{equation}
\frac{V_{\rm atoms}}{V_{\rm electrons}} = \left(\frac{m_e}{Am_p}\right)^{1/2}=\frac{1}{43~A^{1/2}}~,
\end{equation}
so the collision rates scale as
\begin{equation}
\frac{C_{l,u}^{thermal}(\rm atoms)}{C_{l,u}^{thermal}(\rm electrons)} \sim \frac{1}{43~A^{1/2}}\frac{n_a}{n_e}~.
\end{equation}
Collisions with atoms can thus be ignored in astrophysical plasmas of high enough ionization degree ($x_e \gtrsim 1/(43~A^{1/2})$). In addition, collisions with atoms are less efficient than collisions with electrons and ions due to the long-range electric force (and ions are obviously no more common than electrons). 
Still, when $x_e$ falls below $\sim$1\%, hydrogen atom collisions ($A=1$) may become important. 


The thermal collision rates are
\begin{eqnarray}
C_{u,l}^{thermal} &=& 8.629\e{-6} n_e  T^{-1/2}\frac{\Upsilon_{l,u}(T)}{g_u}~, \\
C_{l,u}^{thermal} &=& C_{u,l}^{thermal}\frac{g_l}{g_u}e^{-\Delta E_{ul}/k_BT}~,
\end{eqnarray}
where $\Upsilon_{l,u}$ are the \emph{velocity-averaged collision strengths}, which are slowly varying functions of $T$. When they are unknown, approximations have to be used. For allowed transitions, van Regemorter's formula is commonly used. For forbidden transitions, rates are often computed according to \citet{Allen1973}, where an effective collision strength of 1 is assumed. 

\paragraph{Non-thermal collision rates $C_{l,u}^{non-thermal}$}
For the computation of the non-thermal excitation rates in KF92, the individual levels in the ground multiplets are generally assumed to be populated according to their statistical weights, which is true if $k_BT \gg \Delta E$, where $\Delta E$ is the internal energy scale of the multiplet. For typical $\Delta E$ of $\sim$0.1 eV, this corresponds to $ T \gg 400$ K. For the very late-time models considered in \textbf{papers I} and \textbf{II}, this condition is not fulfilled. We therefore modified all cross sections by factors $e^{-E_i/k_BT}$ in the Spencer-Fano routine, where $E_i$ is the excitation energy of level $i$, which approximates the deviations of the level populations from the statistical weights limit. The result is more energy going into channels from the low-lying states in the ground multiplet. 

\section{Global convergence}
\citet{Axelrod1980} showed that the equation sets \ref{eq:tea}, \ref{eq:approxion}, and \ref{eq:excitation} (with the time-derivative set to zero) may be solved for global convergence by iteration. All systems are non-linear. The most commonly used solution technique for a set of non-linear equations is the \emph{Newton-Raphson method} \citep{NR}. Here we make a starting guess for the solution, linearize all equations at that point, and solve the resulting linear system. 
In other words, we replace $f_i(x,y,..)=0$ with $f_i^{lin}(x,y,..)=0$, where
\begin{equation}
f_i^{lin}(x,y,..) = f_i(x_0,y_0,..) + \frac{\delta f_i}{\delta x}(x-x_0) + \frac{\delta f_i}{\delta y}(y-y_0) + ...
\label{eq:lin}
\end{equation}
We then repeat the procedure for the new solution point. For this scheme to work, the linearized equations must have a solution at least in the \textit{same direction} at the real one. This is true in the vicinity of the true solution, but may not be true far away. The success of the Newton-Raphson method therefore relies on making good enough starting guesses. We use LTE solutions for ionization and excitation as starting guesses in the NLTE computations.

\subsection{Solving a set of linear equations}

The computation of each Newton-Raphson step requires the solution of a linear system of equations.
The techniques for this are well known, and numerical packages are readily
available for the task. But let us have a look at the estimated requirements of disk space and computation times for the problem.\\
\\
A linear problem ($\mathbf{A}\mathbf{x}=\mathbf{b}$) with $10^4$ variables has a matrix $\mathbf{A}$ with $10^{8}$ entries. Even \textit{storing} that many numbers takes $\sim 8\cdot 10^{8}\sim 1$ GigaBytes. The number of floating point operations needed to solve a system of size $N$ using the standard method of LU decomposition\footnote{which is the most efficient technique for large systems.}
is \citep{Heath}
\begin{equation}
N_{fp} \sim \frac{1}{3}N^3~,
\end{equation}
which tells us that it is very expensive to include more atoms in the ionization balance and more levels in the excitation structure. It also tells us that, in general, it is a good idea to split the problem into chunks which are solved iteratively, rather than attempting to solve everything in one big calculation. 
There is however, a problem in that we intuitively can expect convergence difficulties to increase with increasing fragmentation of the equation set. We therefore need to strike a balance in the amount of splitting that is done.

For an atom size of $N=10^3$, we have $N_{fp} = 3 \cdot 10^{8}$.  
On a 1 GHz CPU, the computation time is then $\sim$ 0.03 seconds\footnote{Today's CPU's can do $\sim$ 10 flops per clock cycle.}. Since we need to solve for $\sim 10^2$ atoms in $\sim$10 zones over perhaps $\sim 10^2$ iterations, the total CPU requirements for the statistical equilibrium is then $\sim$1 hour. We parallelize the computations over the zones, reducing the run times to $\sim$ 0.1 hours. 

\subsection{Convergence}

The stability of a Newton-Raphson algorithm often increases if we use a damped step-size rather than the full one. This means that the step implemented is some fraction $h$ of the one computed from the solution to $f_i^{lin}=0$. We then reduce the risk of over-shooting the solution to a non-convergent regime. In the literature, we find that \cite{LucyTPII} uses a damped step size $h=0.8$, and \citet{Lucy1999} uses $h=0.5-0.8$, furthermore with a temperature step size limit of $\Delta T=200~K$. The convergence can be further enhanced by making $h$ and $\Delta T$ adjustable to smaller values if the step sizes are not decreasing for each iteration.
We currently use $h=0.5$ for the temperature steps and $h=0.8$ for the level population steps. 

The partial derivatives needed for Eq. \ref{eq:lin} are computed numerically, with a step-size of $\Delta x/x=0.01$. There are methods for computing the optimal value of this step-size, discussed in \citet{Heath}. 

The criteria for convergence are difficult to specify. We mark convergence when the relative Newton-Raphson step, $|\Delta x_i/x_i|$, is less than 1\% in all directions. For the thermal equilibrium, we require $\Delta T/T < 1$\%, and also that the net heating rate is smaller than 1\% of the total heating rate. For the ionization balance, we similarly require that the net ionization rate is smaller than 1\% of the total ionization rate. 

The excitation structure is more difficult to quantify a convergence criterion for, because there are often relatively large spurious changes for high levels with low populations. \citet{Lucy2003} uses a criteria that the relative change averaged over the five lowest levels should be smaller than $10^{-5}$. The exact choice of convergence criteria is still under development for the code here.

Handling of run-away situations has also been implemented. Temperature steps leading to negative temperatures, or temperatures above $3\e{4}$ K, are halved until they give a value within the $0-3\e{4}$ range. Similar boundary checks on the ionization and excitation solutions have been implemented. When convergence fails for some zone within 100 iterations, the zone is flagged, and the solutions are set to their previous ones. Final models are then checked for flags, and if any are present the error has to be tracked down by more detailed investigation.

\chapter{Radiation transport}
The 20th century witnessed the development of radiative transfer theory and techniques for computing the radiation
field inside and outside astrophysical objects. The basic theory is rather straight-forward, but
the resulting equation systems are often large, complex, and difficult to solve. The fundamental difficulty
is that the radiation emitted at one point in the object may influence the physical conditions at another
point, which emission may influence the conditions at the first point, etc. Much of the literature and research 
is focused on how one may construct approximate equations that are easier
to solve than the exact ones, and how one solves the original equation system in the most efficient way.

\section{Basic theory}
The flow of radiative energy at frequency $\nu$, position $\bf{x}$, time $t$, and in direction $\hat{\textbf{n}}$ is described by the \emph{specific intensity}
\begin{equation}
I_\nu(\textbf{x},t,\hat{\textbf{n}}) = \frac{dE}{dt~dA~d\nu~d\Omega}~~,~\mbox{erg}~\mbox{s}^{-1} \mbox{cm}^{-2} \mbox{Hz}^{-1} \mbox{Ster}^{-1}~,
\label{eq:EORT}
\end{equation}
which is a complete description of the radiation field, apart from polarization. The \emph{mean intensity} is the angle-averaged value of $I_\nu$
\begin{equation}
J_\nu(\textbf{x},t,\hat{\textbf{n}}) = \frac{1}{4\pi}\oint I_\nu(\textbf{x},t,\hat{\textbf{n}}) d\Omega~.
\end{equation}
The change in $I_\nu$ as the beam travels from point \textbf{x} to point \textbf{x+dx} comprises emission, represented by the \emph{emission coefficient} $j_\nu$ (erg s$^{-1}$ cm$^{-3}$ ster$^{-1}$), and absorption, represented by the \emph{absorption coefficient} $\alpha_\nu$ (cm$^{-1}$): 
\begin{equation}
\label{eq:rt0}
I_\nu(\textbf{x}+\textbf{dx}, t+dt,\hat{\textbf{n}}) - I_\nu(\textbf{x},t,\hat{\textbf{n}}) = j_\nu(\textbf{x},t,\hat{\textbf{n}}) ds - \alpha_\nu(\textbf{x},t,\hat{\textbf{n}}) I_\nu(\textbf{x},t,\hat{\textbf{n}})  ds~,
\end{equation}
where we let $ds$ be a distance along the direction $\hat{\textbf{n}}$. 

The ratio of emission and absorption coefficients is called the \emph{source function}
\begin{equation}
S_\nu(\textbf{x},t,\hat{\textbf{n}}) = \frac{j_\nu(\textbf{x},t,\hat{\textbf{n}}) }{\alpha_\nu(\textbf{x},t,\hat{\textbf{n}})}~~,~~\mbox{erg}~\mbox{s}^{-1} \mbox{cm}^{-2} \mbox{Hz}^{-1} \mbox{Ster}^{-1}~,
\end{equation}
which has the same dimension as $I_\nu$. 

Rewriting the LHS of Eq. \ref{eq:rt0} as
\begin{eqnarray}
I_\nu(\textbf{x}+\textbf{dx}, t+dt,\hat{\textbf{n}}) - I_\nu(\textbf{x},t,\hat{\textbf{n}}) = \nonumber \\
= \frac{\partial I_\nu}{\partial t}dt + \frac{\partial I_\nu}{\partial s}ds \nonumber \\
 = \frac{\partial I_\nu}{\partial t}\frac{dt}{ds}ds + \frac{\partial I_\nu}{\partial s}ds\nonumber \\
= \left(\frac{1}{c}\frac{\partial}{\partial t} + \frac{\partial}{\partial s}\right) I_\nu ds~,
\end{eqnarray}
we obtain
\begin{equation}
\left(\frac{1}{c}\frac{\partial}{\partial t} + \frac{\partial}{\partial s}\right) I_\nu(\textbf{x},t,\hat{\textbf{n}}) = j_\nu(\textbf{x},t,\hat{\textbf{n}})  - \alpha_\nu(\textbf{x},t,\hat{\textbf{n}}) I_\nu(\textbf{x},t,\hat{\textbf{n}})~.
\end{equation} 
In flows with $V/c \ll 1$ and no radioactivity, the time-derivative term is much smaller than the space derivative and may be ignored \citep{Mihalas1984}. In SNe, the requirement for dropping the time-derivative is
\begin{equation}
 \frac{V}{c}\ll \mbox{min}[1,\frac{t_{decay}}{t}]~.
\end{equation}

In spherically symmetric models, we can replace $\textbf{x}$ with radius $r$, and direction $\hat{\textbf{n}}$ with the angle $\mu$ relative to the radial direction, so that
\begin{eqnarray}
I_\nu(\textbf{x},t,\hat{\textbf{n}}) \rightarrow I_\nu(r,t,\mu)\nonumber \\
j_\nu(\textbf{x},t,\hat{\textbf{n}}) \rightarrow  j_\nu(r,t,\mu)\nonumber\\
\alpha_\nu(\textbf{x},t,\hat{\textbf{n}}) \rightarrow \alpha_\nu(r,t,\mu)\nonumber~.
\end{eqnarray}

In spherical coordinates, Eq. \ref{eq:EORT} (with the time-derivative dropped) becomes
\begin{equation}
\label{eq:eortss}
\left[\mu\frac{\partial}{\partial r} + \frac{(1-\mu^2)}{r}\frac{\partial}{\partial \mu}\right]I_\nu(r,t,\mu)
= j_\nu(r,t,\mu)  - \alpha_\nu(r,t,\mu) I_\nu(r,t,\mu)~.
\end{equation}
As we shall see, $j_\nu(r,t,\mu)$ generally depends on $\mu$-integrals of $I_\nu$, and this is therefore a partial integro-differential equation in $r$ and $\mu$. Solutions may be approached by \emph{difference equations} of the form in Eq. \ref{eq:eortss}, or by \emph{integral equations}. For moving media like SNe, the integral equations become cumbersome, and the differential form is preferred \citep{Mihalas1978}.


\subsection{Emission coefficient}
\label{sec:ecoff}
The emission coefficient, $j_\nu(r,t,\mu)$, is different for different observers due to Doppler effects (Sect. \ref{sec:mm}). Below, we derive expressions for the coefficient in the rest frame of the emitting gas.

Emission of photons may occur by spontaneous processes, stimulated processes, or by scattering. Because stimulated rates are proportional to $I_\nu$, just as absorptions, they are most easily treated as negative absorptions (Sect. \ref{sec:bb}), so only spontaneous processes and scattering contribute to the emission coefficient.

Spontaneous emission involves a change of energy state for the electron, which at any given time is either free ($E_{\rm kinetic}+E_{\rm potential} > 0$) or bound ($E_{\rm kinetic}+E_{\rm potential} < 0$). Depending on the state of the electron before and after the photon emission, emission processes can then be divided into free-free (ff), free-bound (fb), and bound-bound (bb). In total we have 

\begin{equation}
j_\nu = j_\nu^{ff} +j_\nu^{fb} + j_\nu^{bb} + j_\nu^{scatt}~.
\end{equation}

\paragraph{Free-free emission $ j_\nu^{ff}$}
A free electron can emit photons if it is accelerated in the Coulomb field of an ion. This is called free-free emission, or Bremsstrahlung. The emission coefficient for ion $k$ (representing some element and ionization stage), is
\begin{equation}
j_{\nu}^{ff,k}=\frac{1}{4\pi} 1.42 \cdot 10^{-27}n_e n_{k} Z_k^2 g_{ff}^k T^{1/2}~,
\end{equation}
where $n_e$ is the electron number density, $n_{k}$ is the number density of the ion, and $g_{ff}^k$ is the Gaunt factor. Free-free emission will play a role mainly at early times when ionization and temperature are high. Just as for H II regions, it is not very competitive for radiating thermal energy.

The total free-free emission coefficient is the sum over all ions
\begin{equation}
j_\nu^{ff} = \sum_k j_\nu^{ff,k}~.
\end{equation}

\paragraph{Free-bound emission $j_\nu^{fb}$}
When a free electron is captured into a bound state $l$ of an ion $k$ by radiative recombination, free-bound (recombination) emission occurs. The emission coefficient for reactions with ion $k+1$ is
\begin{eqnarray}
j_{\nu}^{fb,kl} &=& \frac{1}{4\pi}\alpha_{\nu}^{fb,kl}(T)n_e n_{k+1} h\nu,~~\nu> \chi_{kl}\nonumber \\
                &=&0~,~~~~~~~~~~~~~~~~~~~~~~~~~~~~~~~\nu< \chi_{kl}~,
\end{eqnarray} 
where $n_e$ is the electron number density, $n_{k+1}$ is the ion number density, $\alpha_\nu^{fb,kl}(T)$ the reaction rate, and $\chi_{kl}$ is the ionization potential.
Often, the detailed $\nu$-dependence of the reactions rates $\alpha_{\nu}^{fb,kl}$ are unknown. In such instances, an approximate treatment is to distribute the total reaction rate $\alpha_{tot}^{fb,kl}(T)$ (which often \emph{is} known) over a box-shaped profile of width $h \Delta \nu = k_BT$: 
\begin{eqnarray}
j_{\nu}^{fb,kl} &\approx& \frac{1}{4\pi}\frac{\alpha^{fb,kl}_{tot}(T)}{\Delta \nu} n_e n_{k+1} h\nu,~~\chi_{kl} < h\nu < \left(\chi_{kl} +h\Delta \nu\right)\nonumber \\
       &=  & 0~~~~~~~~~~~~~~~~~~~~~~~~~~~~~~~~, ~\mbox{otherwise}~.
\end{eqnarray}
This approach has been used in this thesis for all atoms except hydrogen, where exact free-bound reaction rates are available.

The total free-bound emissivity is obtained by adding up the contribution by all ions and levels 
\begin{equation}
j_{\nu}^{fb} = \sum_{k,l} j_{\nu}^{fb,kl}~.
\end{equation}

\paragraph{Bound-bound emission $j_\nu^{bb}$}
Excited, bound electrons transition spontaneously to lower states by certain transition rates $A_{ul}$ (s$^{-1}$), where $u$ and $l$ denote the upper and lower states. The emission coefficient is
\begin{equation}
j_{\nu}^{bb,kul,exact} = \frac{1}{4\pi}n_{ku} A_{kul} h\nu \psi(\nu-\nu_0^{kul})~,
\end{equation}
where $\psi(\nu-\nu_0^{kul})$ is the emission line profile.
In SNe, the velocity gradients are so large that the typical Doppler shifts become much larger than the thermal line widths. We can therefore approximate the line profile with the Dirac delta function.
\begin{equation}
 \psi(\nu-\nu_0^{kul}) = \delta(\nu-\nu_0^{kul})~.
\end{equation}
As shown in Sect. \ref{sec:lt}, in SNe one may replace the emissivity above with an effective emission coefficient
\begin{equation}
j_{\nu}^{bb,kul} = \frac{1}{4\pi}n_{ku} A_{kul}\beta_{kul}(n_{kl},n_{ku},t) h\nu \delta(\nu-\nu_0^{kul})~,
\end{equation}
where $\beta_{kul}$ is the escape probability of the line, and then skip doing the radiative transfer through the line itself. 

Bound-bound deexcitation can occur also by two-photon (2$\gamma$) emission, where the sum of the photon energies equals the transition energy. The line profile is then a smooth function, starting at $\nu_0^{kul}$ and extending to infinity. In Type II SNe, two such transitions are important; H I $2s-1s (\lambda=1215~\mbox{\AA}-\infty)$ and He I $2s-1s (\lambda=601~\mbox{\AA}-\infty)$. The H I two-photon emission is stronger than Ly$\alpha$ as long as
\begin{eqnarray}
A_{2\gamma} > A_{Ly\alpha}\beta_{Ly\alpha}^S &\rightarrow &\\
t &<& 8~\mbox{years} \left(\frac{M_H}{10~M_\odot}\right)^{1/2}\left(\frac{V}{3000~\mbox{km~s}^{-1}}\right)^{-3}~.
\end{eqnarray}
However, boosts in the escape probability due to overlapping lines may produce other estimates. 

The total bound-bound emissivity is
\begin{equation}
j_{\nu}^{bb} =  \sum_{k,u,l} j_{\nu}^{bb,kul} ~.
\end{equation} 

\paragraph{Scattering $j_\nu^{scatt}$}
\emph{Scattering} refers to photon-matter interactions where the photon continues on (in a new direction) with only a small frequency change. In the code, scattering is not included in the computation of the emissivity function, but is instead treated directly in the radiative transfer simulation (\textbf{Paper III}, Appendix A). However, for completeness I describe the process for constructing scattering emissivity below.

Scatterings in SNe mainly occur on free electrons, called Thomson scattering. The emission coefficient is
\begin{equation}
j_\nu^{scatt}(\mu) = \frac{1}{4\pi}\sigma_T n_e \oint_\Omega \int_0^\infty I_{\nu'}(\mu') R(\nu',\nu,\mu',\mu)d\Omega' d\nu'~,
\end{equation}
where the cross section is $\sigma_T=6.65\e{-25}$cm$^2$ (independent of wavelength), and $R(\nu',\nu,\mu',\mu)$ is the redistribution function. If we ignore the small energy transfer occurring and take $\nu=\nu'$, we can write
\begin{equation}
j_\nu^{scatt}(\mu) = \frac{1}{4\pi}6.65\e{-25}n_e \oint_\Omega I_{\nu'}(\mu') g(\mu',\mu)d\Omega'~,
\end{equation}
where the angular redistribution function is
\begin{equation}
g(\mu',\mu) = \frac{3}{4}\left[1+\cos^2 (\mu'-\mu)\right]~.
\end{equation}
Another scattering process is \emph{Rayleigh scattering}, which however has a much smaller cross section than $\sigma_T$, and is neglegible in SN environments (see Sect. \ref{sec:scattabs}).


\paragraph{Isotropy}
All spontaneous emission processes are isotropic in the rest frame of the gas since they involve thermal distributions of particles. Scattering, on the other hand, is anisotropic since it depends on the angular distribution of the incoming radiation field, and on the angular redistribution function $g(\mu',\mu)$. However, several authors find that this anisotropy introduces quite small effects \citep[e.g.][]{Auer1972}, so one can to good approximation take
\begin{equation}
g(\mu',\mu) = \mbox{constant}~.
\end{equation}
The total emission coefficient is then isotropic:
\begin{equation}
j_\nu(r,t,\mu) \rightarrow j_\nu(r,t)~.
\end{equation}

\subsection{Absorption coefficient}
\label{sec:acoff}
Just as for the emission coefficient, the absorption coefficient $\alpha_\nu(r,t,\mu)$ is different for different observers. Below, we derive the expression for the coefficient in the rest frame of the absorbing gas.

Absorption may occur by free-free processes, bound-free processes (photoionization), bound-bound processes, or by scattering:
\begin{equation}
\alpha_\nu = \alpha_\nu^{ff} + \alpha_\nu^{bf} + \alpha_\nu^{bb} + \alpha_\nu^{scatt}~.
\end{equation}

\paragraph{Free-free absorption  $\alpha_\nu^{ff}$}
The free-free absorption coefficient is
\begin{equation}
\alpha_\nu^{ff} = 3.7\e{8} T^{-1/2} n_e \nu^{-3}\left(1-e^{-h\nu/k_BT}\right)\sum_k Z_k^2 n_k g_{ff}^k~,
\end{equation}
where summation is done over all ions $k$.
The free-free optical depth at 1 $\mu$m is (ignoring the exponential, which is small)
\begin{equation}
\tau^{ff}_{1~\mu m} = 0.13~T_{5000~K}^{-1/2}\left(\frac{M}{10~M_\odot}\right)^2\left(\frac{V}{3000~\mbox{km}~\mbox{s}^{-1}}\right)^{-2}\left(\frac{x_e(t)}{0.1} \right)^2\left(\frac{t}{100~\mbox{days}}\right)^{-2}~.
\end{equation}
It may therefore influence the early nebular phase, in particular the infrared spectra. Because of free-free absorption, we can not expect to see any far-infrared emission lines early on.

\paragraph{Bound-bound  absorption  $\alpha_\nu^{bb}$}
\label{sec:bb}
The line absorption coefficient is, dropping the $k$ index,
\begin{equation}
\alpha_\nu^{bb,lu} = \sigma_{lu}\phi(\nu-\nu_0^{lu}) n_{l} - \sigma_{ul}\psi(\nu-\nu_0^{lu}) n_{u}~,
\end{equation}
where $\phi(\nu-\nu_0^{lu})$ is the absorption line profile, and $\psi(\nu-\nu_0^{lu})$ is the emission line profile (which is the same for spontaneous and stimulated emission). The second term is the correction for stimulated emission. 
The cross sections are related to the Einstein B-coefficients as
\begin{equation}
\sigma_{lu} = \frac{h\nu_0^{lu}}{4\pi} B_{lu}~,
\end{equation}
and
\begin{equation}
\sigma_{ul} = \frac{h\nu_0^{lu}}{4\pi} B_{ul}~.
\end{equation}
Using the Einstein relations $B_{lu}/B_{ul}= g_{u}/g_{l}$ and $A_{ul}/B_{ul} = 2h\nu_{ul}^3/c^2$, and approximating the line profile as a box function of width $\Delta \nu$, we get
\begin{equation}
\alpha_{\nu}^{bb,lu} =  \frac{1}{8\pi} \frac{g_{u}}{g_{l}}A_{ul} {\lambda^2_{ul0}} (n_{l}-\frac{g_{l}}{g_{u}} n_{u})\frac{1}{\Delta \nu}~, |\nu-\nu_{lu0}| < \frac{\Delta \nu}{2}~.
\end{equation}
Since in homologous expansion $dV/ds = 1/t$, and the Doppler formula gives $\Delta V_{lu} = c \Delta \nu/\nu_{lu0} = \Delta \nu \lambda_{lu0}$, a line profile of velocity width $\Delta V_{lu}$ corresponds to a spatial distance of
\begin{equation}
L_{lu} = \Delta V_{lu} / \frac{dV}{ds} = \Delta \nu \lambda_{lu0} t~,
\end{equation}
which is called the \emph{Sobolev length}. Then, assuming a constant value of $\alpha_\nu^{bb,lu}$ over the distance $L_{lu}$, we obtain the line optical depth
\begin{equation}
\tau_\nu^{bb,lu} = \alpha_\nu^{bb,lu} L_{lu} = \frac{1}{8\pi}\frac{g_u}{g_l}A_{lu} \lambda_{lu}^3 t n_{l}\left(1-\frac{g_{l}}{g_{u}}\frac{n_{u}}{n_{l}}\right)~,~~|\nu-\Delta \nu|<\frac{\Delta \nu}{2}~,
\label{eq:taus}
\end{equation}
which is called the \emph{Sobolev optical depth} ($\tau^S$). The result holds for all line profiles. Note that since the density evolves as $\rho(t)\propto t^{-3}\propto R^{-3}$, the Sobolev optical depth scales as $R^{-2}$ (assuming constant ionization/excitation), just as in static media.

The size of the line interaction region relative to the size of the nebula is
\begin{equation}
\frac{L_{\rm line}}{L_{\rm nebula}}=\frac{L_{lu}}{Vt} = \frac{\Delta V_{lu}}{V} \sim \frac{1}{V}\left(\frac{k_BT}{Am_p}\right)^{1/2}~.
\end{equation}
At $T=5000~K$, $\Delta V_{lu}$ is of order 10 \kms~for hydrogen. For $V=10^4$ \kms, we then get $L_{\rm line}/L_{\rm nebula} \sim 10^{-3}$. It is thus a good approximation to assume constant physical conditions over the line interaction region as long as (1) clumping does not occur on a smaller scale than $\sim 10^{-3}$, and (2) the line is not so optically thick that the Lorentz wings are important (then $\Delta V_{lu} > \Delta V^{thermal}$). 

\paragraph{Bound-free absorption $\alpha_\nu^{bf}$}
Whereas calculation of bound-free absorption coefficients are available for the ground-state of all atoms \citep[we use data from][]{Verner1996}, they are more scarce for excited states. 
For hydrogenic atoms, the coefficient scales as

\begin{equation}
\alpha_\nu^{bf,n} = 7.9\e{-18} n_{n} \left(\frac{\nu}{\nu_0^{n}}\right)^{-3}g_{n}Z^4~,
\label{eq:hydrogenic}
\end{equation}
where $n$ is the principal quantum number. For non-hydrogenic ions, this expression is still a good approximation for levels with effective quantum numbers $n\gtrsim 10$. For other levels, the cross sections are more complex, often with resonances at certain frequencies that may differ by several orders of magnitude from the underlying level. However, the effective 'smearing' of frequencies in the differentially expanding SN environment likely allows lower-resolution approximations to be used to good accuracy. We use Eq. \ref{eq:hydrogenic} for all excited levels with unknown bound-free cross sections.

Just as stimulated bound-bound emission is most easily treated as negative absorption, stimulated recombinations are most easily treated as negative bound-free absorptions. Then, the absorption coefficient from level $i$ is \citep{Mihalas1970}
\begin{equation}
\alpha_\nu^{bf,corr} = \alpha_\nu^{bf} \left(1-\frac{n_k/n_k^*}{n_i/n_i^*}e^{-h\nu/k_BT}\right)~,
\end{equation}
where $n_k$ is the number density of the next ion, $n_i$ is the number density of the ionizing level, and starred quantities represent LTE values. At the relevant wavelengths and temperatures ($\sim 10^3$ \AA~and $\sim 10^4$ K), the exponential has value $\sim 10^{-6}$. We therefore ignore stimulated recombinations in the transfer.




\paragraph{Scattering  $\alpha_\nu^{scatt}$}
\label{sec:scattabs}
The scattering coefficient is, assuming contribution only from electron scattering
\begin{equation}
\alpha_\nu^{scatt}=\alpha^{scatt} = 6.65\e{-25}n_e~.
\end{equation}
The optical depth is, for a homogeneous hydrogen sphere
\begin{equation}
\tau^{scatt}  = 2.8 \left(\frac{M}{10~M_\odot}\right)\left(\frac{V}{3000~\mbox{km}~\mbox{s}^{-1}}\right)^{-2}\left(\frac{x_e(t)}{0.01}\right)\left(\frac{t}{100~\mbox{days}}\right)^{-2}~,
\end{equation}
It is indeed electron scattering that traps the radiation for the first $\sim$ 100 days in Type II SNe, causing the photospheric phase.

Another scattering process is \emph{Rayleigh scattering}, where photons scatter on bound-bound transitions of much higher energy than the photon energy ($\nu_{ul} \gg \nu$). The cross section is
\begin{equation}
\sigma_\nu^{scatt,R} = \sigma_T f_{ul}\frac{\nu^4}{\left(\nu_{ul}^2-\nu^2\right)^2}~,
\end{equation}
where $f_{ul}$ is the oscillator strength. The Rayleigh cross section is much smaller than  $\sigma_T$ since $\nu \ll \nu_{ul}$. However, if $x_e \ll 1$, the total absorption coefficient may come to exceed the Thomson one, as Lyman lines or metal resonance lines on neutral elements provide scattering targets. In low-velocity, low \iso{56}Ni-mass SNe, Rayleigh scattering may not be neglegible in early nebular phases.

\paragraph{Isotropy}
All absorption coefficients are isotropic in the rest-frame of the gas:

\begin{equation}
\alpha_\nu(r,t,\mu) \rightarrow \alpha_\nu(r,t)~.
\end{equation}

\begin{figure}[htb]
\includegraphics[width=1\linewidth]{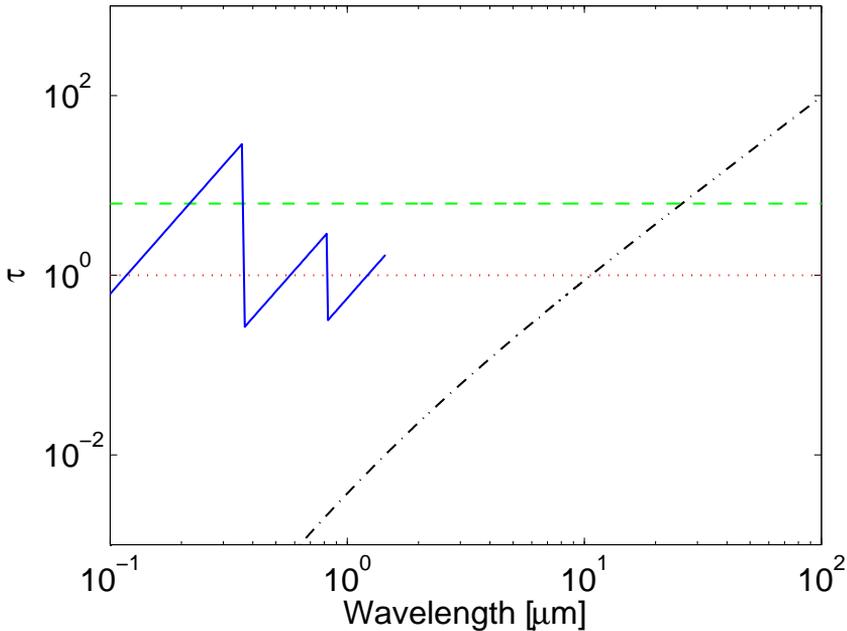}
\caption{The continuum optical depths for free-free absorption (dash-dotted), bound-free absorption (solid), and Thomson scattering (dashed), for a 10 \msun~hydrogen sphere at 150 days, with $V=3000$~\kms, $T=6000$ K and $x_e=0.05$. The bound-free absorption has been computed up to principal number $n=4$, above which LTE becomes a bad approximation.}
\label{fig:contop}
\end{figure}

Fig. \ref{fig:contop} shows the various continuum optical depths at 150 days for a 10 \msun~hydrogen sphere with $V=3000$~\kms, $T=6000$ K, $x_e = 0.05$, and an LTE excitation structure. We see that Thomson scattering (dashed line) and photoionization (solid line) are still important for UVOIR photons, whereas also free-free absorption (dash-dotted line) is important in the FIR. 

\begin{figure}[htb]
\centering
\includegraphics[width=1\linewidth]{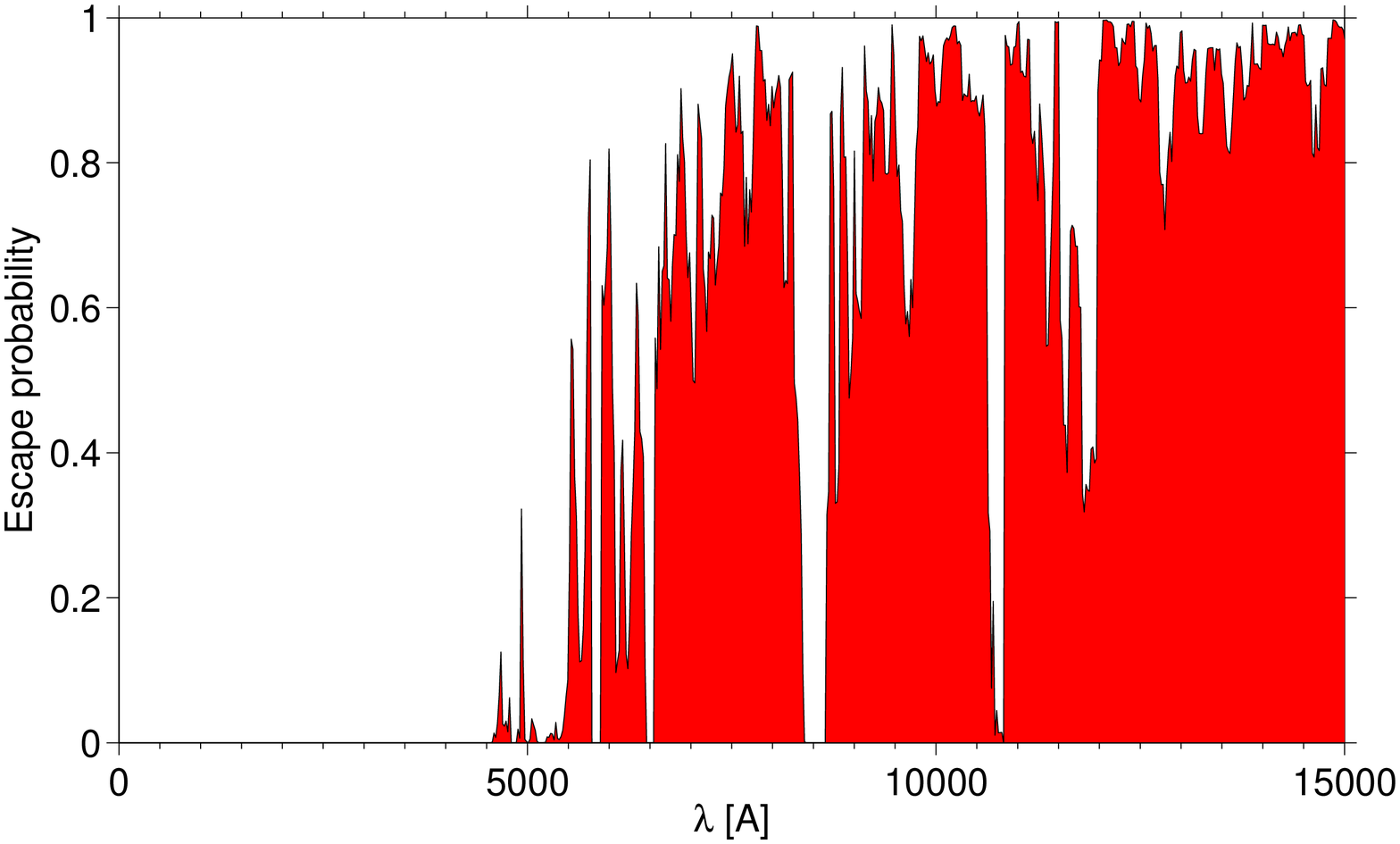}
\caption{The probability for a photon emitted from the center of the SN to escape the ejecta without being absorbed in a line, for a 19 \msun~model at 300 days.}
\label{fig:eprob}
\end{figure}

Fig. \ref{fig:eprob} illustrates the line opacity by plotting the probability for a photon emitted in the core of the 19 \msun~model in \textbf{paper III} to escape the SN, at 300 days. We see that below $\sim$ 5000 \AA~there is complete blocking, between $5000-12,000$ \AA~there is partial blocking, and after that there is little blocking.

\subsection{Lambda iteration}

Clearly, the emission and absorption coefficients, $j_\nu$ and $\alpha_\nu$, depend on the state of the matter (temperature, ionization, excitation), and on $I_\nu$, both directly (scattering terms) and indirectly (through the influence of $I_\nu$ on the state of the matter). Even if LTE is assumed, this coupling remains because the temperature is influenced by the radiation field. Thus, equation \ref{eq:eortss} is more difficult to solve than it may appear, and has to be solved in conjunction with the equations governing the the state of the matter (Chapter 3). 

If scattering is treated as isotropic, we in general have, for a time-independent problem, that $j_\nu$ and $\alpha_\nu$ depend on $J_{\nu'}$ at the same location, for all frequencies $\nu'$.
The mean intensities $J_\nu'$, in turn, depend on the emission and absorption coefficients at all other points and frequencies:
\begin{equation}
J_\nu(r) = \Lambda[j_{\nu'}(r'),\alpha_{\nu'}(r')],~\forall \nu',\forall r'~,
\end{equation}
where the \emph{$\Lambda$-operator} represents computation of $J_\nu(r)$ from $j_{\nu'}(r')$ and $\alpha_{\nu'}(r')$ by some method.

To solve the coupled equation systems, one starts with a guess for $j_\nu(r)$ and $ \alpha_\nu(r)$. A radiation field can then be computed by solving the radiative transfer equation (\ref{eq:eortss}) by some method. With this solution, new evaluations of the gas state can be made through the formulas described in Chapter 3. This allows new $j_\nu$ and $\alpha_\nu$ to be computed, and the procedure is repeated. 

In the stellar atmospheres context, $\Lambda$-iteration has proven to work well in the case of small or moderate optical depths, but converges slowly or not at all for large optical depths \citep{Mihalas1970}. The reason for the latter is that information is propagated only one optical depth per iteration through the grid, so will, in a random walk, take $\sim \tau^2$ steps from boundary-to-boundary. In the thick case, one may still devise tricks to enhance the convergence properties, generally referred to as \emph{accelerated lambda iteration (ALI)} or \emph{operator splitting} procedures (e.g Cannon 1973). The first widely used methods of this kind were developed by \citet{Scharmer1981, Scharmer1984}. 

In the nebular-phase SN context, the $\Lambda$-iteration scheme has quite benign properties since the optical depths are moderate, and in addition the gas state is mainly determined by radioactivity rather than by the UVOIR radiation field. As we saw i Chapter 3, the radioactive ionization, excitation, and heating rates are only weakly coupled to the gas state. The coupling between conditions at different points in the nebula is thus greatly reduced, and the convergence properties are improved. It has been demonstrated, with the code in this thesis and a similar code developed at MPA in Munich, that $\Lambda$-iteration without any strong acceleration schemes is rapidly convergent in nebular phase SN environments \citep{Maurer2011}.

Still, in the early phases radioactivity reaches only the inner parts of the nebula, and conditions in the outer parts are completely determined by the radiation field. If optical depths become high, ALI methods are needed.  \citet{Eastman1993} developed an ALI operator splitting technique for this situation, and \citet{Lucy2002, Lucy2003, Lucy2005}, \citet{Sim2007a}, and \citet{Kromer2009} developed Monte Carlo methods enforcing radiative equilibrium by the use of indestructible energy packets. The radiative equilibrium condition is otherwise just fulfilled asymptotically, and the enforcement from the first iterations speeds up the convergence.



\section{Moving media}
\label{sec:mm}
If the medium is moving, radiation transport is complicated by the fact that the absorption and emission coefficients differ between the rest frame of the gas (which is the one we have worked in so far) and the rest frame of the star (the observer frame), becoming direction-dependent due to the direction-dependence of Doppler shifts, aberration, and advection. Whereas aberration and advection involve effects of order $V/c$, Doppler shifts introduce major changes already when $V/V_{\rm thermal}$ becomes non-significant \citep{Mihalas1978}. It is therefore reasonable to take Doppler effects into account but neglect aberration and advection. 

The relationship between the observer-frame (unprimed) and comoving frame (primed) emission and absorption coefficients are (note that the primes were not written out in Sect. \ref{sec:ecoff}-\ref{sec:acoff}) \citep[e.g.][]{RL}
\begin{equation}
j_\nu(\mu) = j_{\nu'}^{'}\left(\frac{\nu}{\nu'}\right)^2 = j_{\nu\left[1-\frac{V(r)}{c}\cos \mu\right]}^{'}\left[1-\frac{V(r)}{c}\cos \mu\right]^{-2}
\end{equation}
and
\begin{equation}
\alpha_\nu(\mu) = \alpha_{\nu'}^{'}\left(\frac{\nu}{\nu'}\right)^{-1} =  \alpha_{\nu\left[1-\frac{V(r)}{c} \cos{\mu}\right]}^{'}\left[1-\frac{V(r)}{c}\cos \mu\right]~,
\end{equation}
where the last equalities are valid for spherically symmetric flows only. The effect of the motion on the continuum transfer is small, but for the line-transfer the situation may change completely if the frequency shifts become larger than the line widths.
The angular dependencies makes solution of the radiative transfer equation more complicated, and direct solution in the observer's frame is only feasible in certain cases \citep{Mihalas1978}. It is instead advantageous to compute all radiation-matter interaction in the comoving frame, the frame where the interacting matter is at rest, because here emission and scattering coefficients are isotropic. Since specific intensity can be shown to transform as
\begin{equation}
\label{eq:Inutrans}
I_\nu = I_{\nu'}'\left(\frac{\nu}{\nu'}\right)^3~.
\end{equation}
One may thus solve the differential equation of transfer in the comoving frame to update $I_\nu'$, and then use Eq. \ref{eq:Inutrans} to transform to the the lab frame or to the next comoving frame of interest. This is the method preferred in most computer codes, also used in this thesis, even though we transfer Monte Carlo packets rather than $I_\nu$ explicitly. (Sect. \ref{sec:MC}).

Since $V/c \ll 1$ even in SN explosions, it is not a bad approximation to consider only the wavelength shifts and set $j_{\nu} = j_{\nu'}^{'}$ and $\alpha_\nu = \alpha_{\nu'}^{'}$ \citep{Lucy1999,Lucy2005}. This is also justified by the moderate accuracy to which many other physical processes can be modelled in SNe, and the fact that we ignore the time-derivative term in the equation of transfer.

\subsection{Scattering}
A fundamental difference to static media is that scattering now introduces potentially large frequency changes for the photon in the star's frame. Consider a photon that moves in the same direction as the scattering particle, and is scattered 180$^{\rm o}$ backwards. Given that the scattering is coherent in the rest frame of the scattering particle, the photon suffers a redshift of $\Delta \nu/\nu = 2V/c$ in the star's frame, where $V$ is the velocity of the scatterer. If the photon instead moves anti-parallel to the scatterer, there is an energy gain of $2V/c$. Since the nebula is radiating, photons must find themselves moving outwards (same direction as the gas) more often than inwards (opposite direction as the gas). The net effect of scattering is therefore that energy is transferred from the radiation field to the gas. 

Electron scattering will tend to broaden and redshift the line profiles because any photon that escapes will have gone through additional frequency-varying direction changes, with more redshifting than blueshifting ones \citep[see e.g.][]{Auer1972}. The corresponding increase in path length for the photons increases the chance of line or continuum absorption, and so the escape of line radiation from the core will be made more difficult. The \emph{rest-frame} broadening due to thermal motions can be seen for the narrow lines in some Type IIn SNe, but is generally unimportant for the ejecta lines, where the broadening is dominated by the bulk flow.

\subsection{Line radiation}
\label{sec:lt}
In static media, a photon of a given frequency faces absorption by one or a few lines close to that particular frequency. Absorption can happen at any place in the nebula. In a moving medium, on the other hand, each line is only in resonance with the photon at one particular point along the trajectory. This means that a velocity gradient provides an \emph{escape mechanism} for the photon, shifting most of the atoms out of resonance with it.

On the other hand, there may now be other lines that shift \emph{into} resonance with the photon somewhere along the path. Thus, each photon can now be absorbed by \emph{many} lines, but for each line just at \emph{one} point. The effect of many lines being coupled together is called \emph{interlocking}, and their collective opacity is referred to as \emph{line blocking}. 

\paragraph{Sobolev approximation}

\citet{Sobolev1957} derived the solution for the line transfer in a moving medium with expansion velocities much larger than the intrinsic line widths. With the assumption of an isotropic and frequency-independent source function $S_{ul}$, the mean intensity weighted over the line profile becomes \citep{Castor1970}
\begin{equation}
\bar{J}_{ul} = \bar{J}_{ul}^{line} + \bar{J}_{ul}^{ext} = \left(1-\beta^S_{ul}\right)S_{ul} + \beta^S_{ul} J_{ul}^b~,
\end{equation}
where $\bar{J}^{line}_{ul}$ is the contribution to the mean intensity due to emission in the line itself, $\bar{J}^{ext}_{lu}$ is the contribution due to external photons, which approximately equals $J_{ul}^b$, the mean intensity in the blue wing of the line, and $\beta^S_{ul}$ is the \emph{Sobolev escape probability}, which for homologous flows is
\begin{equation}
\beta^S_{ul} = \frac{1-e^{-\tau^S_{ul}}}{\tau^S_{ul}}~,
\end{equation}
where $\tau^S_{ul}$ is defined by Eq. \ref{eq:taus}. The escape probability $\beta_{ul}^S$ is the angle and frequency-averaged probability that a photon emitted in the line will escape rather than be reabsorbed in the same line by another atom. 
The net radiative rate between the levels is
\begin{eqnarray}
\mathcal{R}_{ul} &=& \lefteqn{n_u A_{ul} + n_u B_{ul} \bar{J}_{ul} - n_l B_{lu}\bar{J}_{ul}}\nonumber \\
&=& n_u A_{ul} + n_u B_{ul} \left[(1-\beta^S_{ul})S_{ul} + \beta^S_{ul} J_{ul}^{b}\right] - n_l B_{lu} \left[(1-\beta^S_{ul})S_{ul} + \beta^S_{ul} J_{ul}^{b}\right]\nonumber \\
&=& n_u A_{ul}\beta^S_{ul} + (n_u B_{ul} - n_l B_{lu}) J_{ul}^b\beta^S_{ul}~,\\
\end{eqnarray}
where we have used
\begin{equation}
\label{eq:source}
S_{ul} = \frac{j^{line}_{ul}}{\alpha^{line}_{ul}} = \frac{n_u A_{ul}}{n_l B_{lu} - n_u B_{ul}}~.
\end{equation}
This means that instead of emitting a total flux $n_u A_{ul}$ and doing the radiative transfer through the line, we can emit a flux $n_u A_{ul}\beta_{ul}^S$ and start the transfer \emph{after} the line. We also recover the result stated in Sect. \ref{sec:transrates}, that the photoexcitation rate is given by
\begin{equation}
R_{l,u}^{abs} = B_{lu}J_{ul}^b\beta_{ul}^S~.
\end{equation}
\paragraph{Generalized escape probability}
The actual source function does not look like Eq. \ref{eq:source}, but rather like
\begin{equation}
S_{ul} = \frac{j^{line}_{ul} + j^{\rm other}_{ul}}{\alpha^{line}_{ul} + \alpha^{\rm other}_{ul}}~.
\end{equation}
\citet{Hummer1985} and \citet{Chugai1987} derived expressions for the escape probability when the $\alpha^{\rm other}_{ul}$ term is non-zero, having a contribution from continuum opacity. They find that one can write, denoting $R=\alpha_{ul}^{cont}/\alpha_{ul}^{line}$
\begin{equation}
\beta^{eff}_{ul} =\beta^S_{ul} + R\cdot f(\tau^S_{ul},R)~,
\end{equation}
where $f(\tau^S_{ul},R)$ is a function of order unity. For complete redistribution over the Doppler core
\begin{equation}
f(\tau^S_{ul},R) = 2\left(-\ln \sqrt{\pi}\left(R + \frac{2}{\tau^S_{ul}}\times \left[-\ln \sqrt{\pi}\left(R + \frac{4}{\tau^S_{ul}}\right)\right]^{1/2}\right)\right)^{1/2}~.
\end{equation}
For partial redistribution, \citet{Chugai1987} finds
\begin{equation}
f(\tau^S_{ul},R) = 1.87\left(\frac{a}{R}\right)^{1/4}~,
\end{equation}
where $a$ is the damping parameter. For $T=5000~K$ and $Z=1$:
\begin{equation}
a = \frac{\sum_{l < u} A_{ul}}{4\pi \Delta \nu} = 4\e{-4}\left(\frac{\sum_{l<u} A_{ul}}{10^8~s^{-1}}\right)\left(\frac{\lambda}{5000~\mbox{\AA}}\right)
\end{equation}

\begin{figure}[htb]
\centering
\includegraphics[width=1\linewidth]{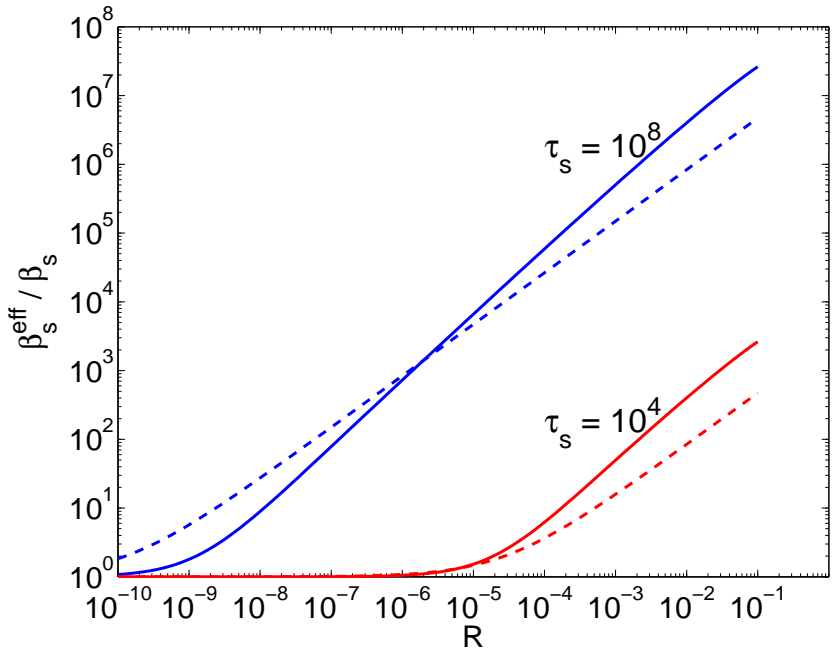}
\caption{The boost in escape probability due to continuum destruction as function of $R=\alpha_{ul}^{cont}/\alpha_{ul}^{line}$. The solid curves are for complete redistribution and the dashed ones for partial redistribution, using $a=10^{-3}$.}
\label{fig:boost}
\end{figure}

As an example, for Ly$\alpha$ in the core hydrogen zone in our model for SN 2004et at 400 days (\textbf{Paper III}), $R=10^{-12}, f_{pr}\sim 300, \mbox{ and }\beta^{eff} = 3\e{-10}\sim 4 \beta^S$. In Fig. \ref{fig:boost} we show the boost in escape probability as function of $R$ for two different values of $\tau^S$.

The escape probability may be further increased by absorption in neighboring lines. This has particular importance for the Ly$\alpha$ line. The $n=2$ population in H I will be affected if $A\beta^{eff}_{21} \gtrsim A_{2\gamma}$, i.e. $\beta^{eff}_{21} \gtrsim 4\e{-9}$. Several metal lines exist around Ly$\alpha$ that may bring $\beta_{21}^{eff}$ above this value. Table \ref{table:lyaneighbors} lists all lines within 2 \AA~ ($\approx 50 \Delta V^{thermal}$ at $T=5000$ K) on the red side of Ly$\alpha$ having $\tau^S > 10^{-10}\tau^S_{Ly\alpha}$ in the core hydrogen zone in the 12 \msun~progenitor model of \textbf{Paper IV}, at 300 days. It is clear from the $\tau^S_{\rm tot}/\tau^S_{\rm Ly\alpha}$ values that $\beta_{21}^{eff}$ may transcend $4\e{-9}$. Preliminary simulations have given $\beta_{Ly\alpha}^{eff}\approx 10^{-8}$ (Jerkstrand \& Fransson, in prep.), which is the value used in paper III and IV. 

\begin{table}[htbp]
\centering
\begin{tabular}{ c c c }
\hline
Line & $x$ & $\tau^S/\tau^S_{Ly\alpha}$\\
\hline
Fe II 1215.85 \AA & 4.5 & $2.7\e{-9}$\\
Fe II 1216.24 \AA& 14 & $1.2\e{-9}$\\
Fe II 1216.27 \AA & 15 & $1.2\e{-9}$\\
Fe II 1216.52 \AA & 21 & $1.3\e{-9}$\\
Fe II 1216.85 \AA & 30 & $1.2\e{-9}$\\
Cr II 1217.14 \AA& 37 & $8.3\e{-10}$\\
Fe II 1217.15 \AA& 37 & $1.5\e{-9}$\\
Fe II 1217.62 \AA& 49 & $1.0\e{-9}$\\
\hline
\end{tabular}
\caption{Lines between 1215.67 and 1217.67 \AA, in the core hydrogen zone in the 12 \msun~model in \textbf{Paper IV}, at 300 days. The $x$ values are the line separations in number of (hydrogen) Doppler widths at 5000 K.}
\label{table:lyaneighbors}
\end{table}

Another possibly important line overlap occurs between Ly$\beta$ (1025.72 \AA) and O I  2p$^3$P-3d$^3$D (1025.76 \AA). The Ly$\beta$ line-width is $\Delta \lambda=0.04$ \AA~at 5000 K, so the line separation is within the Doppler core. 
One may show that the absorption probability in the O I line is 
\begin{equation}
\beta^{\rm Bowen}_{\rm Ly\beta}= 2.8\e{-5}\left(\frac{f_{\rm O}}{f_{\rm O, solar}}\right)
\end{equation}
where $f_{\rm O}$ is the number fraction of O I relative to H I, and $f_{\rm O, solar}$ is the solar value. This is the value we use in paper III and IV.

\citet{Chugai1980} demonstrated that the Sobolev approximation holds also under partial redistribution ($j^{\rm line}_{ul}/\alpha^{\rm line}_{ul}$ then becomes frequency-dependent), which \citet{Hummer1992} verified by numerical simulations. However, they made the important point that the spatial random walk performed by the photon during the scattering process is likely to take it so far away from its starting point that one can hardly assume the physical conditions to stay constant. Earlier, \citet{Osterbrock1962} and \citet{Adams1972} made the important findings that in static nebulae with optical depths $\tau\gtrsim 10^5$, photons escape by random-walking to the edge of the nebula rather than by diffusing to a frequency in the wing where the nebula is optically thin. This situation will occur also in SNe at some optical depth, and the Sobolev approximation will then break down.

\section{Monte-Carlo method}
\label{sec:MC}

The transport of photons (or any kind of particles) may be computed by following them on their trajectories through the grid. The interactions along the way are determined by random draws, and the method is therefore referred to as the \emph{Monte Carlo (MC) method}. The technique is especially attractive in situations of  multiple, complex interactions that are difficult to construct and solve the corresponding equation systems for. 

For photon transport, the emission and absorption coefficients $j_\nu$ and $\alpha_{\nu}$, derived in Sect. \ref{sec:ecoff} and \ref{sec:acoff}, can be used to construct probabilities for the interactions. There is no need to compute $I_\nu$ explicitly, but instead the emergent spectrum is constructed by simply binning the escaping photons in accordance with their wavelengths, angles, and number (if photon \emph{packets} are used). In \textbf{Paper II} we give a detailed account of our implementation of the creation and transport of photons using a MC technique. In summary, we create photons by random sampling of the emissivity function $j_\nu(r)$, send them off in a random direction, and let them be absorbed when the integrated optical depth (computed from the $\alpha(r)$ function) transcends a randomly drawn optical depth. By constantly transforming the photon frequency to the local comoving frame, we can use the rest-frame expressions for the absorption coefficients. Following an absorption, random numbers determine a deexcitation channel, and new directions for the scattered/fluorescent photons.

Apart from its conceptual simplicity, the MC method has the further advantages of being relatively easy to construct and test, suitable for 2D and 3D problems, easy to implement parallel algorithms for, and suitable for anisotropic phase functions as easily as isotropic ones. 
The numerical difficulties often encountered in attempting to calculate $I_\nu$  are to a large extent removed when photons are instead moved in MC schemes \citep{Auer2003}. In the context of moving media, the MC method is especially suitable since the line interlocking (Sect. \ref{sec:lt}) produces long and complex expressions when attempting to solve the equation of transfer by integration. MC transfer, on the other hand, simply moves the photons from line to line in a scheme simple to code. 

Another advantage, exploited in this thesis, is the possibility to include clumping and mixing by direct incorporation into the random sampling machinery for the photon trajectories. This is explained fully in \textbf{paper II} (Sect. 2.1.1).

The disadvantages of the MC method are that computing times can become long, and that rare channels are difficult to study due to statistical noise. If the optical depths are too high, the method becomes inefficient as the photons have to be tracked through a large number of interactions. But with ever-growing computer speeds and increasing availability of multi-processor computing, MC methods are likely to keep increasing in popularity.

\citet{Abbott1985} demonstrated the use of the MC method for dealing with the expanding atmospheres of hot stars. The explosion of SN 1987A prompted the development of MC codes for SNe \citep{Lucy1987, Mazzali1993}. Some of these MC codes use \emph{indestructible} energy packets \citep[e.g.][]{Lucy2005, Kasen2006, Sim2007a}, so that after any radiation-matter interaction, a packet with the same energy as the incoming packet is re-emitted. This ensures that the \emph{radiation field} satisfies the condition of radiative equilibrium (e.g. it is divergence-free), even if the \emph{temperature structure} does not. This, in turn, has been shown to accelerate the convergence properties of the $\Lambda$-iteration \citep{Lucy1999, Lucy2002}. 

\subsection{Probability distribution sampling}
The core of the MC method is to sample probability distributions. To do this, we need the \emph{cumulative probability distribution}
\begin{equation}
\phi(x) = \frac{\int_a^x p(x') dx'}{\int_a^b p(x') dx'}~,
\end{equation}
where $p(x)$ is the (not necessarily normalized) probability distribution over the variable interval $[a,b]$. By sampling $\phi$ from a uniform distribution $z = [0,1]$ and solving for $x$, we sample the $p(x)$ distribution (the \emph{fundamental principle}, e.g. Duderstadt and Martin 1979). For example, isotropic emission corresponds to  $p(\theta) = \sin{\theta}$, where $\theta=[0,\pi]$. We get
\begin{eqnarray}
\phi(\theta) = \frac{\int_0^\theta \sin{\theta'}d\theta'}{\int_0^\pi \sin \theta' d\theta'} \nonumber \\
 = \frac{1-\cos{\theta}}{2} \nonumber \\
\rightarrow \cos{\theta} = 1 - 2z
\end{eqnarray}
Another example is the distance a photon travels before being absorbed. Here $p(\tau) = e^{-\tau}$, so
\begin{eqnarray}
\phi(\tau) = \frac{\int_0^{\tau} e^{-\tau'} d\tau'}{\int_0^\infty e^{-\tau'}d\tau'}= 1-e^{-\tau} \\
\rightarrow \tau  = -\ln\left(1-z\right)
\end{eqnarray} 

Some probability distributions are not analytically integrable. We then accumulate the integral numerically until the drawn $z$ value is reached.


s, but overestimated absolute fluxes.

\subsection{Line interaction}
It is the coupling between radiation field and line transitions that causes the most difficulties in obtaining convergence for the $\Lambda$-iteration process \citep{Mihalas1970}. It is therefore of interest to consider approximations in the radiation-line interaction, and here the MC method allows us great freedom in tailoring specific treatments.

When a photon is absorbed in a line, several things may follow. The atom may de-excite in the same transition, re-emitting an identical photon in a new (random) direction. Since there is no net energy transfer between matter and radiation, this type of interaction gives benign convergence properties. It was used by \citet{Lucy1987} and \citet{Mazzali1993} for computing the transfer through SN atmospheres. 

The atom may also deexcite to another level with the emission of a fluorescent photon. This photon will be of higher energy than the original one if the fluorescence occurs to a state below the original level, and of smaller energy otherwise. Here, demanding a divergence-free radiation field means that the interaction can only be approximately treated, as there is an actual energy gain or loss for the photon. \citet{Lucy1999}, \citet{Mazzali2000} and \citet{Kasen2006} implemented treatments allowing for the initial fluorescence channels by weighting the transition probabilities with the transition energies. This gives the correct \emph{relative} fluxes in the fluorescence lines, but incorrect absolute fluxes. 

The atom may also transition to any other level by a collision with an electron. The probability that a thermalizing deexcitation occurs is
\begin{equation}
p_{\rm therm} = \frac{\sum_{l<u} C_{ul}^{thermal}}{\sum_{l<u} \left[C_{ul}^{thermal} + A_{ul}\beta_{ul}^S \right]}~.
\end{equation}
The collision coefficients $C_{ul}^{thermal}$ are of order $10^{-6}n_e$ s$^{-1}$. If any allowed, optically thin channels exist, $A_{ul}\beta_{ul}^S$ is of order $10^8$ s$^{-1}$, and thermalization cannot occur since it would require $n_e > 10^{14}$ cm$^{-3}$, much higher than in SNe. If only forbidden channels exist, $A_{ul}\beta_{ul}^S$ is of order $10^{-3}-1$ s$^{-1}$, and thermalization occurs if $n_e \gtrsim 10^3-10^6$ cm$^{-3}$. This electron density is achieved in most parts of the ejecta for quite a long time. Consequently, radiative energy absorbed to meta-stable levels will thermalize.


A lower degree of matter-radiation coupling can be achieved if the photoexcitation/de-excitation rates due to non-local radiation are ignored. One may then compute the excitation structure by LTE approximation \citep{Lucy1987, Mazzali1993, Lucy1999, Kasen2006} or by approximate NLTE solutions (\textbf{paper II}). Then, we must continue following the excited atom in the current iteration, since the absorption will not be remembered in the next one. For this thesis, such a decoupled scheme was used in \textbf{paper I} and \textbf{II}. Photon absorptions were then followed by randomly sampling radiative and collisional deexcitations, until the atom had de-excited to the ground state. For \textbf{papers III} and \textbf{IV}, an upgrade to include photoexcitations for a limited number of transitions was implemented.

\section{Implementation}

\subsection{Assumptions}
The major assumptions for the radiative transfer treatment are
\begin{itemize}
\item \emph{Spherical symmetry.} Although the choice of the MC technique is partially motivated by the planned use of multidimensional models in the future, only one-dimensional ones have so far been considered. There is growing evidence that strong asymmetries are present in SN ejecta, but for the Type IIP SNe analysed here, much of the action occurs in the large hydrogen envelope where asymmetries should play a smaller role. 
\item \emph{Sobolev approximation with no line overlaps (in the comoving frame) except for Ly$\alpha$ and Ly$\beta$.}
A detailed discussion about the possible problems associated with the Sobolev approximation can be found in \citet{Kasen2006}. Other uncertainties likely introduce errors more serious than the ones caused by this approximation, a statement supported by test calculations \citep{Eastman1993}. 
Line overlap can be a problem in the photospheric phases when the ejecta are hot and dense, causing a huge number of lines
to be optically thick. In the nebular phase, at lower temperature and density, it is less of a problem.
\item \emph{No time delays (the photons travel instantaneously between points).} 
See Sect. \ref{sec:ssnp} for a discussion about this point.
\item \emph{No effects of order V/c except for the frequency Doppler shifts.}
Relativistic effects may easily be included, but has not been implemented in consideration of neglecting time-delays and time-derivatives. For the SNe considered in this thesis, $V/c$ is small enough for these effects to be neglegible. 
\end{itemize}

\subsection{Transport of free-bound continua}
A common method in astrophysics is to replace the recombination rate $\alpha$ by $\alpha-\alpha_{gs}$, and assume that all free-bound continua to the ground state are reabsorbed on-the-spot. The justification is that these continua usually have very high optical depths. Here, we perform transfer also for these continua, as they may affect the photoionization rates of trace elements as well as be partially absorbed by lines.

\subsection{Packet splitting}
In runs where photons are tracked over their whole trajectories, many MC codes avoid splitting the photon upon matter interaction, but instead choose one of the resultant photons by random draws. This is a matter of coding simplicity \citep{Lucy2002}. However, we find that photon-splitting is not a big problem, and has the advantage that fluorescence is sampled to a higher degree of accuracy. At high enough optical depths, though, the memory requirements for following chains of fluorescence photons become demanding.

\subsection{The randomized grid}
A major idea advanced in this thesis (\textbf{paper II}) is to exploit the nature of the MC transfer by letting clumping and mixing be represented by a randomized ejecta structure. Few objects are more suitable for such a treatment than SNe, where the ejecta consist of several chemically distinct clumps that become mixed around during the explosion. To represent such a structure in a fixed one-dimensional grid, each zone type inevitable has to be represented by a set of shells, completely enclosing the material inside. This may give emergent spectra that differ drastically from spectra from clumpy models. One example is the impossibility to have some optically thick clumps while still allowing for some clear line-of-sights through the ejecta.

\subsection{Spatial resolution}
We assume constant conditions in each zone type within the core. The uniform density and \iso{56}Ni distribution implies that the radiation field is not expected to change by more than a factor $\sim$2 over the core.

For the envelope, we wish to resolve changes by a factor of $\sim$2 in density or radiation field. 
The density changes by a power law no steeper than $\rho(r) \sim r^{-4}$ in most parts of the hydrogen envelope. For a $r^{-n}$ power law, the required resolution is
\begin{equation}
\frac{r_{i+1}}{r_i} = 2^{1/n}~,
\end{equation}
which equals 1.2 for $n=4$. This is the value we use.





\subsection{Line-list}
To reduce run times, it is desirable to consider only lines with optical depth above some threshold value in the MC simulation. This value should be significantly smaller than 1, since many weak lines can produce collective blocking. We typically use $\tau^S_{\rm min}=10^{-4}$. The typical number of lines in the early nebular phase is then $10^4-10^5$. \citet{Lucy1987} uses a cut-off at $\tau^S_{\rm min}=10^{-2}$.

\subsection{Wavelength  resolution and range}
To optimize run-times, as narrow wavelength range as possible should be used for the radiation transport. We certainly want to include the He I ionization edge at 504 \AA. Little emissivity occurs at shorter wavelengths than this, so a lower limit of $\lambda_{\rm min}=$ 400 \AA~is reasonable. The choice of the upper limit has to be at least the wavelength where continuum absorption may occur. Since the Balmer continuum is optically thick for up to a few years, $\lambda_{\rm max}=3650$ \AA~is a reasonable choice. However, at early times both Paschen and free-free absorption may occur, and $\lambda_{\rm max}$ should then be at least $8000$ \AA.

With lower limits $\lambda_{\rm min}$ and $\lambda_{\rm max}$, and wavelength resolution $\Delta \lambda/\lambda=r$, the number of frequency points is
\begin{equation}
N = \frac{\log(\lambda_{\rm max}/\lambda_{\rm min})}{\log (1+r)} \approx \frac{1}{r}\log\left(\frac{\lambda_{\rm max}}{\lambda_{\rm min}}\right)
\end{equation}
It is thus more rewarding to set $r$ as large as possible than to squeeze the frequency range. We certainly want to have $r \ll V/c$ to resolve the lines well. Since $V/c\sim 1\e{-2}$, $r\sim 1\e{-3}$ is a good choice. 

In the UV, lines are separated by less than this in the early phases. However, the requirements of the Sobolev approximation are anyway not fulfilled here, so retaining this approximation while using higher resolution in the transfer is not meaningful. 
\subsection{Code testing}
An extensive library of testing routines was built up alongside the development of the code. This library has verified the basic functionality for all units, and allows an opportunity to redo the testing after any code changes. 

In addition to this, the output of the code was compared to the output of a similar code simultaneously developed at MPA. The results are described in \citet{Maurer2011}.

\chapter{Summary of the papers}
\label{summary}

\section{Paper I: The 3-D structure of SN 1987A's inner ejecta}
\label{paperI}
The explosion mechanism of supernovae is still unknown (Sect. \ref{sec:explm}). 
The main aim of this paper is to diagnose the explosion of SN 1987A by determining the three-dimensional structure of the ejecta. 
To this end, SN 1987A was observed in the near-infrared with the VLT SINFONI integral-field spectrometer, which provides a spectrum for each spatial location in the field of view of the instrument. SINFONI is an adaptive optics assisted instrument and therefore operates in the near-infrared. Identifying strong lines in the spectrum, and assuming that there is no blending with continuum or other lines, the velocity of the gas can then be determined, giving the full 3D structure of the ejecta.

The results revealed an asymmetric morphology, elongated more towards the equatorial ring plane than perpendicular to it. This probably rules out any rotationally-driven explosion mechanism, as one then would expect the latter type of asymmetry. The morphology instead favors the SASI-type of explosion, described in Sect. \ref{sec:rsm}, because this type of instability causes asymmetries in random directions. One may note that a random direction will, with 50\% probability, be within $\pm$30\% of the equatorial plane, producing an elongation apparently quite parallel to the equatorial plane, as observed here.

My most important contribution to this project was to run spectral modeling to determine the origin of the 1.644 $\mu$m and 2.058 $\mu$m lines, i.e. to identify which elements and which zones emit them.
Both lines turned out to have a surprising origin. According to the model, the 1.644 $\mu$m line is not dominated by the [Fe II] 1.644 $\mu$m transition, but by [Si I] 1.645 $\mu$m. This line is emitted by silicon in the Si/S zone, and tells us that the 3D mapping shows the distribution of oxygen-burning ashes in the SN (see Sect. \ref{sec:oxburn} and \ref{sec:explburn}). 
The theoretical predictions for the amount of silicon produced ($\sim 0.1$ \msun), and its mixing with radioactive nuclides (with $\sim$10\% of the \iso{44}Ti), are in good agreement with the observations. 

The iron distribution is instead traced by the He I 2.058 $\mu$m mapping, as the model shows that this line is produced by helium created by freeze-out in the iron core, described in Sect. \ref{sec:explburn}. This conclusion is further strengthened by the fact that the He I 2.058 $\mu$m and [Si I] 1.645 $\mu$m distributions are similar, an unlikely outcome if the He I emission comes from the helium zone and not the Fe/He zone. The results therefore provide confirmation for $\alpha$-rich freeze-out actually having occured in the explosion.

For the rest of the spectrum, a reasonable model agreement was obtained using M(\iso{44}Ti) = $1\e{-4}$ \msun~and and optical depth for the dust in the ejecta $\tau_{\rm dust}=1$, although below $\sim$ 1.4 $\mu$m, the model is too low by a factor of $\sim$ 2. This can likely be explained by the higher \iso{44}Ti-mass found in \textbf{Paper II} ($1.5\e{-4}$ \msun), and possibly also by input by X-rays at late times (Sect. \ref{sec:xrays}). The X-rays do not penetrate to the core material \citep{Larsson2011}, which means that they have no direct influence on the [Si I] 1.646 $\mu$m and He I 2.058 $\mu$m lines, can can thus not explain the observed morphology. However, the X-rays may influence the H, He, and continuum emission from the envelope, or fluorescence thereof.


\section{Paper II: The \iso{44}Ti-powered spectrum of SN 1987A}
\label{paperII}

This paper contains a lot of the basic background for the modeling, and results of application to the late-time optical spectrum of SN 1987A. The spectrum of SN 1987A in the \iso{56}Co and \iso{57}Co-powered phases have been thoroughly modelled before. Less attention has been paid to the \iso{44}Ti-dominated phase ($t \gtrsim$ 5 years). At these late times, the ejecta temperatures have dropped to a few hundred degrees, and the entire UV/optical/NIR spectrum is produced by non-thermal processes. The powering is mainly due to positrons, as most of the $\gamma$-rays escape the ejecta (see Sect. \ref{sec:deposition}). These epochs therefore offer a unique opportunity to see emission from the deepest interiors where the positrons get trapped. 

We choose to analyze the epoch of eight years, where a high-quality HST spectrum as well as a ground-based NIR spectrum exist. Previous attempts to identify lines in the HST spectrum has left question-marks for several of the strongest lines, especially in the UV \citep{Chugai1997}. With the addition of multi-line radiative transfer to our model, we were in a unique position to identify these lines.
After eight years, increasing contamination by the circumstellar interaction makes spectral analysis of the ejecta more difficult. 

 An important question is where in the core the positrons deposit their energy. As described in Sect. \ref{sec:deposition}, even a weak magnetic field will trap the positrons close to the radioactive material. If no magnetic field is present, they can instead travel substantial distances, but will still be absorbed somewhere in the core. We find a model without a magnetic field to produce too strong emission lines from the oxygen and hydrogen components, and therefore conclude that a magnetic field is indeed present. Energy deposition into the mainly neutral iron clumps produces prominent Fe I emission lines in the spectrum, matching several of the previously unidentified lines.

From investigation of the line formation process, we could show that multi-line transfer plays an important role, even at eight years after explosion. As described in Sect. \ref{sec:acoff}, the velocity gradients in SNe lead to very efficient line blocking, and a large fraction of the emergent spectrum is made of scattered light. 

The dust clumps that formed at $\sim$ 500 days (Sect. \ref{sec:specan87A}) are still optically thick at eight years, damping the red sides of the emission lines. With the best estimate for the effect of the dust on the UVOIR field, we determined the \iso{44}Ti-mass to $1.5_{-0.5}^{+0.5}\e{-4}$ \msun. Interestingly, this is close to the value in Cas A ($1.6_{-0.3}^{+0.6}\e{-4}$ \msun), the only SN
where the $\gamma$-ray decay lines by \iso{44}Ti have been directly detected. As we discuss in the paper, the determined \iso{44}Ti mass is higher than obtained in spherically symmetric models, but enhanced production occurs in asymmetric explosions. The results here therefore ties in with the findings in \textbf{Paper I}, that the SN 1987A explosion was quite asymmetric.

\section{Paper III: The progenitor mass of the Type IIP supernova 2004et from late-time spectral modeling}
\label{paperIII}
In Chapter 2, we describe how the amount of nucleosynthesis occurring in a star is strongly dependent on its main-sequence mass. 
The aim of this paper is to constrain the amount of nucleosynthesis products present in the bright Type IIP SN 2004et, thereby providing constraints on its progenitor mass. This SN is interesting because the progenitor was detected in pre-explosion images, and analysis of these suggests a significantly less massive star ($14$ \msun) than obtained by hydrodynamical models of the SN light-curve ($> 25~$\msun). With an excellent UV to mid-infrared nebular phase data-set, spectral modeling can add a crucial third diagnosis for which type of star SN 2004et was.

To analyze the nebular-phase spectra, we apply the spectral synthesis code,  
finding the nebular phase spectrum to best agree with models of a $\sim$15~\msun~(non-rotating) progenitor (as described in Sect. \ref{sec:rotation}, rotation would result in the progenitor mass being lower for a given amount of metals produced, so 15 \msun~can be considered an upper limit). We identify  Mg I] 4571 \AA, Na I 5890, 5896 \AA, [O I] 6300, 6364 \AA, blends of metal lines at $1.15-1.2$ $\mu$m and $1.4-1.6$ $\mu$m, and [Ne II] 12.81 $\mu$m, as the most useful lines for constraining nucleosynthesis in Type IIP SNe. We can therefore conclude that our spectral modeling is in good agreement with the progenitor analysis, but in disagreement with the hydrodynamical modeling.

We can also confirm that silicate dust emission, as well as SiO and CO fundamental band emission, is present in the Spitzer spectra taken by \citet{Kotak2009}. SN 1987A did not have any significant dust continuum until after $\sim$ 500 days, whereas in SN 2004et it is present already at day $350$. The CO emission is in agreement with the total heating of the CO zone in the 15 \msun~model.


\section{Paper IV: Constraining the physical properties of Type IIP supernovae using nebular phase spectra}
\label{paperIV}
This paper looks at a sample of Type IIP SNe with progenitor detections as well as spectroscopic data sets in the nebular phase, investigating the variety in line profiles and flux ratios. How large is the diversity within this class of SNe, and what can the differences tell us? Standard stellar evolution models predict all stars in the $8-30~$\msun~range to end as red supergiants and explode as Type IIP SNe. But all the progenitors detected in the last decade or two tend to cluster in a smaller range of $8-17$~\msun~\citep{Smartt2009b}. Can spectral analysis settle which of the ranges is the correct one?

A first step is to understand the formation of the strongest lines in the nebular phase spectra. As most SNe lack infrared spectra, we concentrated on the optical [O I] 6300, 6364 \AA~, H$\alpha$, [Fe II] 7155, 7172 \AA, [Ca II] 7291, 7323 \AA, and Ca II IR triplet lines. As most of the SNe in the sample have identified progenitors between $8-15$ \msun, we interpret the evolution of various lines by the use of a 12 \msun~model.
 
For the calcium lines, we could show that the conclusion drawn by \citet{Li1993} for SN 1987A, that they originate in primordial calcium in the hydrogen envelope, is valid for all epochs. From the discussion in Sect. \ref{sec:massloss}, it is clear that the final hydrogen envelope mass varies little with the progenitor mass, and the envelope lines therefore show little variation within the sample. Furthermore, none of the SNe in the sample show any flat-topped emission lines, which means that the hydrogen-zone material has been mixed down towards zero velocity. This occurs by the hydrodynamical instabilities described in Sect. \ref{sec:rsm}, and the results therefore suggest that these instabilities invariably occur when the star has a significant hydrogen envelope left at the time of explosion.

We also found that the [O I] \wll 6300, 6364 doublet has more than half of its flux coming from natal oxygen in the hydrogen envelope, a possibility not previously recognized. This fact helps to explain why the [O I] \wll 6300, 6364 to H$\alpha$ line ratio does not vary much within the sample, even though the synthesized oxygen mass varies greatly with progenitor mass.


My most important contributions to the paper consists of modeling the line flux ratios (Sect. 5), as well as modeling SN 2008bk to identify the emission lines (Sect. 6), but I was also involved in the development and formulation of all other chapters.

\backmatter

\chapter{Publications not included in this thesis}

\begin{romanlist}
\item {\bf NERO - A Post Maximum Supernova Radiation Transport Code} \\
	Maurer~I., Jerkstrand~A., Mazzali, P.A., Taubenberger~S., Hachinger~S., Kromer~M., Sim~S., Accepted for publication in {\em MNRAS}. ArXiv: 1105.3049. \\
\vspace{3mm}
\item {\bf X-ray illumination of the ejecta of SN 1987A} \\
	Larsson~J., [3 authors], Jerkstrand~A, [23 authors], 2011, {\em Nature}, 474, 484 \\
\vspace{3mm}
\end{romanlist}

\include{ack}

\nocite{*}
\bibliography{references_avh}

\begin{thebibliography}{134}
\expandafter\ifx\csname natexlab\endcsname\relax\def\natexlab#1{#1}\fi

\bibitem[{{Abbott} \& {Lucy}(1985)}]{Abbott1985}
{Abbott}, D.~C. \& {Lucy}, L.~B. 1985, \apj, 288, 679

\bibitem[{{Adams}(1972)}]{Adams1972}
{Adams}, T.~F. 1972, \apj, 174, 439

\bibitem[{{Aitken} {et~al.}(1988){Aitken}, {Smith}, {James}, {Roche}, {Hyland},
  \& {McGregor}}]{Aitken1988}
{Aitken}, D.~K., {Smith}, C.~H., {James}, S.~D., {et~al.} 1988, \mnras, 231, 7P

\bibitem[{{Alburger} \& {Wesselborg}(1990)}]{Alburger1990}
{Alburger}, D.~E. \& {Wesselborg}, C. 1990, \prc, 42, 2728

\bibitem[{{Allen}(1973)}]{Allen1973}
{Allen}, C.~W. 1973, {Astrophysical quantities}, ed. {Allen, C.~W.}

\bibitem[{{Arnett} {et~al.}(1989){Arnett}, {Bahcall}, {Kirshner}, \&
  {Woosley}}]{Arnett1989}
{Arnett}, W.~D., {Bahcall}, J.~N., {Kirshner}, R.~P., \& {Woosley}, S.~E. 1989,
  \araa, 27, 629

\bibitem[{{Auer}(2003)}]{Auer2003}
{Auer}, L. 2003, in Astronomical Society of the Pacific Conference Series, Vol.
  288, Stellar Atmosphere Modeling, ed. {I.~Hubeny, D.~Mihalas, \& K.~Werner},
  405--+

\bibitem[{{Auer} \& {van Blerkom}(1972)}]{Auer1972}
{Auer}, L.~H. \& {van Blerkom}, D. 1972, \apj, 178, 175

\bibitem[{{Axelrod}(1980)}]{Axelrod1980}
{Axelrod}, T.~S. 1980, PhD thesis, AA(California Univ., Santa Cruz.)

\bibitem[{{Bethe}(1930)}]{Bethe1930}
{Bethe}, H. 1930, Annalen der Physik, 397, 325

\bibitem[{{Bethe} \& {Pizzochero}(1990)}]{Bethe1990}
{Bethe}, H.~A. \& {Pizzochero}, P. 1990, \apjl, 350, L33

\bibitem[{{Blondin} \& {Shaw}(2007)}]{Blondin2007}
{Blondin}, J.~M. \& {Shaw}, S. 2007, \apj, 656, 366

\bibitem[{{Bouchet} {et~al.}(1991){Bouchet}, {Phillips}, {Suntzeff},
  {Gouiffes}, {Hanuschik}, \& {Wooden}}]{Bouchet1991a}
{Bouchet}, P., {Phillips}, M.~M., {Suntzeff}, N.~B., {et~al.} 1991, \aap, 245,
  490

\bibitem[{{Castor}(1970)}]{Castor1970}
{Castor}, J.~I. 1970, \mnras, 149, 111

\bibitem[{{Cherchneff} \& {Lilly}(2008)}]{Cherchneff2008}
{Cherchneff}, I. \& {Lilly}, S. 2008, \apjl, 683, L123

\bibitem[{{Cherchneff} \& {Sarangi}(2011)}]{Cherchneff2011}
{Cherchneff}, I. \& {Sarangi}, A. 2011, in IAU Symposium, Vol. 280, IAU
  Symposium

\bibitem[{{Chevalier}(1976)}]{Chevalier1976}
{Chevalier}, R.~A. 1976, \apj, 207, 872

\bibitem[{{Chevalier} \& {Klein}(1978)}]{ChevalierKlein1978}
{Chevalier}, R.~A. \& {Klein}, R.~I. 1978, \apj, 219, 994

\bibitem[{{Chugai}(1980)}]{Chugai1980}
{Chugai}, N.~N. 1980, Soviet Astronomy Letters, 6, 91

\bibitem[{{Chugai}(1987)}]{Chugai1987}
{Chugai}, N.~N. 1987, Astrofizika, 26, 89

\bibitem[{{Chugai} {et~al.}(1997){Chugai}, {Chevalier}, {Kirshner}, \&
  {Challis}}]{Chugai1997}
{Chugai}, N.~N., {Chevalier}, R.~A., {Kirshner}, R.~P., \& {Challis}, P.~M.
  1997, \apj, 483, 925, {(C97)}

\bibitem[{{Clayton} {et~al.}(2001){Clayton}, {Deneault}, \&
  {Meyer}}]{Clayton2001}
{Clayton}, D.~D., {Deneault}, E.~A.-N., \& {Meyer}, B.~S. 2001, \apj, 562, 480

\bibitem[{{Clayton} {et~al.}(1992){Clayton}, {Leising}, {The}, {Johnson}, \&
  {Kurfess}}]{Clayton1992}
{Clayton}, D.~D., {Leising}, M.~D., {The}, L.-S., {Johnson}, W.~N., \&
  {Kurfess}, J.~D. 1992, \apjl, 399, L141

\bibitem[{{Colgate} {et~al.}(1980){Colgate}, {Petschek}, \&
  {Kriese}}]{Colgate1980}
{Colgate}, S.~A., {Petschek}, A.~G., \& {Kriese}, J.~T. 1980, \apjl, 237, L81

\bibitem[{{Dalgarno} \& {McCray}(1972)}]{Dalgarno1972}
{Dalgarno}, A. \& {McCray}, R.~A. 1972, \araa, 10, 375

\bibitem[{{Danziger} {et~al.}(1991){Danziger}, {Lucy}, {Bouchet}, \&
  {Gouiffes}}]{Danziger1991P}
{Danziger}, I.~J., {Lucy}, L.~B., {Bouchet}, P., \& {Gouiffes}, C. 1991, in
  Supernovae, ed. {S.~E.~Woosley}, 69

\bibitem[{{de Kool} {et~al.}(1998){de Kool}, {Li}, \& {McCray}}]{deKool1998}
{de Kool}, M., {Li}, H., \& {McCray}, R. 1998, \apj, 503, 857

\bibitem[{{Dessart} {et~al.}(2010){Dessart}, {Livne}, \&
  {Waldman}}]{Dessart2010}
{Dessart}, L., {Livne}, E., \& {Waldman}, R. 2010, \mnras, 408, 827

\bibitem[{{Eastman} \& {Pinto}(1993)}]{Eastman1993}
{Eastman}, R.~G. \& {Pinto}, P.~A. 1993, \apj, 412, 731

\bibitem[{{Endt}(1990)}]{Endt1990}
{Endt}, P.~M. 1990, Nuclear Physics A, 521, 1

\bibitem[{{Fransson} {et~al.}(1989){Fransson}, {Cassatella}, {Gilmozzi},
  {Kirshner}, {Panagia}, {Sonneborn}, \& {Wamsteker}}]{Fransson1989UV}
{Fransson}, C., {Cassatella}, A., {Gilmozzi}, R., {et~al.} 1989, \apj, 336, 429

\bibitem[{{Fransson} \& {Chevalier}(1989)}]{Fransson1989}
{Fransson}, C. \& {Chevalier}, R.~A. 1989, \apj, 343, 323

\bibitem[{{Fransson} \& {Kozma}(1993)}]{Fransson1993}
{Fransson}, C. \& {Kozma}, C. 1993, \apjl, 408, L25, (FK93)

\bibitem[{{Fryxell} {et~al.}(1991){Fryxell}, {Arnett}, \&
  {Mueller}}]{Fryxell1991}
{Fryxell}, B., {Arnett}, D., \& {Mueller}, E. 1991, \apj, 367, 619

\bibitem[{{Fukuda}(1982)}]{Fukuda1982}
{Fukuda}, I. 1982, \pasp, 94, 271

\bibitem[{{Gearhart} {et~al.}(1999){Gearhart}, {Wheeler}, \&
  {Swartz}}]{Gearhart1999}
{Gearhart}, R.~A., {Wheeler}, J.~C., \& {Swartz}, D.~A. 1999, \apj, 510, 944

\bibitem[{{Gilmozzi} {et~al.}(1987){Gilmozzi}, {Cassatella}, {Clavel},
  {Fransson}, {Gonzalez}, {Gry}, {Panagia}, {Talavera}, \&
  {Wamsteker}}]{Gilmozzi1987}
{Gilmozzi}, R., {Cassatella}, A., {Clavel}, J., {et~al.} 1987, \nat, 328, 318

\bibitem[{{Hammer} {et~al.}(2010){Hammer}, {Janka}, \&
  {M{\"u}ller}}]{Hammer2010}
{Hammer}, N.~J., {Janka}, H., \& {M{\"u}ller}, E. 2010, \apj, 714, 1371

\bibitem[{{Hashimoto} {et~al.}(1989){Hashimoto}, {Nomoto}, \&
  {Shigeyama}}]{Hashimoto1989}
{Hashimoto}, M., {Nomoto}, K., \& {Shigeyama}, T. 1989, \aap, 210, L5

\bibitem[{{Heath}(1997)}]{Heath}
{Heath}, M. 1997, {Scientific Computing - An Introductory Survey} (McGraw-Hill)

\bibitem[{{Heger} {et~al.}(2000){Heger}, {Langer}, \& {Woosley}}]{Heger2000}
{Heger}, A., {Langer}, N., \& {Woosley}, S.~E. 2000, \apj, 528, 368

\bibitem[{{Herant} \& {Benz}(1991)}]{Herant1991}
{Herant}, M. \& {Benz}, W. 1991, \apjl, 370, L81

\bibitem[{{Herant} \& {Woosley}(1994)}]{Herant1994}
{Herant}, M. \& {Woosley}, S.~E. 1994, \apj, 425, 814

\bibitem[{{Hillebrandt} \& {Meyer}(1989)}]{Hillebrandth1989}
{Hillebrandt}, W. \& {Meyer}, F. 1989, \aap, 219, L3

\bibitem[{{Hirschi} {et~al.}(2004){Hirschi}, {Meynet}, \&
  {Maeder}}]{Hirschi2004}
{Hirschi}, R., {Meynet}, G., \& {Maeder}, A. 2004, \aap, 425, 649

\bibitem[{{Hummer} \& {Rybicki}(1985)}]{Hummer1985}
{Hummer}, D.~G. \& {Rybicki}, G.~B. 1985, \apj, 293, 258

\bibitem[{{Hummer} \& {Rybicki}(1992)}]{Hummer1992}
{Hummer}, D.~G. \& {Rybicki}, G.~B. 1992, \apj, 387, 248

\bibitem[{{Imshennik} \& {Popov}(1992)}]{Imshennik1992}
{Imshennik}, V.~S. \& {Popov}, D.~V. 1992, \azh, 69, 497

\bibitem[{{Junde}(1992)}]{Junde1992}
{Junde}, H. 1992, Nuclear Data Sheets, 67, 523

\bibitem[{{Kasen} {et~al.}(2006){Kasen}, {Thomas}, \& {Nugent}}]{Kasen2006}
{Kasen}, D., {Thomas}, R.~C., \& {Nugent}, P. 2006, \apj, 651, 366

\bibitem[{{Kervella} {et~al.}(2011){Kervella}, {Perrin}, {Chiavassa},
  {Ridgway}, {Cami}, {Haubois}, \& {Verhoelst}}]{Kervella2011}
{Kervella}, P., {Perrin}, G., {Chiavassa}, A., {et~al.} 2011, \aap, 531, A117+

\bibitem[{{Kifonidis} {et~al.}(2003){Kifonidis}, {Plewa}, {Janka}, \&
  {M{\"u}ller}}]{Kifonidis2003}
{Kifonidis}, K., {Plewa}, T., {Janka}, H.-T., \& {M{\"u}ller}, E. 2003, \aap,
  408, 621

\bibitem[{{Kifonidis} {et~al.}(2006){Kifonidis}, {Plewa}, {Scheck}, {Janka}, \&
  {M{\"u}ller}}]{Kifonidis2006}
{Kifonidis}, K., {Plewa}, T., {Scheck}, L., {Janka}, H.-T., \& {M{\"u}ller}, E.
  2006, \aap, 453, 661

\bibitem[{{Kotak} {et~al.}(2009){Kotak}, {Meikle}, {Farrah}, {Gerardy},
  {Foley}, {Van Dyk}, {Fransson}, {Lundqvist}, {Sollerman}, {Fesen},
  {Filippenko}, {Mattila}, {Silverman}, {Andersen}, {H{\"o}flich}, {Pozzo}, \&
  {Wheeler}}]{Kotak2009}
{Kotak}, R., {Meikle}, W.~P.~S., {Farrah}, D., {et~al.} 2009, \apj, 704, 306,
  (K09)

\bibitem[{{Kozma} \& {Fransson}(1992)}]{Kozma1992}
{Kozma}, C. \& {Fransson}, C. 1992, \apj, 390, 602, (KF92)

\bibitem[{{Kozma} \& {Fransson}(1998{\natexlab{a}})}]{Kozma1998I}
{Kozma}, C. \& {Fransson}, C. 1998{\natexlab{a}}, \apj, 496, 946, (KF98 a)

\bibitem[{{Kozma} \& {Fransson}(1998{\natexlab{b}})}]{Kozma1998II}
{Kozma}, C. \& {Fransson}, C. 1998{\natexlab{b}}, \apj, 497, 431, (KF98 b)

\bibitem[{{Kromer} \& {Sim}(2009)}]{Kromer2009}
{Kromer}, M. \& {Sim}, S.~A. 2009, \mnras, 398, 1809

\bibitem[{{Lamers}(1981)}]{Lamers1981}
{Lamers}, H.~J.~G.~L.~M. 1981, \apj, 245, 593

\bibitem[{{Larsson} {et~al.}(2011){Larsson}, {Fransson}, {Ostlin},
  {Groningsson}, {Kozma}, \& {Sollerman}}]{Larsson2011}
{Larsson}, J., {Fransson}, C., {Ostlin}, G., {et~al.} 2011, \nat, submitted

\bibitem[{{Lepp} {et~al.}(1990){Lepp}, {Dalgarno}, \& {McCray}}]{Lepp1990}
{Lepp}, S., {Dalgarno}, A., \& {McCray}, R. 1990, \apj, 358, 262

\bibitem[{{Li} \& {McCray}(1993)}]{Li1993}
{Li}, H. \& {McCray}, R. 1993, ApJ, 405, 730

\bibitem[{{Li} {et~al.}(2011){Li}, {Leaman}, {Chornock}, {Filippenko},
  {Poznanski}, {Ganeshalingam}, {Wang}, {Modjaz}, {Jha}, {Foley}, \&
  {Smith}}]{Li2011}
{Li}, W., {Leaman}, J., {Chornock}, R., {et~al.} 2011, \mnras, 412, 1441

\bibitem[{{Liu} \& {Dalgarno}(1995)}]{Liu1995}
{Liu}, W. \& {Dalgarno}, A. 1995, \apj, 454, 472

\bibitem[{{Liu} \& {Dalgarno}(1996)}]{Liu1996}
{Liu}, W. \& {Dalgarno}, A. 1996, \apj, 471, 480

\bibitem[{{Liu} {et~al.}(1992){Liu}, {Dalgarno}, \& {Lepp}}]{Liu1992}
{Liu}, W., {Dalgarno}, A., \& {Lepp}, S. 1992, \apj, 396, 679

\bibitem[{{Lodders}(2003)}]{Lodders2003}
{Lodders}, K. 2003, \apj, 591, 1220

\bibitem[{{Lucy}(1987)}]{Lucy1987}
{Lucy}, L.~B. 1987, \aap, 182, L31

\bibitem[{{Lucy}(1991)}]{Lucy1991}
{Lucy}, L.~B. 1991, \apj, 383, 308

\bibitem[{{Lucy}(1999{\natexlab{a}})}]{Lucy1999static}
{Lucy}, L.~B. 1999{\natexlab{a}}, \aap, 344, 282

\bibitem[{{Lucy}(1999{\natexlab{b}})}]{Lucy1999}
{Lucy}, L.~B. 1999{\natexlab{b}}, \aap, 345, 211

\bibitem[{{Lucy}(2002)}]{Lucy2002}
{Lucy}, L.~B. 2002, \aap, 384, 725

\bibitem[{{Lucy}(2003{\natexlab{a}})}]{LucyTPII}
{Lucy}, L.~B. 2003{\natexlab{a}}, aap, 403, 261

\bibitem[{{Lucy}(2003{\natexlab{b}})}]{Lucy2003}
{Lucy}, L.~B. 2003{\natexlab{b}}, \aap, 403, 261

\bibitem[{{Lucy}(2005)}]{Lucy2005}
{Lucy}, L.~B. 2005, \aap, 429, 19

\bibitem[{{Lucy} {et~al.}(1989){Lucy}, {Danziger}, {Gouiffes}, \&
  {Bouchet}}]{Lucy1989}
{Lucy}, L.~B., {Danziger}, I.~J., {Gouiffes}, C., \& {Bouchet}, P. 1989, in
  Lecture Notes in Physics, Berlin Springer Verlag, Vol. 350, IAU Colloq. 120:
  Structure and Dynamics of the Interstellar Medium, ed. G.~{Tenorio-Tagle},
  M.~{Moles}, \& J.~{Melnick}, 164

\bibitem[{{Lucy} \& {Solomon}(1970)}]{Lucy1970}
{Lucy}, L.~B. \& {Solomon}, P.~M. 1970, \apj, 159, 879

\bibitem[{{Maeda} {et~al.}(2007){Maeda}, {Kawabata}, {Tanaka}, {Nomoto},
  {Tominaga}, {Hattori}, {Minezaki}, {Kuroda}, {Suzuki}, {Deng}, {Mazzali}, \&
  {Pian}}]{Maeda2007}
{Maeda}, K., {Kawabata}, K., {Tanaka}, M., {et~al.} 2007, \apjl, 658, L5

\bibitem[{{Mashonkina}(1996)}]{Mashonkina1996}
{Mashonkina}, L.~J. 1996, in Astronomical Society of the Pacific Conference
  Series, Vol. 108, M.A.S.S., Model Atmospheres and Spectrum Synthesis, ed.
  {S.~J.~Adelman, F.~Kupka, \& W.~W.~Weiss}, 140--+

\bibitem[{{Maurer} {et~al.}(2011){Maurer}, {Jerkstrand}, {Mazzali},
  {Taubenberger}, {Hachinger}, {Kromer}, {Sim}, \& {Hillebrandt}}]{Maurer2011}
{Maurer}, I., {Jerkstrand}, A., {Mazzali}, P.~A., {et~al.} 2011, ArXiv e-prints

\bibitem[{{Mazzali}(2000)}]{Mazzali2000}
{Mazzali}, P.~A. 2000, \aap, 363, 705

\bibitem[{{Mazzali} \& {Lucy}(1993)}]{Mazzali1993}
{Mazzali}, P.~A. \& {Lucy}, L.~B. 1993, \aap, 279, 447

\bibitem[{{McCray}(1993)}]{McCray1993}
{McCray}, R. 1993, \araa, 31, 175

\bibitem[{{McCray}(2007)}]{McCray2007}
{McCray}, R. 2007, in American Institute of Physics Conference Series, Vol.
  937, Supernova 1987A: 20 Years After: Supernovae and Gamma-Ray Bursters, ed.
  {S.~Immler, K.~Weiler, \& R.~McCray}, 3--14

\bibitem[{{Meikle} {et~al.}(1993){Meikle}, {Spyromilio}, {Allen}, {Varani}, \&
  {Cumming}}]{Meikle1993}
{Meikle}, W.~P.~S., {Spyromilio}, J., {Allen}, D.~A., {Varani}, G., \&
  {Cumming}, R.~J. 1993, \mnras, 261, 535

\bibitem[{{Melius}(1974)}]{Melius1974}
{Melius}, C.~F. 1974, Journal of Physics B Atomic Molecular Physics, 7, 1692

\bibitem[{{Mihalas}(1970)}]{Mihalas1970}
{Mihalas}, D. 1970, {Stellar atmospheres}, ed. {Mihalas, D.}

\bibitem[{{Mihalas}(1978)}]{Mihalas1978}
{Mihalas}, D. 1978, {Stellar atmospheres /2nd edition/}, ed. {Hevelius, J.}

\bibitem[{{Mihalas} \& {Weibel Mihalas}(1984)}]{Mihalas1984}
{Mihalas}, D. \& {Weibel Mihalas}, B. 1984, {Foundations of radiation
  hydrodynamics}, ed. {Mihalas, D.~\& Weibel Mihalas, B.}

\bibitem[{{Morrison} \& {McCammon}(1983)}]{Morrison1983}
{Morrison}, R. \& {McCammon}, D. 1983, \apj, 270, 119

\bibitem[{{Mowlavi} \& {Forestini}(1994)}]{Mowlavi1994}
{Mowlavi}, N. \& {Forestini}, M. 1994, \aap, 282, 843

\bibitem[{{Nieuwenhuijzen} \& {de Jager}(1990)}]{Nieu1990}
{Nieuwenhuijzen}, H. \& {de Jager}, C. 1990, \aap, 231, 134

\bibitem[{{Nomoto} {et~al.}(1997){Nomoto}, {Hashimoto}, {Tsujimoto},
  {Thielemann}, {Kishimoto}, {Kubo}, \& {Nakasato}}]{Nomoto1997}
{Nomoto}, K., {Hashimoto}, M., {Tsujimoto}, T., {et~al.} 1997, Nuclear Physics
  A, 616, 79

\bibitem[{{Nomoto} {et~al.}(1984){Nomoto}, {Thielemann}, \&
  {Yokoi}}]{Nomoto1984}
{Nomoto}, K., {Thielemann}, F.-K., \& {Yokoi}, K. 1984, \apj, 286, 644

\bibitem[{{Nussbaumer} \& {Storey}(1983)}]{Nussbaumer1983}
{Nussbaumer}, H. \& {Storey}, P.~J. 1983, \aap, 126, 75

\bibitem[{{Osterbrock}(1962)}]{Osterbrock1962}
{Osterbrock}, D.~E. 1962, \apj, 135, 195

\bibitem[{{Osterbrock} \& {Ferland}(2006)}]{OB}
{Osterbrock}, D.~E. \& {Ferland}, G.~J. 2006, {Astrophysics of gaseous nebulae
  and active galactic nuclei} (Astrophysics of gaseous nebulae and active
  galactic nuclei, 2nd.~ed.~by D.E.~Osterbrock and G.J.~Ferland.~Sausalito, CA:
  University Science Books, 2006)

\bibitem[{{Park} {et~al.}(2011){Park}, {Zhekov}, {Burrows}, {Racusin}, {Dewey},
  \& {McCray}}]{Park2011}
{Park}, S., {Zhekov}, S.~A., {Burrows}, D.~N., {et~al.} 2011, \apjl, 733, L35+

\bibitem[{{Pinto} \& {Eastman}(2000)}]{Pinto2000b}
{Pinto}, P.~A. \& {Eastman}, R.~G. 2000, \apj, 530, 757

\bibitem[{{Podsiadlowski} {et~al.}(1990){Podsiadlowski}, {Joss}, \&
  {Rappaport}}]{Podsiadlowski1990}
{Podsiadlowski}, P., {Joss}, P.~C., \& {Rappaport}, S. 1990, \aap, 227, L9

\bibitem[{{Press} {et~al.}(1992){Press}, {Teukolsky}, {Vetterling}, \&
  {Flannery}}]{NR}
{Press}, W.~H., {Teukolsky}, S.~A., {Vetterling}, W.~T., \& {Flannery}, B.~P.
  1992, {Numerical recipes in FORTRAN. The art of scientific computing}
  (Cambridge: University Press, |c1992, 2nd ed.)

\bibitem[{{Puls} {et~al.}(2008){Puls}, {Vink}, \& {Najarro}}]{Puls2008}
{Puls}, J., {Vink}, J.~S., \& {Najarro}, F. 2008, \aapr, 16, 209

\bibitem[{Rybicki \& Lightman(1979)}]{RL}
Rybicki, G. \& Lightman, A. 1979, Radiative Processes in Astrophysics (Wiley)

\bibitem[{{Schaller} {et~al.}(1992){Schaller}, {Schaerer}, {Meynet}, \&
  {Maeder}}]{Schaller1992}
{Schaller}, G., {Schaerer}, D., {Meynet}, G., \& {Maeder}, A. 1992, \aaps, 96,
  269

\bibitem[{{Scharmer}(1981)}]{Scharmer1981}
{Scharmer}, G.~B. 1981, \apj, 249, 720

\bibitem[{{Scharmer}(1984)}]{Scharmer1984}
{Scharmer}, G.~B. 1984, {Accurate solutions to non-LTE problems using
  approximate lambda operators}, ed. {Kalkofen, W.}, 173--210

\bibitem[{{Scheck} {et~al.}(2008){Scheck}, {Janka}, {Foglizzo}, \&
  {Kifonidis}}]{Scheck2008}
{Scheck}, L., {Janka}, H.-T., {Foglizzo}, T., \& {Kifonidis}, K. 2008, \aap,
  477, 931

\bibitem[{{Sedov}(1959)}]{Sedov1959}
{Sedov}, L.~I. 1959, {Similarity and Dimensional Methods in Mechanics}, ed.
  {Sedov, L.~I.}

\bibitem[{{Seitenzahl}(2011)}]{Seitenzahl2011}
{Seitenzahl}, I.~R. 2011, Progress in Particle and Nuclear Physics, 66, 329

\bibitem[{{Shigeyama} \& {Nomoto}(1990)}]{Shigeyama1990}
{Shigeyama}, T. \& {Nomoto}, K. 1990, \apj, 360, 242

\bibitem[{{Sim}(2007)}]{Sim2007a}
{Sim}, S.~A. 2007, \mnras, 375, 154

\bibitem[{{Smartt} {et~al.}(2009){Smartt}, {Eldridge}, {Crockett}, \&
  {Maund}}]{Smartt2009b}
{Smartt}, S.~J., {Eldridge}, J.~J., {Crockett}, R.~M., \& {Maund}, J.~R. 2009,
  \mnras, 395, 1409

\bibitem[{{Sobolev}(1957)}]{Sobolev1957}
{Sobolev}, V.~V. 1957, Soviet Astronomy, 1, 678

\bibitem[{{Soderberg} {et~al.}(2008){Soderberg}, {Berger}, {Page}, {Schady},
  {Parrent}, {Pooley}, {Wang}, {Ofek}, {Cucchiara}, {Rau}, {Waxman}, {Simon},
  {Bock}, {Milne}, {Page}, {Barentine}, {Barthelmy}, {Beardmore}, {Bietenholz},
  {Brown}, {Burrows}, {Burrows}, {Byrngelson}, {Cenko}, {Chandra}, {Cummings},
  {Fox}, {Gal-Yam}, {Gehrels}, {Immler}, {Kasliwal}, {Kong}, {Krimm},
  {Kulkarni}, {Maccarone}, {M{\'e}sz{\'a}ros}, {Nakar}, {O'Brien}, {Overzier},
  {de Pasquale}, {Racusin}, {Rea}, \& {York}}]{Soderberg2008}
{Soderberg}, A.~M., {Berger}, E., {Page}, K.~L., {et~al.} 2008, \nat, 453, 469

\bibitem[{{Spyromilio} {et~al.}(1988){Spyromilio}, {Meikle}, {Learner}, \&
  {Allen}}]{Spyromilio1988}
{Spyromilio}, J., {Meikle}, W.~P.~S., {Learner}, R.~C.~M., \& {Allen}, D.~A.
  1988, \nat, 334, 327

\bibitem[{{Stritzinger} {et~al.}(2002){Stritzinger}, {Hamuy}, {Suntzeff},
  {Smith}, {Phillips}, {Maza}, {Strolger}, {Antezana}, {Gonz{\'a}lez},
  {Wischnjewsky}, {Candia}, {Espinoza}, {Gonz{\'a}lez}, {Stubbs}, {Becker},
  {Rubenstein}, \& {Galaz}}]{Stritzinger2002}
{Stritzinger}, M., {Hamuy}, M., {Suntzeff}, N.~B., {et~al.} 2002, \aj, 124,
  2100

\bibitem[{{Suntzeff} \& {Bouchet}(1990)}]{Suntzeff1990}
{Suntzeff}, N.~B. \& {Bouchet}, P. 1990, \aj, 99, 650

\bibitem[{{Swartz} {et~al.}(1995){Swartz}, {Sutherland}, \&
  {Harkness}}]{Swartz1995}
{Swartz}, D.~A., {Sutherland}, P.~G., \& {Harkness}, R.~P. 1995, \apj, 446, 766

\bibitem[{{Thielemann} {et~al.}(1996){Thielemann}, {Nomoto}, \&
  {Hashimoto}}]{Thielemann1996}
{Thielemann}, F., {Nomoto}, K., \& {Hashimoto}, M. 1996, \apj, 460, 408

\bibitem[{{Tsuruta} {et~al.}(2002){Tsuruta}, {Teter}, {Takatsuka}, {Tatsumi},
  \& {Tamagaki}}]{Tsuruta2002}
{Tsuruta}, S., {Teter}, M.~A., {Takatsuka}, T., {Tatsumi}, T., \& {Tamagaki},
  R. 2002, \apjl, 571, L143

\bibitem[{{Verner} {et~al.}(1996){Verner}, {Ferland}, {Korista}, \&
  {Yakovlev}}]{Verner1996}
{Verner}, D.~A., {Ferland}, G.~J., {Korista}, K.~T., \& {Yakovlev}, D.~G. 1996,
  \apj, 465, 487

\bibitem[{{Wanajo} {et~al.}(2009){Wanajo}, {Nomoto}, {Janka}, {Kitaura}, \&
  {M{\"u}ller}}]{Wanajo2009}
{Wanajo}, S., {Nomoto}, K., {Janka}, H.-T., {Kitaura}, F.~S., \& {M{\"u}ller},
  B. 2009, \apj, 695, 208

\bibitem[{{Wooden} {et~al.}(1993){Wooden}, {Rank}, {Bregman}, {Witteborn},
  {Tielens}, {Cohen}, {Pinto}, \& {Axelrod}}]{Wooden1993}
{Wooden}, D.~H., {Rank}, D.~M., {Bregman}, J.~D., {et~al.} 1993, \apjs, 88, 477

\bibitem[{{Woosley}(1988)}]{Woosley1988}
{Woosley}, S.~E. 1988, \apj, 330, 218

\bibitem[{{Woosley} {et~al.}(1972){Woosley}, {Arnett}, \&
  {Clayton}}]{Woosley1972}
{Woosley}, S.~E., {Arnett}, W.~D., \& {Clayton}, D.~D. 1972, \apj, 175, 731

\bibitem[{{Woosley} \& {Heger}(2007)}]{Woosley2007}
{Woosley}, S.~E. \& {Heger}, A. 2007, \physrep, 442, 269, (WH07)

\bibitem[{{Woosley} {et~al.}(2002){Woosley}, {Heger}, \&
  {Weaver}}]{Woosley2002}
{Woosley}, S.~E., {Heger}, A., \& {Weaver}, T.~A. 2002, Reviews of Modern
  Physics, 74, 1015

\bibitem[{{Woosley} \& {Weaver}(1988)}]{WW88}
{Woosley}, S.~E. \& {Weaver}, T.~A. 1988, \physrep, 163, 79

\bibitem[{{Woosley} \& {Weaver}(1995)}]{WW95}
{Woosley}, S.~E. \& {Weaver}, T.~A. 1995, \apjs, 101, 181, (WW95)

\bibitem[{{Woosley} {et~al.}(1980){Woosley}, {Weaver}, \& {Taam}}]{Woosley1980}
{Woosley}, S.~E., {Weaver}, T.~A., \& {Taam}, R.~E. 1980, in Texas Workshop on
  Type I Supernovae, ed. {J.~C.~Wheeler}, 96--112

\bibitem[{{Wright} \& {Barlow}(1975)}]{Wright1975}
{Wright}, A.~E. \& {Barlow}, M.~J. 1975, \mnras, 170, 41

\bibitem[{{Xu} \& {McCray}(1991)}]{Xu1991}
{Xu}, Y. \& {McCray}, R. 1991, \apj, 375, 190

\bibitem[{{Xu} {et~al.}(1992){Xu}, {McCray}, {Oliva}, \& {Randich}}]{Xu1992}
{Xu}, Y., {McCray}, R., {Oliva}, E., \& {Randich}, S. 1992, \apj, 386, 181

\bibitem[{{Yoon} \& {Cantiello}(2010)}]{Yoon2010}
{Yoon}, S.-C. \& {Cantiello}, M. 2010, \apjl, 717, L62

\end{thebibliography}
\bibliographystyle{/Users/andersj/latex/aa}

\end{document}